\documentclass{article}

\usepackage{amssymb}
\usepackage{graphicx}
\newcommand{\Timesa}[1]{\raisebox{-1ex}{{$\!\!\!\begin{array}{c}\xcross \\[-1.5ex]
                                            \scriptstyle #1 \end{array}\!\!\!$}}}
\newcommand{\Timesb}[2]{\raisebox{.4ex}[4ex]{{$\!\!\!\begin{array}{c}\scriptstyle #1
                                            \\[-.5ex] \xcross \\[-1.5ex]
                                            \scriptstyle #2 \end{array}\!\!$}}}
\newcommand{\Timesai}[3]{\raisebox{-.8ex}{$\!\!\!\begin{array}{c}\scriptstyle #1 \\[-.5ex]
                                            \xcross \\[-1.5ex] \scriptstyle #2 \\[-1.5ex]
                                            \scriptstyle #3 \end{array}\!\!$}}
\newcommand{\Timesbi}[2]{\raisebox{-2.2ex}{$\!\!\!\begin{array}{c} \xcross
                                            \\[-1.5ex] \scriptstyle #1 \\[-1.5ex]
                                            \scriptstyle #2 \end{array}\!\!$}}
\newcommand{\xcross}{\mbox{\begin{picture}(10,10)\thicklines
    \put(0,0){\line(1,1){10}}
    \put(0,10){\line(1,-1){10}}\end{picture}}}
\newcommand{\EV}{\mbox{\,\,\tt EV}}
\newcommand{\ED}{\mbox{\,\,\tt ED}}
\newcommand{\OD}{\mbox{\,\,\tt OD}}
\newcommand{\EVA}{\mbox{\,\,\tt EVA}}
\newcommand{\ODA}{\mbox{\,\,\tt ODA}}
\newcommand{\EDA}{\mbox{\,\,\tt EDA}}
\newcommand{\DATA}{\mbox{\tt DATA}}
\newcommand{\ACK}{\mbox{\tt ACK}}
\newcommand{\RELE}{\mbox{\tt RELE}}
\newcommand{\addsp}{\rule[-2mm]{0cm}{6mm}}    
\newcommand{\addspu}{\rule[-2mm]{0cm}{7mm}}   
\newcommand{\addspd}{\rule[-3mm]{0cm}{7mm}}   
\newcommand{\trans}[3]{\mbox{$#1$ $\stackrel{#2}{\rightarrow}$ $#3$}}
\newcommand{\vldash}{\mid \!\!\! - \!\!\! -}
\newcommand{\Trans}[3]{\mbox{$#1$ $\stackrel{#2}{\vldash}$ $#3$}}
\newcommand{\Transi}[1]{\mbox{$\stackrel{#1}{\vldash}$}}
\newcommand{\TTrans}[2]{\mbox{$#1$ $\vldash^*$ $#2$}}
\newcommand{\looongarrow}
{- \!\!\! - \!\!\! - \!\!\! - \!\!\! - \!\!\! - \!\!\! - \!\!\! - \!\!\! \longrightarrow}
\newcommand{\prot}{{\bf P}}
\newcommand{\ZZ}{{\bf Z}}
\newcommand{\TT}{{\bf T}}
\newcommand{\RR}{{\bf R}}
\newcommand{\LL}{\mbox{\rm\bf L}}
\newcommand{\WW}{\mbox{\rm\bf W}}
\newcommand{\la}{$\lambda$}
\newcommand{\tags}{\mbox{$( \Sigma, g, w_0 ) $}}
\newcommand{\und}{\mbox{$\nabla G$}}
\newcommand{\twograph}{$\; 0
    {\raisebox{.3ex}[2ex][2ex]{$\begin{array}{c} \scriptstyle\alpha\;\\[-10pt] \rightarrow\\[-10pt]
       \leftarrow\\[-10pt] \;\scriptstyle\beta \end{array}$}} 1 \;$}
\newcommand{\twographii}{$\; 0
    {\raisebox{.3ex}[2ex][2ex]{$\begin{array}{c} \scriptstyle\alpha\;\\[-10pt] \rightarrow\\[-10pt]
       \rightarrow\\[-10pt] \;\scriptstyle\beta \end{array}$}} 1 \;$}
\newcommand{\twographi}{$\; 0 \stackrel{\textstyle\rightarrow}{\leftarrow} 1 \;$}
\newcommand{\gl}{\mbox{$(S,C)$}}
\newcommand{\gli}{\mbox{$(S',C')$}}
\newcommand{\glii}{\mbox{$(S'',C'')$}}
\newcommand{\glo}{\mbox{$(S^0,C^0)$}}
\newcommand{\pstate}{\mbox{$(p_j :j \!\in\! N)$}}
\newcommand{\qstate}{\mbox{$(q_j :j \!\in\! N)$}}
\newcommand{\xchannel}{\mbox{$(x_\xi \! :\xi \!\in\! E)$}}
\newcommand{\ychannel}{\mbox{$(y_\xi \! :\xi \!\in\! E)$}}
\newcommand{\Image}[1]{\mbox{${\rm Im}_i ( #1 )$}}
\newcommand{\Imagei}[2]{\mbox{${\rm Im}_#1 ( #2 )$}}
\newcommand{\inn}{\!\in\!}
\newcommand{\qed}{\hfill $\sqcap \!\!\!\! \sqcup $\\}
\newcommand{\proof}[1]{{\bf Proof#1. }}
\newcommand{\proofnodot}[1]{{\bf Proof#1 }}

\newtheorem{theorem}{Theorem}[section]
\newtheorem{lemma}[theorem]{Lemma}
\newtheorem{corollary}[theorem]{Corollary}
\newtheorem{example}[theorem]{Example}
\newtheorem{definition}[theorem]{Definition}
\newtheorem{problem}[theorem]{Open problem}

\newcommand{\fourgraph}{\begin{picture}(280,120)(-140,-60)  
    \thicklines
    \put(-140,-10){\framebox(60,20){Process 0}}
    \put(80,-10){\framebox(60,20){Process 1}}
    \put(-30,-60){\framebox(60,20){Demon 3}}
    \put(-30,40){\framebox(60,20){Demon 2}}
    \put(-110,10){\vector(2,1){80}}
    \put(-80,30){\makebox(10,10){$\alpha$}}
    \put(30,50){\vector(2,-1){80}}
    \put(70,30){\makebox(10,10){$\beta$}}
    \put(110,-10){\vector(-2,-1){80}}
    \put(70,-40){\makebox(10,10){$\gamma$}}
    \put(-30,-50){\vector(-2,1){80}}
    \put(-80,-40){\makebox(10,10){$\delta$}}
\end{picture}}

\title{\large REACHABILITY PROBLEMS FOR COMMUNICATING \\
    FINITE STATE MACHINES*}

\author{Jan K. Pachl  \\
    Department of Computer Science \\
    University of Waterloo \\
    Waterloo, Ontario, Canada \\
    N2L 3G1 \\[8pt]
    Research Report CS-82-12}

\date{May 1982}

\begin{document}

\maketitle
\thispagestyle{empty}  
\vfill

\footnoterule
\vspace{2mm}
\noindent
* This research was supported by the
Natural Sciences and Engineering Research Council of Canada
under grant No. A7403.


\newpage
{\large\bf Table of contents} \\[8pt]

\begin{enumerate}
\item
    Introduction
\item
    Introductory examples
\item
    Communicating finite state machines
\item
    Reachability properties
\item
    Reachability analysis and abstract flow control
\item
    Affine SR-machines
\item
    Undecidable problems
\item
    Rational channels for cyclic protocols
\item
    Recognizable channels for general protocols
\item
    Abstract flow control in general graphs
\item
    Recapitulation and conclusions
\item[]
    Appendix: Post's tag systems
\end{enumerate}

\vfill

\footnoterule
\vspace{2mm}
\noindent
This is a newly formatted version of the report.
Page numbers differ from those in the original version.
Several typographical errors have been corrected.


\newpage
\section{Introduction}
    \label{sec-1}

This paper is about a state transition model for communication protocols.

The protocols governing data communication in computer systems are
becoming ever more complex, and therefore more difficult to design,
understand and analyze.
This leads a number of researchers to advocate the use of formal methods
for description and analysis of protocols \cite{bib:bo1, bib:bo2}.

State transition models are often used to describe formally
(certain aspects of) communication protocols.
This paper is concerned with a state transition model in which stations
(modelled by finite state machines) communicate by exchanging messages,
which are subjected to {\em unpredictable and unbounded delays.}
(Thus transitions in the finite state machines are loosely coupled,
in contrast to the directly coupled transitions of \cite{bib:bo2}.)
The communication channels function as {\em potentially unbounded\/}
FIFO queues.

An attractive feature of state transition models is that various
general properties (called ``syntactic properties'' in \cite{bib:zaf})
can be automatically verified if the queues (channels) are bounded.
On the other hand, Brand and Zafiropulo \cite{bib:bra} show that
the verification of the same properties {\em cannot\/} be automated
for general collections of communicating finite state machines connected
by unbounded queues.

This paper investigates the question of decidability (algorithm existence)
in some detail, and concentrates on a class of communicating finite state
machines in which certain general properties are algorithmically decidable,
although the queues are not necessarily bounded.
(Thus our goal is similar to that of \cite{bib:bra}, but our methods
and results are different.)
The technique proposed in this paper is the third stage in the following
hierarchy of formalisms for protocol description.
(All three stages will be exemplified in the next section.)

\begin{itemize}
\item
The list of all interactions.
\item
Communicating finite state machines (CFSM).
\item
CFSM augmented with channel expressions.
\end{itemize}

The paper is organized as follows:
Section 2, which is a continuation of this introduction, contains several examples.
In section 3, where the formalism begins, communicating finite state machines
(CFSM) are defined.
Section 4 lists various properties that can be defined in the CFSM model.
Section 5 introduces two basic techniques for analyzing CFSM protocols,
the exhaustive reachability analysis and abstract flow control.
Section 6 shows that certain properties of SR-machines are decidable,
although they are seemingly similar to the properties proved undecidable
in section 7.
In section 7 we shall see that most of the interesting properties
in the CFSM model are undecidable (cf. \cite{bib:bra}).
For example, there is no algorithm to decide whether a protocol is
deadlock-free.

It is then natural to ask: When can we {\em prove\/} that a protocol is deadlock-free?
A simple proof formalism is offered and investigated in sections 8, 9 and 10.
Its virtue is its simplicity, which allows straightforward automatic
proof checking.
Not every deadlock-free protocol can be proved to be deadlock-free in
the formalisms
(nor in any other formalism, in view of the undecidability result),
but the method applies to the protocols that
``use their channels in a simple manner''.
Section 8 presents a simple version of the formalism,
applicable to the protocols consisting of finite state machines arranged in a circle.
A more general theory is presented in section 9.
Section 10 generalizes the results of section 5 about abstract flow control,
and concludes with several decidability results.


\section{Introductory examples}
    \label{sec-2}

This section presents three examples to illustrate the three methods
of protocol description listed in the introduction.

\subsection{Description by listing all interactions}

A simple access authorization protocol (adapted from \cite{bib:zaf}, p. 652),
allowing only two interactions (communication histories), is depicted in
Fig. 2.1(a) and Fig. 2.1(b).

\begin{figure}[p]
\begin{center}
\begin{picture}(230,190)(-100,10)
\thicklines

\put(-110,190){\makebox(60,20)[c]{\tt \underline{Process 0}}}
\put(50,190){\makebox(60,20)[c]{\tt \underline{Process 1}}}

\put(-80,20){\line(0,1){160}}
\put(80,20){\line(0,1){160}}

\put(-80,170){\circle*{4}}
\put(-80,170){\vector(4,-1){155}}
\put(-80,170){\line(4,-1){160}}
\put(-35,155){\makebox(100,20)[l]{\tt ACCESS\_REQUEST}}
\put(80,130){\circle*{4}}

\put(80,120){\circle*{4}}
\put(80,120){\vector(-4,-1){155}}
\put(80,120){\line(-4,-1){160}}
\put(-35,105){\makebox(100,20)[l]{\tt GRANTED\_ACCESS}}
\put(-80,80){\circle*{4}}

\put(-80,70){\circle*{4}}
\put(-80,70){\vector(4,-1){155}}
\put(-80,70){\line(4,-1){160}}
\put(-40,55){\makebox(100,20)[l]{\tt RELINQUISHED\_ACCESS}}
\put(80,30){\circle*{4}}

\put(120,150){\vector(0,-1){40}}
\put(126,120){\makebox(60,20)[l]{\tt time}}
\end{picture}

\addspu {\bf Fig. 2.1(a).}
\end{center}
\end{figure}

\begin{figure}[p]
\begin{center}
\begin{picture}(230,140)(-100,60)
\thicklines

\put(-110,190){\makebox(60,20)[c]{\tt \underline{Process 0}}}
\put(50,190){\makebox(60,20)[c]{\tt \underline{Process 1}}}

\put(-80,70){\line(0,1){110}}
\put(80,70){\line(0,1){110}}

\put(-80,170){\circle*{4}}
\put(-80,170){\vector(4,-1){155}}
\put(-80,170){\line(4,-1){160}}
\put(-35,155){\makebox(100,20)[l]{\tt ACCESS\_REQUEST}}
\put(80,130){\circle*{4}}

\put(80,120){\circle*{4}}
\put(80,120){\vector(-4,-1){155}}
\put(80,120){\line(-4,-1){160}}
\put(-35,105){\makebox(100,20)[l]{\tt REFUSED\_ACCESS}}
\put(-80,80){\circle*{4}}

\put(120,150){\vector(0,-1){40}}
\put(126,120){\makebox(60,20)[l]{\tt time}}
\end{picture}

\addspu {\bf Fig. 2.1(b).}
\end{center}
\end{figure}

The description method is straightforward and easy to understand,
and a simple matching algorithm will discover deadlocks,
unspecified receptions etc. However, the protocols that allow
infinitely many (or a very large number of) communication
histories cannot be completely described.

\subsection{Description by communicating finite state machines}

Stations (processes) are represented by finite state machines whose
transitions correspond to transmissions and receptions of messages.
E.g. the protocol of Fig. 2.1 can be described as shown in Fig. 2.2
(cf. Fig. 1 in \cite{bib:zaf}).

\begin{figure}[tbp]
\begin{center}
\begin{picture}(410,270)(-205,0)
\thicklines

\put(-150,260){\makebox(40,20)[c]{\tt \underline{Process 0}}}

\put(-130,10){\circle{20}}
\put(-105,85){\circle{20}}
\put(-130,160){\circle{20}}
\put(-130,240){\circle{20}} \put(-160,240){\vector(1,0){20}}
    \put(-202,230){\makebox(40,20)[r]{\tt start}}

\put(-109,75){\vector(-1,-3){18}}
    \put(-115,35){\makebox(40,20)[l]{\tt SEND}}
    \put(-115,25){\makebox(40,20)[l]{\tt RELINQUISHED\_ACCESS}}

\put(-126,150){\vector(1,-3){18}}
    \put(-110,115){\makebox(40,20)[l]{\tt RECEIVE}}
    \put(-110,105){\makebox(40,20)[l]{\tt GRANTED\_ACCESS}}

\put(-130,230){\vector(0,-1){60}}
    \put(-124,195){\makebox(40,20)[l]{\tt SEND}}
    \put(-124,185){\makebox(40,20)[l]{\tt ACCESS\_REQUEST}}

\put(-130,150){\vector(0,-1){130}}
    \put(-204,85){\makebox(40,20)[l]{\tt RECEIVE}}
    \put(-204,75){\makebox(40,20)[l]{\tt REFUSED\_ACCESS}}

\put(110,260){\makebox(40,20)[c]{\tt \underline{Process 1}}}

\put(130,10){\circle{20}}
\put(105,85){\circle{20}}
\put(130,160){\circle{20}}
\put(130,240){\circle{20}} \put(100,240){\vector(1,0){20}}
    \put(58,230){\makebox(40,20)[r]{\tt start}}

\put(109,75){\vector(1,-3){18}}
    \put(20,35){\makebox(40,20)[l]{\tt RECEIVE}}
    \put(20,25){\makebox(40,20)[l]{\tt RELINQUISHED\_ACCESS}}

\put(126,150){\vector(-1,-3){18}}
    \put(38,115){\makebox(40,20)[l]{\tt SEND}}
    \put(38,105){\makebox(40,20)[l]{\tt GRANTED\_ACCESS}}

\put(130,230){\vector(0,-1){60}}
    \put(54,195){\makebox(40,20)[l]{\tt RECEIVE}}
    \put(54,185){\makebox(40,20)[l]{\tt ACCESS\_REQUEST}}

\put(130,150){\vector(0,-1){130}}
    \put(136,85){\makebox(40,20)[l]{\tt SEND}}
    \put(136,75){\makebox(40,20)[l]{\tt REFUSED\_ACCESS}}

\end{picture}

\addspu

{\bf Fig. 2.2.}
\end{center}
\end{figure}

Since the finite state machines can contain cycles, some protocols
that allow infinitely many message sequences can be described this way.
Deadlock-freedom and other general properties are algorithmically verifiable,
provided there is an upper bound on the number of messages that
can be simultaneously in transit.
This finiteness condition, which is far weaker than the one in 2.1,
is further substantially relaxed in 2.3 below, at the cost of
making the description more elaborate.

\subsection{Communicating finite state machines augmented by channel expressions}

This is an extension of the model in 2.2.
The protocol designer is required to provide not only the finite state
machines representing the processes,
but also a complete description of channel content for each combination of states.
In this paper we consider such a model, in which the channel content is
described by rational expressions.

{\bf Example.}
A simple alternating-bit protocol for transmission over unreliable channels
can be described as in Fig. 2.3.
There are six message types used in the protocol:
\begin{tabbing}
xxxxx\= $EV$ xxxxx\= \kill
\>\EV \>even data packet \\
\>\OD \>odd data packet \\
\>\ED \>end of data \\
\>\EVA \> acknowledgement of \EV \\
\>\ODA \> acknowledgement of \OD \\
\>\EDA \> acknowledgement of \ED
\end{tabbing}
Receptions are denoted by $+$ and transmissions by $-$.
Following the suggestion in \cite{bib:zaf},
we describe the unreliable channels by two additional finite state machines,
depicted in Fig. 2.4.
We think of all errors on the channel as being concentrated in one place,
under the control of a demon.
The rest of the channel then functions as a perfect FIFO queue.

\begin{figure}[p]
\begin{center}
\includegraphics[width=4.5in]{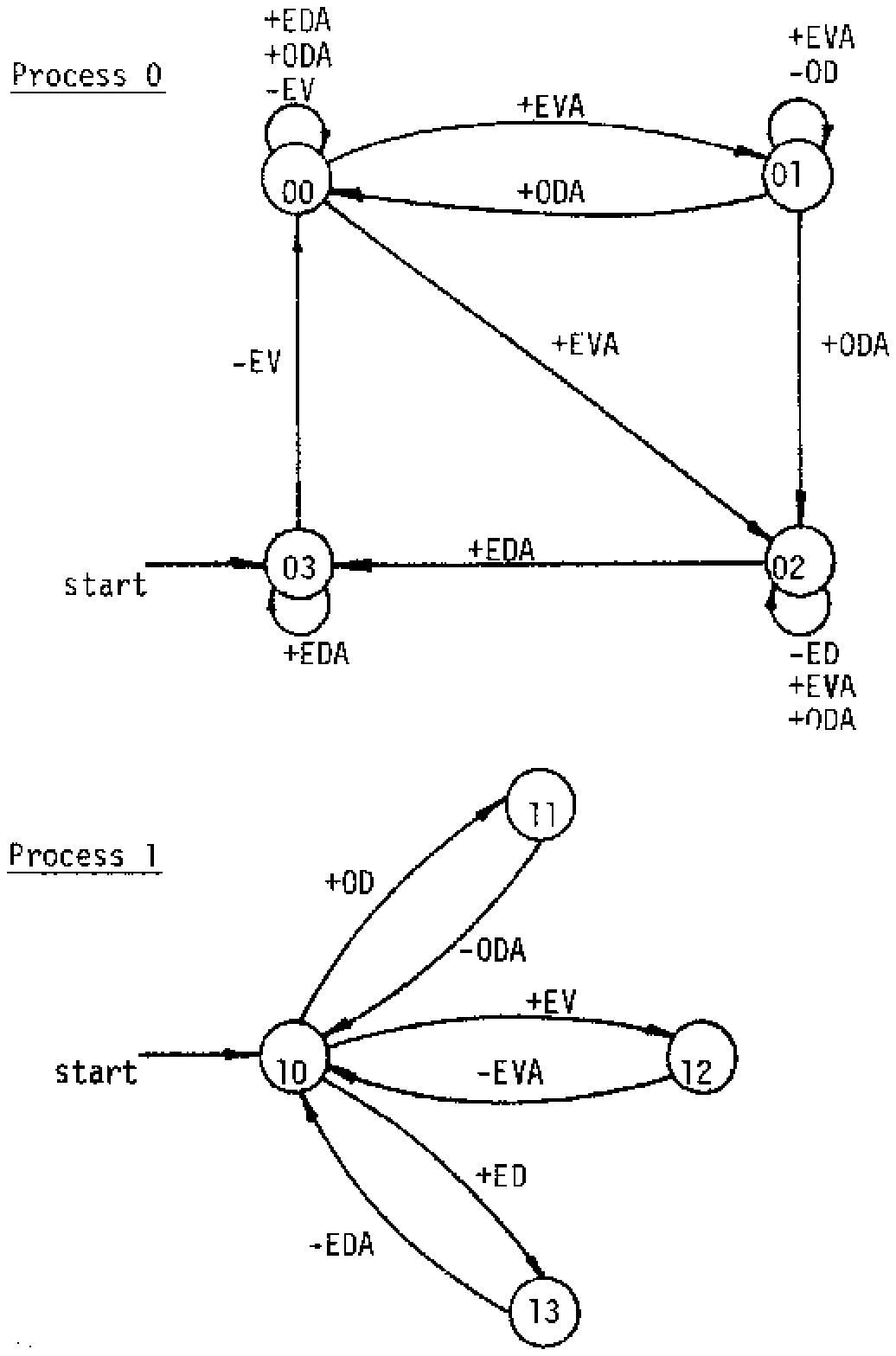}
\addspu\\[1cm]
{\bf Fig. 2.3. A simple alternating-bit protocol.}
\end{center}
\end{figure}

Fig. 2.4 makes precise what we mean by an unreliable channel:
The demon retransmits some of the messages it receives,
and ignores (deletes) others.

The complete model now consists of four finite state machines
connected by four channels, as in Fig. 2.5.

\begin{figure}[p]
\begin{center}
\includegraphics[width=4.5in]{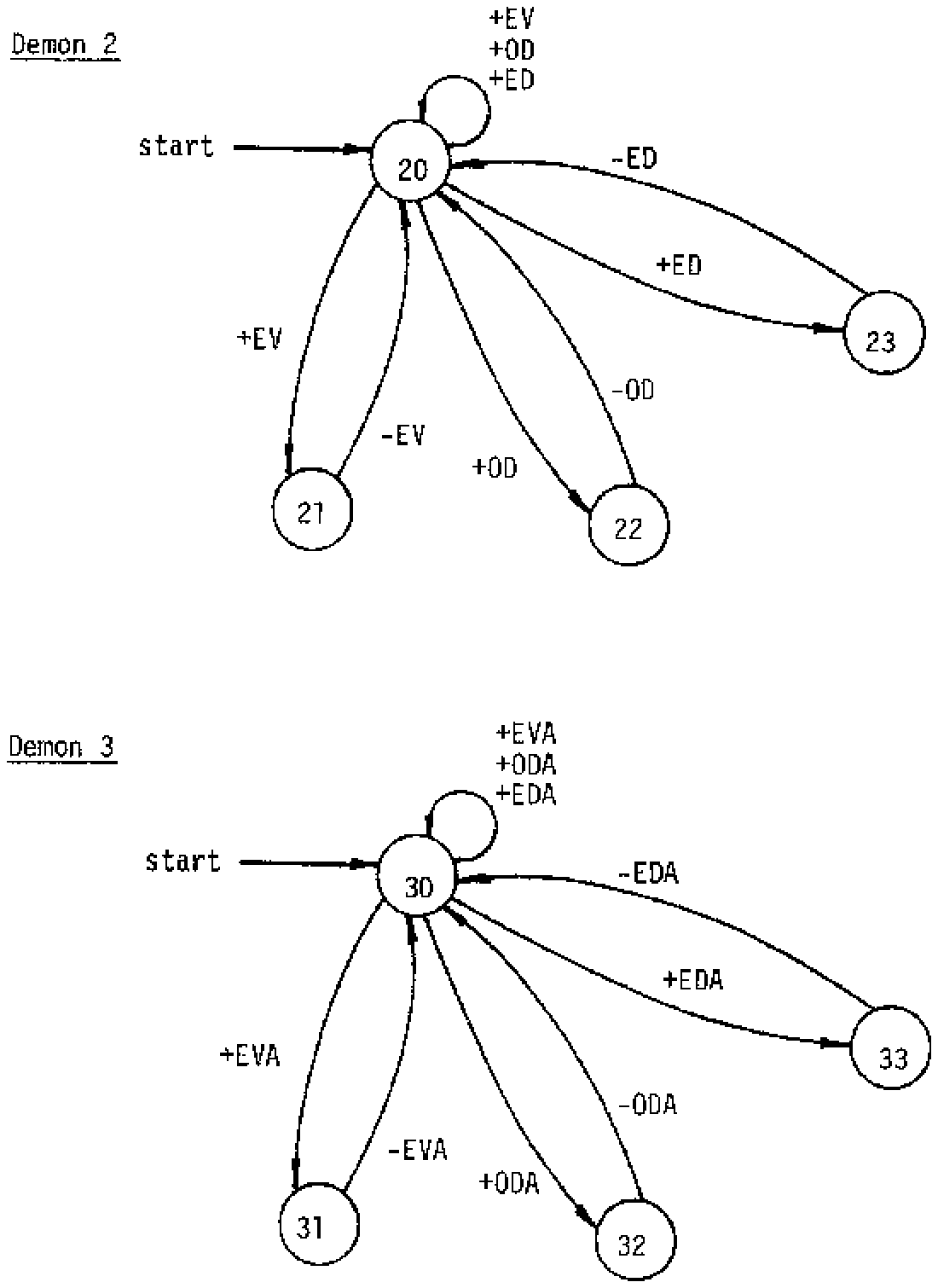}
\addspu\\[1cm]
{\bf Fig. 2.4. Unreliable channels modelled by demons.}
\end{center}
\end{figure}

\begin{figure}[htbp]
\begin{center}
\fourgraph

\addspu

{\bf Fig. 2.5. The communication graph.}
\end{center}
\end{figure}

Since Process 0 can repeatedly send the message \EV, \OD\, or \ED,
there is no upper bound on the number of messages that can be
simultaneously in transit.
Thus the description developed so far, although completely specifying
all interactions, does not easily submit to analysis.
We will aid the analysis by describing all the channel contents that
can occur for each combination of states.
Since the model has four state machines with four states each,
the additional information will be in the form of a table with
$4 \times 4 \times 4 \times 4 = 256$ entries
(one for each state combination), each entry being the set of all channel
contents that can coexist with the state combination.
As the model has four channels, a set of channel contents is a 4-ary
{\em relation\/}.
All 256 relations in this example are rational, i.e. they can be described
by rational expressions.
Fig. 2.6 lists four of the 256 relations in question, namely those for the
state combinations $00/10/20/30$, $01/10/20/30$, $02/10/20/30$ and $03/10/20/30$.
In fact, it is sufficient to specify these four entries;
the remaining 252 can be automatically computed.

In Fig. 2.6, $\ED_\alpha$ is the symbol \ED\ in the channel $\alpha$,
$\ED_\beta$ is the symbol \ED\ in the channel $\beta$, etc.
By using the subscripts we make the channel alphabets disjoint,
and avoid ambiguity in the channel expressions.

\begin{figure}[htp]
\begin{center}
\begin{tabular}{|c|l|} \hline
\addspu\addspd Composite state & \hspace{25mm}Channel contents \\ \hline\hline

\addspu $ 00/10/20/30 $ &
$ \:( \ED^*_\alpha \EV^*_\alpha \ED^*_\beta
\:\cup \EV^*_\alpha \ED^*_\beta \EV^*_\beta \:)
\EDA^*_\gamma \EDA^*_\delta \:\:\cup $ \\
\addsp &
$ \EV^*_\alpha \EV^*_\beta \:( \EDA^*_\gamma \EVA^*_\gamma \EDA^*_\delta
\:\cup \EVA^*_\gamma \EDA^*_\delta \EVA^*_\delta \:) \:\:\cup $  \\
\addsp &
$ \:( \OD^*_\alpha \EV^*_\alpha \OD^*_\beta \:\cup
\EV^*_\alpha \OD^*_\beta \EV^*_\beta \:) \ODA^*_\gamma \ODA^*_\delta \:\:\cup $ \\
\addspd &
$ \EV^*_\alpha \EV^*_\beta \:( \ODA^*_\gamma \EVA^*_\gamma \ODA^*_\delta \:\:\cup
\EVA^*_\gamma \ODA^*_\delta \EVA^*_\delta \:) $  \\ \hline

\addspu $ 01/10/20/30 $ &
$ \:( \EV^*_\alpha \OD^*_\alpha \EV^*_\beta \:\cup
\OD^*_\alpha \EV^*_\beta \OD^*_\beta \: ) \EVA^*_\gamma \EVA^*_\delta \: \cup $ \\
\addspd &
$ \OD^*_\alpha \OD^*_\beta \: ( \EVA^*_\gamma \ODA^*_\gamma \EVA^*_\delta \: \cup
\ODA^*_\gamma \EVA^*_\delta \ODA^*_\delta \:) $ \\ \hline
\addspu $ 02/10/20/30 $ &
$ \:( \EV^*_\alpha \ED^*_\alpha \EV^*_\beta \:\cup
\ED^*_\alpha \EV^*_\beta \ED^*_\beta \:) \EVA^*_\gamma \EVA^*_\delta \:\:\cup $ \\
\addsp &
$ \ED^*_\alpha \ED^*_\beta \:( \EVA^*_\gamma \EDA^*_\gamma \EVA^*_\delta \:\cup
\EDA^*_\gamma \EVA^*_\delta \EDA^*_\delta \:) \:\:\cup $ \\
\addsp &
$ \:( \OD^*_\alpha \ED^*_\alpha \OD^*_\beta \:\cup
\ED^*_\alpha \OD^*_\beta \ED^*_\beta \:) \ODA^*_\gamma \ODA^*_\delta \:\:\cup $ \\
\addspd &
$ \ED^*_\alpha \ED^*_\beta \:( \ODA^*_\gamma \EDA^*_\gamma \ODA^*_\delta \:\cup
\EDA^*_\gamma \ODA^*_\delta \EDA^*_\delta \:) $  \\ \hline

\addspu\addspd $ 03/10/20/30 $ &
$ \ED^*_\alpha \ED^*_\beta \EDA^*_\gamma \EDA^*_\delta $ \\ \hline
\end{tabular}

\addspu

{\bf Fig. 2.6. Rational expressions for channel contents.}
\end{center}
\end{figure}


\section{Communicating finite state machines}
\label{sec-3}

The present paper treats communicating finite state machines as mathematical objects.
They are formally defined in this section.
The formalism is fairly close to that in \cite{bib:bra}.

A {\em directed graph\/} is a pair $ G = ( N, E ) $ where $N$ and $E$ are two sets
(the set of {\em nodes\/} and the set of {\em edges\/}),
together with two maps, denoted $ \xi \mapsto - \xi$ and $ \xi \mapsto + \xi $,
from $E$ to $N$.
We say that $-\xi$ is the {\em tail\/} and $+\xi$ the {\em head\/}
of the edge $\xi$;
when $ i = -\xi $ and $j= +\xi$, we sometimes write \trans{i}{\xi}{j}.
We say that $G$ is finite if both $N$ and $E$ are finite.

A {\em protocol\/} (or, more explicitly, a {\em CFSM protocol\/}) \prot\ consists of
a finite directed graph $ G = ( N , E ) $
(the {\em communication graph\/} of \prot),
a collection of pairwise disjoint finite sets $ M_\xi $ indexed by $ \xi \inn E $,
and a collection of finite state machines $F_j$ indexed by $ j \inn N$.
Each $F_j$ operates over the alphabet
\[
\Sigma_j = \{ \; +b \;\; | \;\; b \inn M_\xi , \;j = +\xi \;\} \cup
    \{ \; -b \;\; | \;\; b \inn M_\xi , \; j = - \xi \; \} \; .
\]
Specifically, $ F_j = ( K_j , \Sigma_j, T_j, h_j ) $ where $ K_j$ is the finite
{\em set of states\/}, $ h_j \inn K_j $ is the {\em initial\/}
(or {\em home\/}) {\em state\/}, and
$ T_j \subseteq K_j \times \Sigma_j \times K_j $ is the {\em set of transitions\/}.

We write \trans{p}{e}{q} in $F_j$ (or simply \trans{p}{e}{q}, if no
misunderstanding is possible) when $ (p, e, q) \inn T_j $.
(Here $ e = -b $ or $ e = +b $, for some $\xi$ and $ b \inn M_\xi$.)
The {\em transition diagram of\/} $F_j$ is the labelled directed graph
with nodes $K_j$ and labelled edges \trans{p}{e}{q} for $(p,e,q) \inn T_j$.

Write \trans{p}{w}{q}, for $ w \inn \Sigma^*_j $, if there is a directed path
from $p$ to $q$, in the transition diagram of $F_j$,
such that the labels on the edges of the path form the string $w$
(in the order from $p$ to $q$).
Sometimes we write \trans{p}{e}{} instead of ``\trans{p}{e}{q} for some $q$'',
and similarly \trans{p}{w}{} for $ w \inn \Sigma^*_j $.

The model corresponds to reality in this way:
The graph $G$ describes the protocol configuration
(the edges are unidirectional communication channels);
we say that the machines $F_j$ in \prot\ {\em communicate according to $G$}.
The set $M_\xi$ is the set of messages that can be sent along the channel $\xi$
(in practice these sets need not be disjoint, but the assumption that they are
causes no loss generality and is technically useful).
The machine $F_j$ represents a process located at $j \inn N$ and capable
of sending messages to the channels $\xi$ such that $j=-\xi$ and of receiving
messages from the channels $\xi$ such that $j=+\xi$.
Message transmissions and receptions match transitions in the state machines:
\trans{p}{+b}{q} in $F_j$ means $ b \inn M_\xi$ received, and
\trans{p}{-b}{q} in $F_j$ means $ b \inn M_\xi$ sent (at $j \inn N$ along $\xi \inn E$).

In the sequel we shall have an opportunity to deal with CFSM protocols
of a special form, the SR-machines of Gouda \cite{bib:gou}:

A state $ p \inn K_j $ is a {\em send state\/} if \trans{p}{+b}{} for no $b$;
similarly $p$ is a {\em receive state\/} if \trans{p}{-b}{} for no $b$.
Say that $F_j$ is an {\em SR-machine\/} if

(a) $K_j$ has only send and receive states,

(b) the transition diagram of $F_j$ is strongly connected, and

(c) if \trans{p}{e}{q_1} and \trans{p}{e}{q_2} in $F_j$ then $q_1 = q_2$.

\noindent
A {\em pair of communicating SR-machines\/} is a protocol with two SR-machines
$F_0$ and $F_1$ communicating according to the graph \twograph \,
(i.e. $ N = \{ 0, 1 \} $, $ E = \{ \alpha , \beta \} $,
$ - \alpha = + \beta = 0 $ and $ + \alpha = - \beta = 1 $ ).

Other variations of communicating finite state machines have been employed to
describe and analyze communication protocols, but the differences between them are not
essential in the present context.
The popularity of the model stems from the fact that,
while being simple and abstract, it is rich enough to embrace
some general communication properties (sometimes called syntactic properties).
Several such properties are enumerated in the next section.
They are all defined in terms of the global state space,
which we now proceed to describe.

In our basic model, we assume that the channels function as perfect FIFO queues.
That is, they are error-free (imperfect channels are modelled indirectly, by demons),
and in each channel messages are received in the same order as sent.
We place no a priori bound on the queue lengths;
the intention is to model unpredictable and unbounded communication delays.

Let \prot\ be a CFSM protocol, with the notation as above.
A {\em composite state\/} of \prot\ is a vector $ S = \pstate $ of states
$ p_j \inn K_j $.
A {\em channel content\/} (or ``composite channel state'') is a vector
$ C = \xchannel $ of strings $ x_\xi \inn M^*_\xi $
(each $x_\xi$ is a string over the alphabet $M_\xi$).
A {\em global state\/} is a pair \gl\ where $S$ is a composite state and
$C$ is a channel content.
The {\em initial global state\/} is \glo\ where
$ S^0 = ( h_j\, : \, j\inn N ) $ and $ C^0 = ( x_\xi \, : \, \xi \inn E ) $
with each $x_\xi $ being the empty string $\lambda$.

Our aim is to define a labelled directed graph whose nodes will be
global states of \prot\ and which will have two kinds of labelled edges
(write $S = \pstate$, $S' = \qstate$, $C = \xchannel$, $C' = \ychannel$):

(1) (Receive from channel $\beta$)
\[
\Trans{\gl}{+b}{\gli}
\]
if there are $i$ and $\beta$ with $i=+\beta$, such that
$p_j=q_j$ for $j\neq i$, \trans{p_i}{+b}{q_i} in $F_i$,
$x_\xi = y_\xi$ for $\xi \neq \beta$, and $x_\beta = b y_\beta$.

(2) (Send to channel $\beta$)
\[
\Trans{\gl}{-b}{\gli}
\]
if there are $i$ and $\beta$ with $i=-\beta$, such that
$p_j=q_j$ for $j\neq i$, \trans{p_i}{-b}{q_i} in $F_i$,
$x_\xi = y_\xi$ for $\xi \neq \beta$, and $y_\beta = x_\beta b$.

Write \Trans{\gl}{}{\gli} if \Trans{\gl}{+b}{\gli} or \Trans{\gl}{-b}{\gli}
for some $b$.
Let $\vldash^*$ be the reflexive and transitive closure of $\vldash$.
Say that a global state \gli\ is {\em reachable\/} from a global state \gl\ if
\TTrans{\gl}{\gli}.

Say that a global state is {\em reachable\/} if it is reachable from \glo.
The {\em global state space\/} of the protocol \prot\ is the labelled directed
graph whose nodes are all the reachable global states of \prot,
with labelled edges $\stackrel{+b}{\vldash}$ and $\stackrel{-b}{\vldash}$
defined above.


\section{Reachability properties}
    \label{sec-4}

The general reachability problem, in its simplest form, is
``Given a (possibly infinite) directed graph and two of its nodes,
can one node be reached from the other along a path in the graph?''
One may wish to construct an algorithm to answer the question;
this leads to a decidability problem:
Is there an algorithm to decide, for any given graph and two nodes,
whether one can be reached from the other?
In other words, is the reachability problem (algorithmically) decidable?

Algorithms to solve two problems of this kind have been found recently,
after a prolonged research effort:
Kannan and Lipton \cite{bib:kan} constructed an algorithm to solve
Harrison's orbit problem,
and Mayr \cite{bib:may} constructed an algorithm for the Petri net
reachability problem.
The CFSM model brings up another reachability problem, which is, unlike
the previous two, undecidable (see section 7).
However, it is worthwhile to investigate restrictions on the problem
that make it decidable; this is the chief subject of the present paper.

In fact, there is not one but a number of reachability problems in the CFSM model.
The (possibly infinite) directed graph where they all reside is the global
state space defined in the previous section.

A {\em simple reachability problem\/}
(or a reachability problem of the first order) has the form
``Is a given global state reachable (from \glo)?''
For example, the problem of finding stable composite states
can be treated as a simple reachability problem:
A composite state $S$ is called {\em stable\/} if $(S,C^0)$
is reachable; cf. \cite{bib:zaf}, \cite{bib:bra}.
Since there are only finitely many composite states, the problem
of listing all stable ones is solved by answering finitely many
simple reachability problems.

A global state \gl\ is said to be {\em deadlocked\/}
if every state in $S$ is a receive state and $C= C^0$.
The protocol \prot\ is {\em deadlock-free\/} if no deadlocked
global state is reachable.
The question whether \prot\ is deadlock-free is a special case
of the stable composite state problem in the previous paragraph.

However, there are other pertinent reachability problems that are not
simple in this sense (or at least it is not immediately obvious
if they are).
Say that $ b \inn M_\beta$ {\em can arrive at the state $p\inn K_i$}
if $i=+\beta$ and there is a reachable global state $(\pstate,\xchannel)$
such that $p=p_i$ and $x_\beta = b y_\beta$ for some $y_\beta \inn M^*_\beta$.
The problem of finding all pairs $(p,b)$ such that $b$ can arrive at $p$
(``executable receptions'' in the terminology of \cite{bib:bra})
is of the form ``Is at least one element of a given {\em set\/}
of global states reachable?''
Let us call this a {\em second order reachability problem.}
Of course, the set of global states in question can be described in various ways;
that can make the problem more or less difficult (or even
decidable or undecidable).

Say that a global state \gl\ is {\em globally blocked\/}
if \Trans{\gl}{}{\gli} for no global state \gli.
(Every deadlocked global state is globally blocked but not vice versa.)

A global state $ \gl = ( S, \xchannel ) $ is {\em blocked on channel $\beta \inn E$}
if $x_\beta = b y $, $ b \inn M_\beta$, $y \inn M^*_\beta$,
and there are no global states \gli\ and \glii\ satisfying
\[
\gl \vldash^* \Trans{\gli}{+b}{\glii} \; .
\]

The property that no reachable global state is blocked on any channel
(that is, every transmitted message can be {\em eventually\/} received)
should be compared with the following stronger property,
defined in \cite{bib:bra}.
The protocol is {\em well-formed\/} if for any $p \inn K_j$ and $b \inn M_\beta$
we have:
$b$ can arrive at $p$ if and only if \trans{p}{+b}{} in $F_j$.
This means that the protocol is able to receive every message {\em immediately\/}
upon arrival and, moreover, the transition diagram of $F_j$ has no useless edges.

Another example of a second order reachability property:
A protocol with the communication graph \twograph\ is said to have the
{\em half-duplex property\/} if every reachable global state
$ (( p_0 , p_1 ) , ( x_\alpha , x_\beta )) $ satisfies
$ x_\alpha = \lambda $ or $ x_\beta = \lambda $.

Finally, certain useful reachability properties are neither first nor
second order.
The protocol \prot\ has the {\em bounded channel property\/}
if there is an upper bound on the total length of all strings in $C$,
over all reachable global states \gl.
Obviously \prot\ has this property if and only if the global state space is finite.

We can see that, although the CFSM model is very simple and general,
it allows us to formulate a number of meaningful protocol properties.
Moreover, the properties are all described in a uniform manner,
as reachability properties in a certain (potentially infinite) graph.
The next question is whether the properties can be algorithmically decided.
In this paper we concentrate on the deadlock problem
(``Is the protocol deadlock-free?''), and the stable composite state problem
(``Is a given composite state stable?''), two representatives of simple
(first order) reachability problems.
Occasionally we also note how the results apply to other reachability problems.


\section{Reachability analysis and abstract flow control}
    \label{sec-5}

When the global state space is finite, all reachability problems can be,
at least in principle, algorithmically solved.
Indeed, one can explicitly construct the global state space
(as a finite directed graph) and search it to decide any reachability problem.
We refer to this method as the {\em exhaustive reachability analysis.}

The method presents a number of implementation and complexity problems,
because the global state space tends to be very large and exhaustive search
is expensive.
Nevertheless, the question of algorithm {\em existence\/} is,
in the case when the global state space is finite, uninteresting:
All problems are decidable for trivial reasons.
The chief aim of this paper is to investigate what can be done
when the global state space is not (or is not known to be) finite.

The global state space has a highly redundant structure.
Concurrent execution is modelled by a set of shuffles of sequential
executions in the participating nodes.
Thus if one global state is reachable from another then there are usually
many paths between them.
We can reduce the redundancy by restricting the order in which concurrent
transmissions and receptions occur.
This is the idea of the {\em abstract flow control.\/}
Its special case was studied (under a different name) by Rubin and West \cite{bib:rub}.

Every path in the global state space defines ``local paths'' in the transition
diagrams of the individual state machines.
These will be called the {\em images\/} of the global path.
In the notation of section~\ref{sec-3}, the image can be defined formally.
Let $ \Gamma = (S_0 , C_0 ) \Transi{e_1} \ldots \Transi{e_k} ( S_k , C_k ) $
be a path in the global state space, and let $ i \inn N$.
If $k=0$ (i.e. the length of $\Gamma$ is 0) and $ S_0 = \pstate $ then
\Image{\Gamma}, the image of $\Gamma$ in $F_i$, is the path of length 0
from $p_i$ to $p_i$.
If $ k > 0 $, $ S_{k-1} = \pstate $ and $ S_k = \qstate $ then \Image{\Gamma}
is defined in terms of
$ \Gamma' = (S_0 , C_0 ) \Transi{e_1} \ldots \Transi{e_{k-1}} ( S_{k-1} , C_{k-1} ) $
as follows:
If $ e_k \not\in \Sigma_i $ then $ \Image{\Gamma} = \Image{\Gamma'} $;
if $ e_k \inn \Sigma_i $ then $ \Image{\Gamma} $ is the concatenation
of $ \Image{\Gamma'} $ with the path \trans{q_i}{e_k}{p_i} (of length 1).

Say that two paths $\Gamma$ and $\Gamma'$ in the global state space are
{\em locally equal\/} if $ \Image{\Gamma} = \Image{\Gamma'} $ for each $ i \inn N $.
The following self-evident lemma is a basis of most that follows.

\begin{lemma}
\label{res-5-1}
If two paths in the global state space are locally equal and start
in the same global state, then they also terminate in the same global state.
\end{lemma}

The aim of the abstract flow control, in the sense used in this paper,
is to reduce the number of the locally equal paths that the reachability
analysis must examine.
Rubin and West \cite{bib:rub} have shown how to select exactly one path in every
set of locally equal paths, in the special case of two-party protocols and paths
between global states of the form $(S , C^0 )$.
The problem can be viewed as a scheduling problem:
For a given path $\Gamma$ in the global state space, the local images of $\Gamma$
are concurrent sequential processes which must share a single processor.
In this terminology, the Rubin and West method uses the round-robin scheduling.
The methods explored in this paper are based on priority scheduling.
They yield particularly simple results when the finite state machines are
arranged in a circle;
to have a short name for such CFSM protocols, we say that a protocol is {\em cyclic\/}
if its communication graph is a directed cycle.

\begin{theorem}
    \label{res-5-2}
\samepage
Let \prot\ be a cyclic CFSM protocol.
Let $\Gamma$ be a path in the global state space from a global state
$(S, C^0 )$ to a global state $ ( S', C^ 0 ) $.
If $\beta$ is any edge in $E$ then there exists a path $\Gamma'$ such that \\
(a) $\Gamma$ and $\Gamma'$ are locally equal; and \\
(b) every global state $ ( S , \xchannel ) $ on the path $\Gamma'$ satisfies
\[
\sum_{\stackrel{\scriptstyle \xi \in E}{\scriptstyle \xi \neq \beta}} \; | x_\xi | \; \leq \; 1 \; .
\]
\end{theorem}

A more general result will be proved in section 10.

\proof{}
Label the edges of the communication graph $G$ as
$ E = \{ \alpha_0 , \alpha_1 , \ldots , \alpha_m \} $
and assume that $ - \alpha_0 = + \alpha_m$,
$ - \alpha_1 = + \alpha_0$,
$ - \alpha_2 = + \alpha_1$,
$\ldots$,
$ \beta = \alpha_0 $:

\begin{center}
\begin{picture}(90,80)
\thicklines
\put(24,72){\vector(1,0){24}}
\put(52,72){\vector(1,-1){20}}
\put(72,48){\vector(0,-1){24}}
\put(72,20){\vector(-1,-1){20}}
\put(0,24){\vector(0,1){24}}
\put(0,52){\vector(1,1){20}}
\put(12,12){\circle*{2}}
\put(20,6){\circle*{2}}
\put(28,0){\circle*{2}}
\put(36,79){\makebox(0,0){$\alpha_0$}}
\put(70,64){\makebox(0,0){$\alpha_1$}}
\put(82,36){\makebox(0,0){$\alpha_2$}}
\put(2,64){\makebox(0,0){$\alpha_m$}}
\end{picture}
\end{center}

Rearrange the execution described by $\Gamma$ as follows:
Assign the highest priority to the process running at the node $+\alpha_m$,
the next highest to the process at $+\alpha_{m-1}$, etc.,
with the lowest priority at $+\alpha_0$.
Thus a process can execute only if all processes with higher priorities are blocked
(which means that their local images of $\Gamma$ call for receptions
and their input channels are empty).
Let $\Gamma'$ be the path corresponding to the priority execution.
It follows that at most one among the channels
$\alpha_1$, $\alpha_2$, $\ldots$, $\alpha_m$ is non-empty at any point along
$\Gamma'$, and that none can grow longer than one symbol.
\qed

Theorem~\ref{res-5-2} (as well the more general results to come)
simplifies the reachability algorithm.
When looking for a deadlock, the algorithm can ignore the global states
in which $ \sum_{\xi \neq \beta} | x_\xi | > 1 $.
The following immediate corollary of Theorem~\ref{res-5-2} generalizes
a result of Brand and Zafiropulo \cite{bib:bra}.

\begin{corollary}
The stable composite state problem is decidable in the class of all cyclic
CFSM protocols with this property:
There is an edge $\beta$ and a constant $c$ such that every reachable global state
$ ( S, \xchannel ) $ satisfies $ | x_\beta | \leq c $.
\end{corollary}

It follows that deadlock-freedom is also decidable in this class.


\section{Affine SR-machines}
    \label{sec-6}

In this section we are going to see that certain properties of a pair
of communicating SR-machines are algorithmically decidable,
although they are superficially similar to the undecidable properties
that we shall encounter later on.

Let \prot\ be a CFSM protocol consisting of two SR-machines
$ F_0 = ( K_0 , \Sigma_0, T_0 , h_0 ) $ and
$ F_1 = ( K_1 , \Sigma_1, T_1 , h_1 ) $ communicating according to the graph
\twograph.
Recall that
\[
\Sigma_0 \; = \; \{ \; -b \; | \; b \inn M_\alpha \;\} \;\cup\; \{ \; +b \; | \; b \inn M_\beta \;\}
\]
and
\[
\Sigma_1 \; = \; \{ \; -b \; | \; b \inn M_\beta \;\} \;\cup\; \{ \; +b \; | \; b \inn M_\alpha \;\}
\]
If $w$ is a string in $\Sigma^*_0$ or $\Sigma^*_1$, denote by $\pi_\alpha(w)$
the string of all $M_\alpha$ symbols in $w$, in the same order;
thus $\pi_\alpha$ erases all the symbols in $w$ that belong to $M_\beta$, and also all $+$ and $-$
($\pi_\alpha$ is the ``projection'' from $ \Sigma^*_0 \cup \Sigma^*_1 $ onto $M^*_\alpha$).
The projection $\pi_\beta$ onto $M^*_\beta$ is defined similarly.
For example, if $d_1 , d_2 \inn M_\alpha$ and $b_1 , b_2 \inn M_\beta$ then
$ \pi_\alpha ( +d_1 +d_2 -b_1 +d_1 -b_2 -b_2 ) = d_1 d_2 d_1 $ and
$ \pi_\beta ( +d_1 +d_2 -b_1 +d_1 -b_2 -b_2 ) = b_1 b_2 b_2 $.

The machine $F_0$ defines a subset $\ZZ_0$ of $ M^*_\alpha \times M^*_\beta$ :
\[
\ZZ_0 \; = \; \{ \; (\pi_\alpha ( w ) , \pi_\beta ( w ) ) \; |
\; \trans{h_0}{w}{h_0} \;\; \mbox{\rm in} \;\; F_0 \; \} \; .
\]
Similarly,
\[
\ZZ_1 \; = \; \{ \; (\pi_\alpha ( w ) , \pi_\beta ( w ) ) \; |
\; \trans{h_1}{w}{h_1} \;\; \mbox{\rm in} \;\; F_1 \; \} \; .
\]
Say that $F_0$ and $F_1$ are {\em affine\/} (or that the protocol \prot\ is affine)
if $ \ZZ_0 = \ZZ_1 $.

Thus two SR-machines are affine if and only if for every sequence of sends and receives
(beginning and ending in the ``home state'') in one machines there is a matching sequence
in the other.
However, the matching is a weak one because, intuitively, it allows a symbol to be received
before it has been sent.

There are interesting connections between affinity and certain desirable protocol properties.
At the same time, unlike the other properties, affinity is decidable;
a minor modification of Bird's algorithm \cite{bib:bir} establishes the following result.

\begin{theorem}
    \label{res-6-1}
There is an algorithm to decide whether an arbitrary pair of SR-machines is affine.
\end{theorem}

Now we consider the bounded channel property for affine SR-machines.
No protocol in which at least one machine can go through a cycle consisting
of send transitions has the bounded channel property;
the machine can repeat the sending cycle any number of times before the other
machine begins receiving.
The forthcoming theorem shows that for affine SR-machines the channel can grow
large {\em only\/} if there is such a cycle.

Say that a state machine $F_j$ has a {\em send cycle\/} if the transition diagram of $F_j$
contains a directed cycle whose all labels are negative
(i.e. of the form $-b$, $b \inn M_\xi$, $j=-\xi$);
a {\em receive cycle\/} is defined analogously.

\begin{lemma}
   \label{res-6-2}
Let $F_0$ and $F_1$ be two affine SR-machines.
For $j=0,1$, let $k_j$ be the number of states in $F_j$
(=the cardinality of $K_j$).
If there is a reachable global state $((p_0 , p_1), (x_\alpha , \lambda ))$
such that $ | x_\alpha | \geq k_0 ( k_1 - 1 ) + 1 $ then $F_1$ has a receive cycle
and $F_0$ has a send cycle.
\end{lemma}

This yields a new automatically verifiable {\em sufficient\/} condition for bounded
channels, namely affinity and absence of send cycles;
cf. \cite{bib:bra} and \cite{bib:gou} for other conditions of this kind.
The condition is also {\em necessary\/} if the protocol is affine and deadlock-free:

\begin{theorem}
    \label{res-6-3}
Let $F_0$ and $F_1$ be two affine SR-machines.
If the protocol is deadlock-free then it has the bounded channel property if and only if
neither $F_0$ nor $F_1$ has a send cycle.
\end{theorem}

\begin{theorem}
    \label{res-6-4}
There is an algorithm to decide, for an arbitrary given pair of affine SR-machines,
whether the protocol is deadlock-free and has the bounded channel property.
\end{theorem}

Another corollary of \ref{res-6-2}, to be proved later in this section,
is the following:

\begin{theorem}
    \label{res-6-5}
There is an algorithm to decide, for an arbitrary given pair of SR-machines with no send cycles,
whether the protocol is affine and deadlock-free.
\end{theorem}

Now we prove the results in this section.
Recall that we deal with a protocol \prot\ with the communication graph \twograph
and two SR-machines
$ F_j = ( K_j , \Sigma_j , T_j , h_j ) , j = 0,1$.
The channel alphabets are $M_\alpha$ and $M_\beta$.

\proof{ of \ref{res-6-2}}
Recording how the global state $ ( (p_0 , p_1 ) , (x_\alpha , \lambda ) ) $
has been reached,
we find two strings $ w_0 \inn \Sigma^*_0 $ and $ w_1 \inn \Sigma^*_1 $
such that \trans{h_0}{w_0}{p_0}, \trans{h_1}{w_1}{p_1},
$\pi_\alpha ( w_0 ) = \pi_\alpha ( w_1 ) x_\alpha $ and
$ \pi_\beta ( w_0 )= \pi_\beta ( w_1 ) $.
Since the transition graph of $F_0$ is strongly connected and has $k_0$ nodes,
\trans{p_0}{u_0}{h_0} for some $u_0 \inn \Sigma^*_0 $ such that $ | u_0 | \leq k_0 - 1 $.

By affinity, \trans{p_1}{u_1}{h_1} for some $ u_1 \inn \Sigma^*_1 $ such that
$ \pi_\alpha ( w_0 u_0 ) = \pi_\alpha ( w_1 u_1 ) $ and
$ \pi_\beta ( w_0 u_0 ) = \pi_\beta ( w_1 u_1 ) $.
This yields
$ \pi_\alpha ( u _1 ) = x_\alpha \pi_\alpha ( u_0 ) $
and $ \pi_\beta ( u _1 ) =  \pi_\beta ( u_0 ) $.
Therefore $u_1$ contains at most $ | u_0 | \leq k_0 - 1 $ symbols of the form
$-b$, $b \inn M_\beta$.
At the same time, the length of $ \pi_\alpha ( u_1 ) $ is
\[
| \pi_\alpha ( u_1 ) | \geq | x_\alpha | \geq k_0 ( k_1 - 1 ) + 1 \; ,
\]
and hence $u_1$ contains a (contiguous) subsequence $v_1$ of length
$ | v_1 | \geq k_1 $ that has no symbols $-b$, $b \inn  M_\beta$.
Since $F_1$ has $k_1$ states, the path corresponding to $v_1$ contains a cycle.
Hence $F_1$ has a receive cycle.

The second assertion in \ref{res-6-2} now follows from the following lemma.

\begin{lemma}
    \label{res-6-6}
Let $F_0$ and $F_1$ be two affine SR-machines.
If $F_1$ has a receive cycle then $F_0$ has a send cycle.
\end{lemma}

\proof{ of \ref{res-6-6}}
Again let $k_j$ be the cardinality of $K_j$, for $j=0,1$.
Since the transition diagram of $F_1$ is strongly connected and has a receive cycle,
\trans{h_1}{w_1}{h_1} for some $ w_1 \inn \Sigma^*_1 $ such that
$ | \pi_\beta ( w_1 ) | \leq 2 ( k_1 - 1 ) $ and
$ | \pi_\alpha ( w_1 ) | \geq ( 2 k_1 - 1 ) ( k_0 - 1 ) + 1 $.
By affinity, \trans{h_0}{w_0}{h_0} for some $w_0 \inn \Sigma^*_0$ such that
$ \pi_\alpha ( w_0 ) = \pi_\alpha ( w_ 1 ) $ and
$ \pi_\beta ( w_0 ) = \pi_\beta ( w_ 1 ) $.
It follows that $w_0$ contains a substring $v_0$ of length $ | v_0 | \geq k_0 $
that has no symbol $+b$.
Since $F_0$ has $k_0$ states, the path corresponding to $v_0$ contains a cycle;
hence $F_0$ has a send cycle.
\qed

The proof of \ref{res-6-3} uses the following two lemmas.

\begin{lemma}
    \label{res-6-7}
If there is a reachable global state $ ( ( p_0 , p_1 ) , ( x_\alpha , x_\beta ) ) $
with $ | x_\alpha | \geq k $ then there is a reachable global state
$ ( ( p_0 , q_1 ) , ( y_\alpha , \lambda ) ) $ with $ | y_\alpha | \geq k $.
\end{lemma}

\begin{lemma}
    \label{res-6-8}
If the pair of affine machines is deadlock-free then for any $w_0$
such that \trans{h_0}{w_0}{p_0} there exists a path in the global state space,
starting in \glo, whose image in $F_0$ is labelled $w_0$.
\end{lemma}

\proof{ of \ref{res-6-7}}
There are \trans{h_0}{w_0}{p_0} and \trans{h_1}{w_1}{p_1} such that
$ \pi_\alpha ( w_0 ) = \pi_\alpha ( w_1 ) x_\alpha $ and
$ \pi_\beta ( w_0 ) x_\beta = \pi_\beta ( w_1 )$.
Find a prefix $v_1$ of $w_1$ such that
$ \pi_\beta ( w_1 ) = \pi_\beta ( v_1 ) x_\beta $.
We have \trans{h_1}{v_1}{q_1} for some $ q_1 \inn K_1 $.
Since $ \pi_\alpha ( v_1 ) $ is a prefix of $ \pi_\alpha ( w_1 ) $,
we can write $ \pi_\alpha ( w_1 ) = \pi_\alpha ( v_1 ) y'_\alpha $
for some $ y'_\alpha \inn M^*_\alpha $.
Set $ y_\alpha = y'_\alpha x_\alpha $;
then $ \pi_\beta ( w_0 ) = \pi_\beta ( v_1 ) $ and
$ \pi_\alpha ( w_0 ) = \pi_\alpha ( v_1 ) y_\alpha $.
Therefore $ ( ( p_0 , q_1 ) , ( y_\alpha , \lambda ) ) $ is reachable.
\qed

\proof{ of \ref{res-6-8}}
First observe that we can assume, without loss of generality,
that $ p_0 = h_0 $ (because the path can be extended to $ h_0 $).
Now, by affinity, there is $ w_1 $ such that \trans{h_1}{w_1}{h_1},
$ \pi_\alpha ( w_1 ) = \pi_\alpha ( w_0 ) $ and
$ \pi_\beta ( w_1 ) = \pi_\beta ( w_0 ) $.
In the global state space, find the longest path that starts in \glo\ and
whose image in $F_j$ is labelled by a prefix $v_j$ of $w_j$, for $j=0,1$;
denote by $ \gl = ( ( q_0 , q_1 ) , ( x_\alpha , x_\beta ) ) $ the end node
of the path.
We want to show that $ v_0 = w_0 $.

Assume $ w_0 \neq v_0 $, i.e. $ w_0 = v_0 e u_0 $ for some
$ e \inn \Sigma_0 $ and $ u_0 \inn \Sigma^*_0 $.
Distinguish several cases: \\
I. $ q_0 $ is a send state; then $ e = -b $ for some $ b\inn M_\alpha $, and
\trans{q_0}{-b}{q'_0} in $F_0$.
Thus \gl\Transi{-b} $(( q'_0 , q_1 ) , ( x_\alpha b , x_\beta )) $,
which contradicts the maximality of the path. \\
II. $ q_0 $ is a receive state and $ x_\beta \neq \lambda $.
Then $ e = +b $, $ b \inn M_\beta $, and $b$ is the first symbol in $ x_\beta $.
Thus \trans{q_0}{+b}{q'_0} in $ F_0 $, and again the path in the global
state space is not maximal. \\
III. $ q_1 $ is a receive state and $ x_\alpha \neq \lambda $.
This leads to a contradiction as in case II. \\
IV. Both $q_0$ and $q_1$ are receive states and $ x_\alpha = \lambda = x_\beta $;
this contradicts the assumption that the protocol is deadlock-free. \\
V. $ q_0 $ is a receive state, $ x_\beta = \lambda $ and $ q_1 $ is a send state.
Then $ e = +b $, $ b \inn M_\beta $ and by affinity \trans{q_1}{-b}{q'_1}.
Again, the path is not maximal.

Thus in each case the assumption $ w_0 \neq v_0 $ leads to a contradiction.
We conclude that $ w_0 = v_0 $.
\qed

\proof{ of \ref{res-6-3}}
By \ref{res-6-2} and \ref{res-6-7}, if the protocol has no send cycles
then it has the bounded channel property.

Conversely, assume that, for example, $F_0$ has a send cycle.
Thus there are $ p_0 \inn K_0 $ and $ u_0 \inn \Sigma^*_0 $ such that
\trans{p_0}{u_0}{p_0}, $ u_0 \neq \lambda $ and $ \pi_\beta ( u_0 ) = \lambda $.
Denote $ y_\alpha = \pi_\alpha ( u_0 ) $.
By Lemma~\ref{res-6-8}, there are $p_1 $, $x_\alpha$ and $x_\beta$ such that the global state
$ ( ( p_0 , p_1 ) , ( x_\alpha , x_\beta ) ) $ is reachable.
It follows that, for every integer $ i \geq 0 $, the global state
$ ( ( p_0 , p_1 ) , ( x_\alpha y^i_\alpha , x_\beta ) ) $ is reachable,
and therefore the protocol has not the bounded channel property.
\qed

\proof{ of \ref{res-6-4}}
This algorithm solves the problem: \\
1. Check whether there are any send cycles. \\
2. If there are no send cycles, then (by~\ref{res-6-2}) the protocol has the bounded
channel property.
Apply the exhaustive reachability analysis to decide deadlock-freedom. \\
3. If there is a send cycle then, by Theorem~\ref{res-6-3}, the protocol is
not deadlock-free or has not the bounded channel property.
\qed

\proof{ of \ref{res-6-5}}
Use the exhaustive reachability analysis.
If any global state $ ( S , ( x_\alpha , \lambda ) ) $ with
$ | x_\alpha | \geq k_0 ( k_1 - 1 ) + 1 $ is reachable then, by~\ref{res-6-2},
the protocol is not affine and deadlock-free
(i.e. it is not affine or it is not deadlock-free).

If no such global state is reachable,
then the protocol has the bounded channel property,
and deadlock-freedom can be decided.
Then affinity can be decided by Bird's algorithm;
or alternatively it can be decided by a modified reachability analysis,
since the two state machines differ by a ``finite balance''.
\qed


\section{Undecidable problems}
    \label{sec-7}

We have seen in the previous section that the following problems are
algorithmically decidable:
\begin{itemize}
\item
Given any pair of SR-machines, is it affine?
\item
Given any pair of affine SR-machines,
is it deadlock-free and has it the bounded channel property?
\item
Given any pair of SR-machines with no send cycles,
is it affine and deadlock-free?
\end{itemize}

In this section we shall see that, in contrast to the previous results,
some very similar problems are undecidable.
Brand and Zafiropulo~\cite{bib:bra} prove the undecidability
of several problems of this kind by reduction to the halting problem
for Turing machines.
The proofs in this section are somewhat similar to those in~\cite{bib:bra},
but it will be more convenient for us to use Post's tag systems
instead of Turing machines.
Every tag system can be encoded as a pair of SR-machines;
the known undecidability results about tag systems yield the following theorem.

\begin{theorem}
    \label{res-7-1}
For pairs of communicating SR-machines, these problems are undecidable: \\
(a) Given any protocol with no send cycles, is it deadlock-free? \\
(b) Given any deadlock-free protocol with no send cycles,
has it the bounded channel property? \\
(c) Given any affine protocol, is it deadlock-free? \\
(d) Given any affine protocol, has it the bounded channel property?
\end{theorem}

Theorem~\ref{res-7-1} and the results in the previous section pinpoint
the frontier between the decidable and the undecidable for pairs
of communicating SR-machines.
Next we turn to more general protocols, and explore connections between
the decidability properties and the topology of the underlying communication graph.

For a directed graph $G$, denote by \und\ the corresponding undirected graph.
Consider first any CFSM protocol (with communication graph $G$) for which \und\ has
no cycles.
(Of course, such protocols are hardly of any use. They allow no feedback.)
As in Theorem~\ref{res-5-2}, one can show that every path in the global state space
starting and ending in global states with empty channels is locally equal to a path
that uses only global states of the form
$ ( S, \xchannel ) $, $ \sum_\xi | x_\xi | \leq 1 $.
It follows that the stable composite state problem
(and, in particular, the deadlock problem) is decidable for these protocols.

On the other hand, all ``practical'' communication graphs lead to undecidable problems.
The claim is made precise, for the stable composite state problem,
in the following theorem.

\begin{theorem}
    \label{res-7-2}
If $G$ is a directed graph such that \und\ has a cycle then the stable
composite state problem is undecidable for the CFSM protocols with the communication
graph $G$.
\end{theorem}

The forthcoming proofs of~\ref{res-7-1} and~\ref{res-7-2} are based on known results
about Post's tag systems; the results are collected in the appendix.

The principal steps in the proof of~\ref{res-7-1} are stated and proved separately
in~\ref{res-7-3}, \ref{res-7-4} and \ref{res-7-5}.

\begin{lemma}
    \label{res-7-3}
\samepage
For every tag system $ \TT = \tags $ there is a protocol of two
communicating SR-machines $F_0$ and $F_1$ with no send cycles
such that \\
(a) the protocol is deadlock-free if and only if $ s_n ( \TT ) \neq \lambda $
for all $n$;  \\
(b) the protocol has the bounded channel property if and only if there is
a constant $c$ such that $ | s_n ( \TT ) | \leq c $ for all $n$.
\end{lemma}

\proof{ of \ref{res-7-3}}
For each $ b \inn \Sigma $ create two new symbols $b_\alpha$ and $b_\beta$;
define
$ M_\alpha = \{ \; b_\alpha \; | \; b \inn \Sigma \;\} \cup \{ f \} $
and
$ M_\beta = \{ \; b_\beta \; | \; b \inn \Sigma \;\} $, where $f$ is a new symbol.
The machine $F_0$ has a single receive state $h_0$, which is also its initial state,
and one send state $p_b$ for every $b\inn\Sigma$, with transitions
\trans{h_0}{+b_\beta}{p_b} and \trans{p_b}{-b_\alpha}{h_0}.
Thus $F_0$ is a repeater (or a perfect transmission demon):
it sends $b_\alpha$ whenever it receives $b_\beta$.

The machine $F_1$ simulates the tag system.
It first transmits the string $w_0$ (subscripted by $\beta$), and then
it alternately receives any $d_\alpha$, receives any $b_\alpha$, and
transmits $g(d)$ subscripted by $\beta$.
The transition diagram of $F_1$ is schematically depicted in Fig. 7.1,
where $ g(d) = g_{d0} g_{d1} \ldots g_{dm(d)} $ for every $ d \inn \Sigma $,
and $ w_0 = d_0 d_1 \ldots d_m $.
There is a transition \trans{q}{+d_\alpha}{q_d} for every $ d \inn \Sigma $.
Note also the ``dummy'' transition \trans{q}{+f}{h_1}; it will never be used,
but it makes the transition diagram strongly connected.
Neither $F_0$ nor $F_1$ has a send cycle and if $ |g|^- > 0 $
then they have no receive cycles.

\begin{figure}[p]
\begin{center}
\begin{picture}(200,500)
\thicklines

\put(20,20){\circle{30}}
\put(20,148){\circle{30}}
\put(92,184){\circle{30}} \put(82,174){\makebox(20,20)[c]{$q$}}

\put(164,148){\circle{30}} \put(154,138){\makebox(20,20)[c]{$q_d$}}
\put(164,68){\circle{30}}
\put(100,20){\circle{30}}

\put(149,241){\circle{30}}
\put(149,341){\circle{30}}
\put(149,421){\circle{30}}
\put(92,478){\circle{30}} \put(82,468){\makebox(20,20)[c]{$h_1$}}

\put(20,36){\vector(0,1){29}}
\put(20,75){\circle*{3}}
\put(20,84){\circle*{3}}
\put(20,93){\circle*{3}}
\put(20,103){\vector(0,1){29}}

\put(34,155){\vector(2,1){44}} \put(25,165){\makebox(40,20)[r]{$-(g_{dm(d)})_\beta$}}
\put(106,177){\vector(2,-1){44}} \put(125,165){\makebox(40,20)[l]{$+d_\alpha$}}
\put(164,132){\vector(0,-1){48}} \put(170,98){\makebox(40,20)[l]{$+b_\alpha$}}
\put(152,56){\vector(-4,-3){38}} \put(130,22){\makebox(40,20)[l]{$-(g_{d0})_\beta$}}
\put(84,20){\vector(-1,0){48}} \put(40,0){\makebox(40,20)[c]{$-(g_{d1})_\beta$}}

\put(46,478){\vector(1,0){30}} \put(6,468){\makebox(38,20)[r]{\tt start}}
\put(104,466){\vector(1,-1){33}} \put(120,450){\makebox(40,20)[l]{$-(d_0)_\beta$}}
\put(149,405){\vector(0,-1){48}} \put(153,370){\makebox(40,20)[l]{$-(d_1)_\beta$}}
\put(149,325){\vector(0,-1){20}}
\put(149,298){\circle*{3}}
\put(149,291){\circle*{3}}
\put(149,284){\circle*{3}}
\put(149,277){\vector(0,-1){20}}
\put(137,229){\vector(-1,-1){33}} \put(120,196){\makebox(40,20)[l]{$-(d_m)_\beta$}}
\put(92,200){\vector(0,1){262}} \put(70,321){\makebox(20,20){$+f$}}

\end{picture}

\addspu

{\bf Fig. 7.1. The transition diagram of $F_1$.}
\end{center}
\end{figure}

The pair $ ( F_0 , F_1 ) $ simulates the tag system \TT\ in the following sense:
For $ w \neq \lambda $ we have $ w = s_n ( \TT ) $ for some $n$ if and only
if the global state $ (( h_0 , q ) , ( w_\alpha , \lambda )) $ is reachable;
and $ \lambda = s_n ( \TT ) $ for some $n$ if and only if either
$ (( h_0 , q ) , ( \lambda , \lambda )) $ or
$ (( h_0 , q_d ) , ( \lambda , \lambda )) $ for some $ d \inn \Sigma $ is reachable.

This proves (a) and, in view of Lemma~\ref{res-6-7}, also (b).
\qed

\begin{lemma}
    \label{res-7-4}
For every pair of communicating SR-machines $F'_0$ and $F'_1$
we can construct an affine pair $F_0$, $F_1$ such that either both pairs
are deadlock-free or none is.
\end{lemma}

\proof{ of \ref{res-7-4}}
Let the channel alphabets be $M'_\alpha$ and $M'_\beta$.
Let $\#_\alpha$ and $\#_\beta$ be two new symbols
(not in $ M'_\alpha \cup M'_\beta $) and define
$ M_\alpha = M'_\alpha \cup \{ \#_\alpha \} $
and
$ M_\beta = M'_\beta \cup \{ \#_\beta \} $.
We construct $F_0$ and $F_1$, with the corresponding relations
$\ZZ_0$ and $\ZZ_1$ (defined in section 6) both equal to
\[
\{ \; ( u_\alpha \#_\alpha , u_\beta \#_\beta ) \; |
   \; u_\alpha \inn M'^*_\alpha \; , \; u_\beta \inn M'^*_\beta \; \}^* \; .
\]
First we modify $F'_0$ and $F'_1$ so that no transitions lead to
the initial states $h'_0$ and $h'_1$.
This is arranged as follows in $F'_0$ (and similarly in $F'_1$):
Add a new state $p_0$.
Add the transition \trans{p}{e}{p_0} whenever \trans{p}{e}{h'_0} in $F'_0$,
and add \trans{p_0}{e}{p} whenever \trans{h'_0}{e}{p} in $F'_0$.
Then delete all transitions leading to $h'_0$.
The resulting diagram is not strongly connected, but otherwise it
satisfies all the properties of an SR-machine.
The deadlock-freedom is not changed by the modification.

The next step in the construction of $F_0$ is illustrated in Fig.~7.2
(it is again the same for $F_1$).
Add two new send states $s$ and $s'$ and a new receive state $r$.
For each send state $p$ (including $h'_0$ if it is a send state)
add the transition \trans{p}{-\#_\alpha}{r},
and add \trans{p}{-b}{s'} whenever \trans{p}{-b}{} not in $F'_0$.
For each receive state $p$ (including $h'_0$ if it is a receive state)
add the transition \trans{p}{+\#_\beta}{s},
and add \trans{p}{+b}{s'} whenever \trans{p}{+b}{} not in $F'_0$.
Also, add \trans{s'}{-\#_\alpha}{r}, \trans{r}{+\#_\beta}{h'_0},
\trans{s}{-\#_\alpha}{h'_0};
\trans{s'}{-b}{s'} and \trans{s}{-b}{s} for every $b \inn M'_\alpha$;
and \trans{r}{+b}{r} for every $ b \inn M'_\beta$.
Call the resulting SR-machine $F_0$, and call $F_1$ that constructed
in the same way from $F'_1$.

\begin{figure}[p]
\begin{center}
\includegraphics[height=5in]{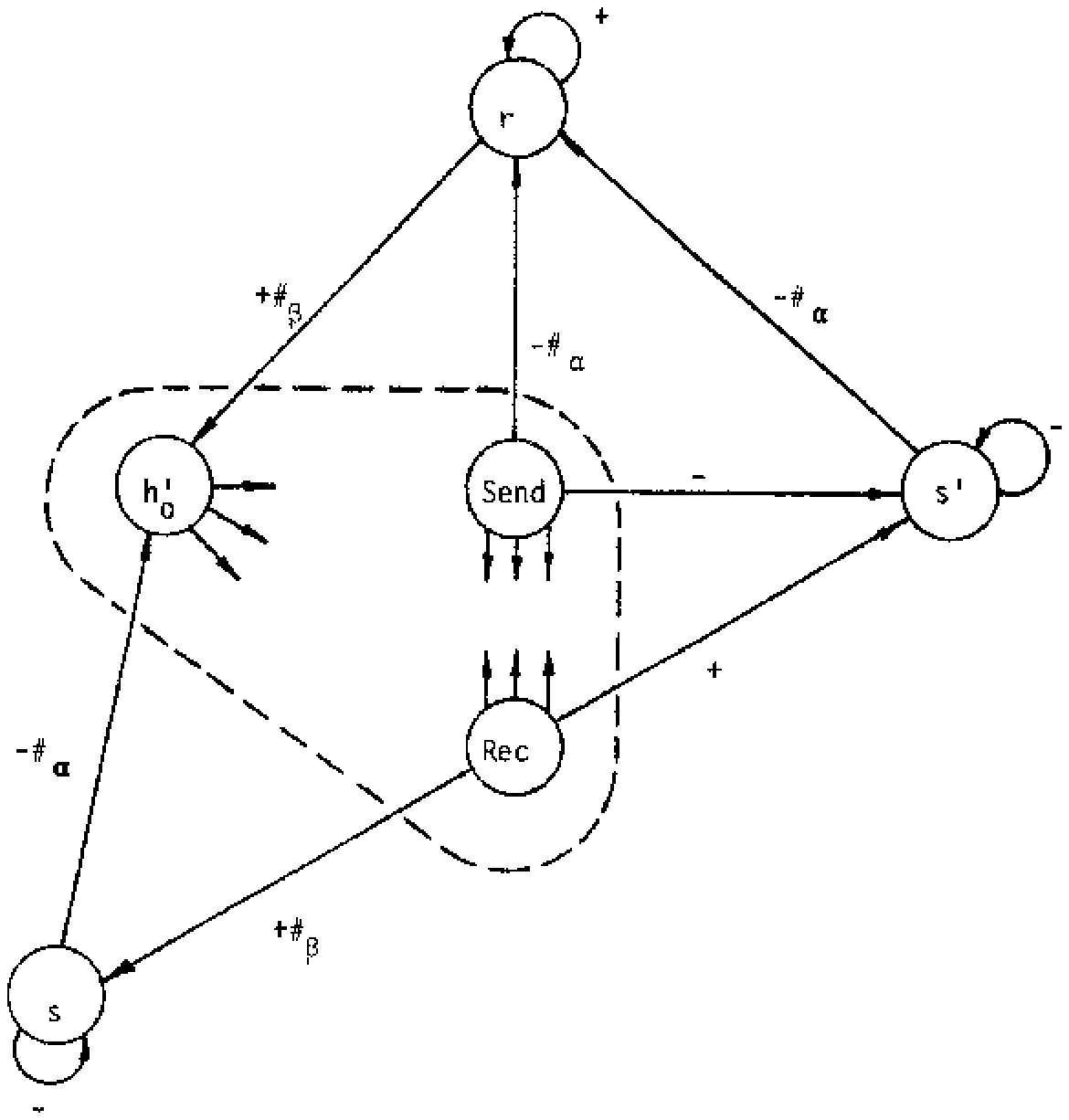}
\addspu\\[1cm]
{\bf Fig. 7.2. The construction of $F_0$ in the proof of~\ref{res-7-4}.}
\end{center}
\end{figure}

If \trans{h'_0}{w}{h'_0} in $F_0$, $ w \neq \lambda $, and if
\trans{h'_0}{u}{h'_0} for no proper nonempty prefix $u$ of $w$,
then $ \pi_\alpha ( w ) = u_\alpha \#_\alpha $ for some
$ u_\alpha \inn M'^*_\alpha $ and
$ \pi_\beta ( w ) = u_\beta \#_\beta $ for some $ u_\beta \inn M'^*_\beta $.
Conversely, for any $ u_\alpha \inn M'^*_\alpha $ and $ u_\beta \inn M'^*_\beta $
there exists $w$ such that \trans{h'_0}{w}{h'_0},
$ \pi_\alpha ( w ) = u_\alpha \#_\alpha $ and $ \pi_\beta ( w ) = u_\beta \#_\beta $.
Therefore
\[
\ZZ_0 \; = \; \{ \; (u_\alpha \#_\alpha , u_\beta \#_\beta ) \; |
    \; u_\alpha \inn M'^*_\alpha \; , \; u_\beta \inn M'^*_\beta \; \}^* \; .
\]
For the same reason, $\ZZ_1$ is equal to the same relation.
Hence $F_0$ and $F_1$ are affine.

Every reachable deadlocked global state for the pair $ ( F'_0 , F'_1 ) $ is
reachable for $ ( F_0 , F_1 ) $.
At the same time, no additional deadlocked global states are reachable for
$ ( F_0 , F_1 ) $;
if, for example, $F_0$ is in its state $r$ and the channels are empty
then $F_1$ must be in its state $s$, which is not a receive state.
\qed

\begin{lemma}
    \label{res-7-5}
For every pair of communicating SR-machines $F'_0$ and $F'_1$ we can construct
an affine pair $F_0$, $F_1$ such that either both pairs have the bounded channel
property or none has.
\end{lemma}

(Note that, in view of \ref{res-6-4}, the constructions in \ref{res-7-4} and
\ref{res-7-5} cannot be combined.
More precisely, it is not true that for every $F'_0$ and $F'_1$ we can construct
an affine pair $F_0$, $F_1$ such that both the deadlock-freedom and
the bounded-channel property are shared by the two pairs.)

\proof{ of \ref{res-7-5}}
As in the proof of~\ref{res-7-4}, we define
$ M_\alpha = M'_\alpha \cup \{ \#_\alpha \} $
and
$ M_\beta = M'_\beta \cup \{ \#_\beta \} $.
We construct $F_0$ and $F_1$ such that the corresponding relations
$\ZZ_0$ and $\ZZ_1$ are both equal to
\[
\{ \; (u_\alpha \#_\alpha \#_\alpha , u_\beta \#_\beta \#_\beta ) \; |
    \; u_\alpha \inn M'^*_\alpha \; , \; u_\beta \inn M'^*_\beta \; \}^* \; .
\]

Again we first arrange that no transitions lead to the initial states $h'_0$ and $h'_1$.
The next step is shown, for $F'_0$, in Fig.~7.3.
\begin{figure}[p]
\begin{center}
\includegraphics[height=5in]{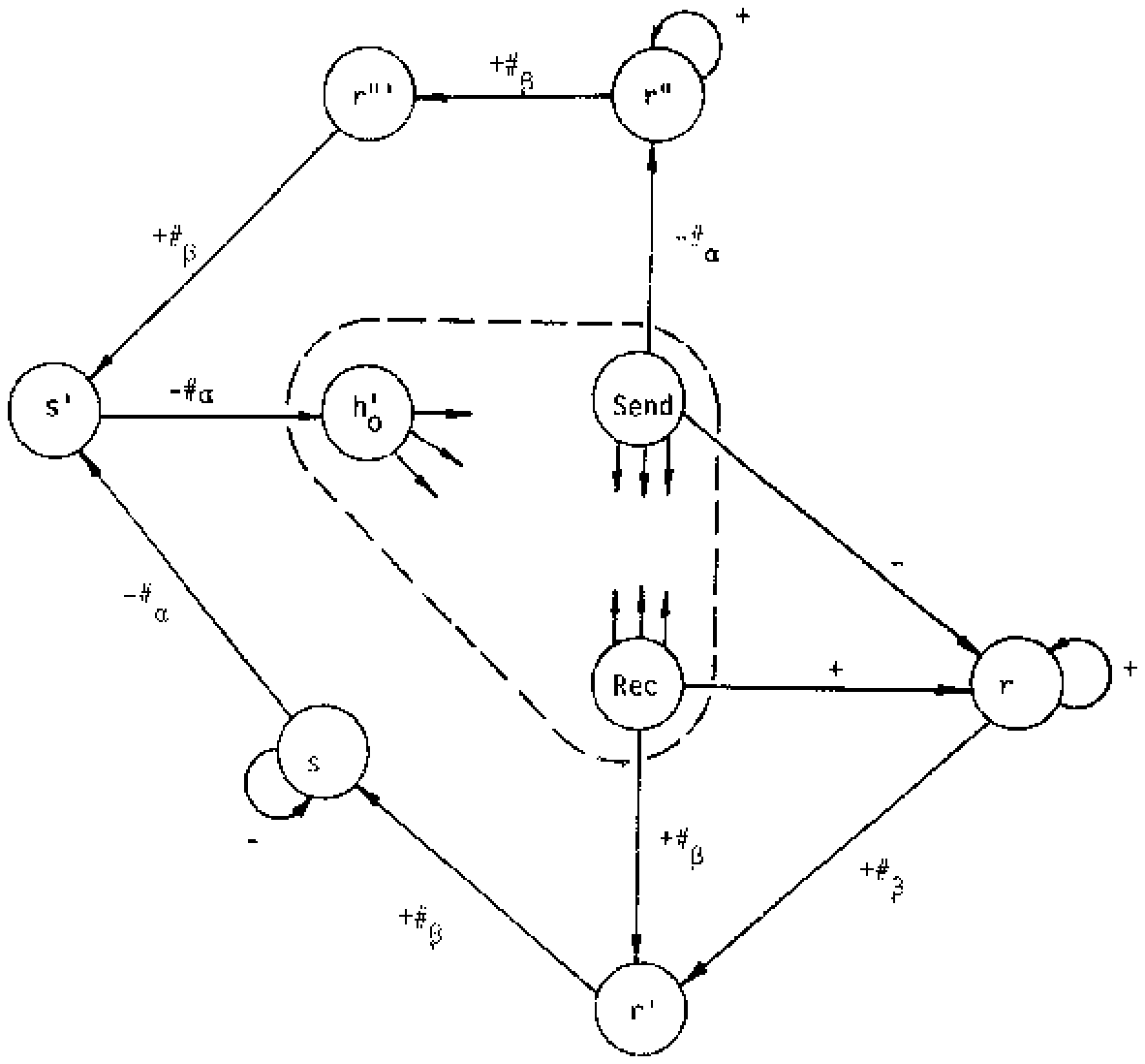}
\addspu\\[1cm]
{\bf Fig. 7.3. The construction of $F_0$ in the proof of~\ref{res-7-5}.}
\end{center}
\end{figure}
Add four new receive states $r$, $r'$, $r''$ and $r'''$ and two new
send states $s$ and $s'$.
For each send state $p$ (including $h'_0$ if it is a send state $p$)
add the transition \trans{p}{-\#_\alpha}{r''}, and add \trans{p}{-b}{r}
whenever \trans{p}{-b}{} not in $F'_0$.
For each receive state $p$ (including $h'_0$ if it is a receive state $p$)
add the transition \trans{p}{+\#_\beta}{r'}, and add \trans{p}{+b}{r}
whenever \trans{p}{+b}{} not in $F'_0$.
Also, add \trans{r}{+\#_\beta}{r'}, \trans{r'}{+\#_\beta}{s},
\trans{s}{-\#_\alpha}{s'}, \trans{r''}{+\#_\beta}{r'''}, \trans{r'''}{+\#_\beta}{s'},
\trans{s'}{-\#_\alpha}{h'_0};
\trans{r}{+b}{r} and \trans{r''}{+b}{r''} for every $ b \inn M'_\beta $;
and \trans{s}{-b}{s} for every $ b \inn M'_\alpha$.
Call $F_0$ the resulting SR-machine, and call $F_1$ that constructed the same way
from $F'_1$.
As in the proof of~\ref{res-7-4} it now follows that
\[
\ZZ_0 = \ZZ_1 = \; \{ \; (u_\alpha \#_\alpha \#_\alpha , u_\beta \#_\beta \#_\beta ) \; |
    \; u_\alpha \inn M'^*_\alpha \; , \; u_\beta \inn M'^*_\beta \; \}^* \; .
\]
The construction creates new reachable deadlocked global states.
In fact, the protocol will never get over the states $r'$ and $r'''$;
hence no global state containing $s$ or $s'$ is reachable.
It follows that the loop at $s$ will never be entered and, therefore,
the pair $ ( F_0 , F_1 ) $ has the bounded channel property if and only if
$ ( F'_0 , F'_1 ) $ has.
\qed

\proof{ of \ref{res-7-1}}
(a) follows directly from Theorem~A.1 (in the appendix) and 7.3(a).

Similarly, (b) follows from A.3, 7.3(a) and 7.3(b).
(Observe that the construction~\ref{res-7-3} is such that if the tag system
$ \TT = \tags $ satisfies $ | g |^- > 0 $ then the protocol has no receive cycles.
Hence the problems (a) and (b) are undecidable even for the protocols
with no send and no receive cycles.)

To prove the undecidability of (c), we combine the already proved case (a)
with~\ref{res-7-4}.
Similarly, (d) follows from (b) and~\ref{res-7-5}.
\qed

The forthcoming Lemma~\ref{res-7-6} will simplify the proof of~\ref{res-7-2}.
Say that two finite directed graphs are {\em homeomorphic\/} if one can be
transformed to the other by a finite sequence of elementary replacements,
each of which either replaces an edge $0 \rightarrow 1$ by two edges
$0 \rightarrow 2 \rightarrow 1$ (where $2$ is a new vertex) or vice versa.
For example, the two graphs in Fig.~\ref{res-7-4} are homeomorphic.

\begin{figure}[p]
\begin{center}
\includegraphics[height=5in]{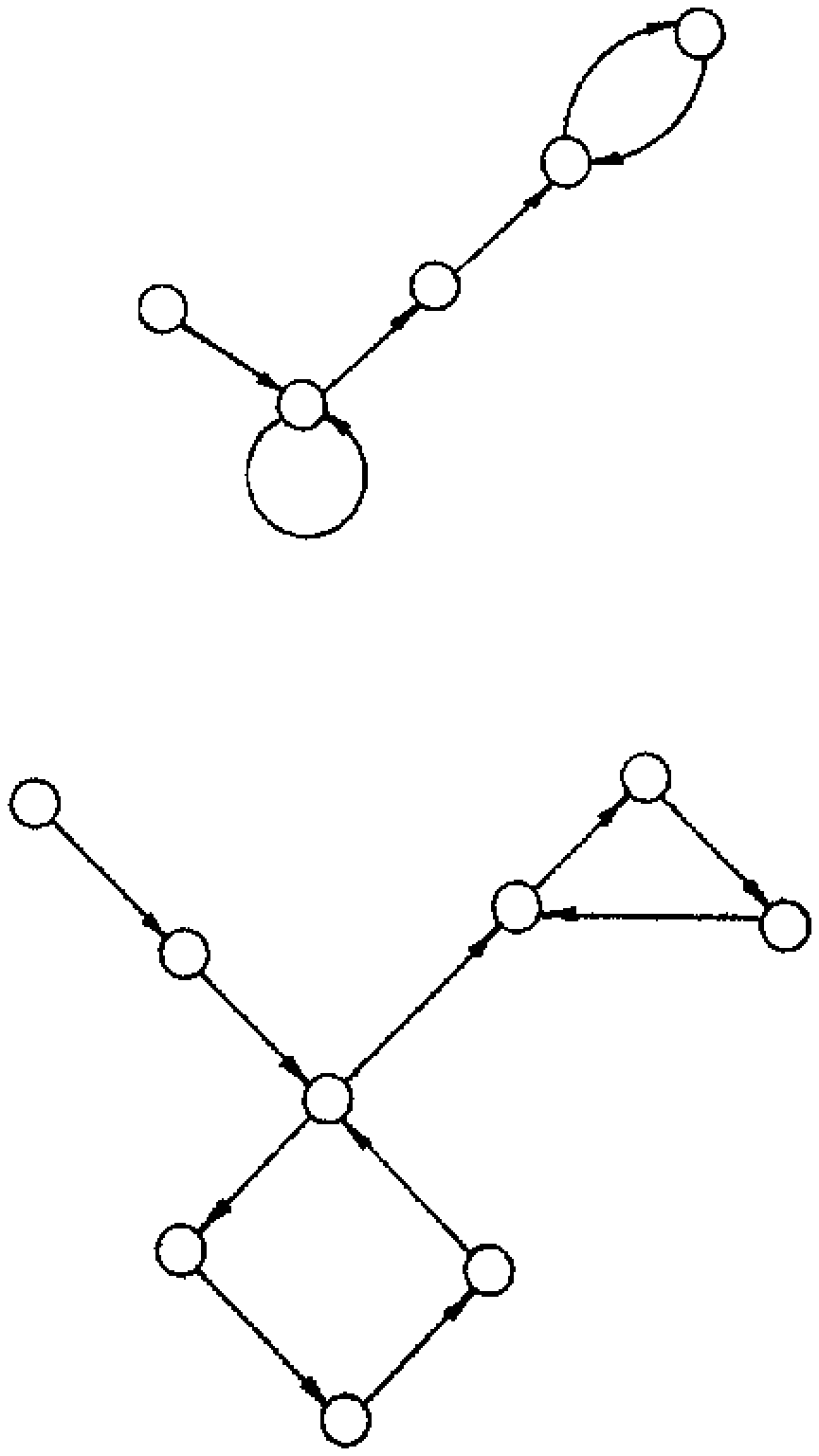}
\addspu\\[1cm]
{\bf Fig. 7.4. Two homeomorphic graphs.}
\end{center}
\end{figure}

\begin{lemma}
    \label{res-7-6}
Let $G$ and $G'$ be two homeomorphic graphs.
The problem ``Is a given composite state stable?'' is decidable for every
CFSM protocol with the communication graph $G$ if and only if it is decidable
for every CFSM protocol with the communication graph $G'$.
\end{lemma}

It will be obvious from the proof of~\ref{res-7-6} that the same result
holds for the deadlock problem, the bounded-channel problem, etc.

\proof{ of \ref{res-7-6}}
It is enough to prove the result under the assumption that $G'$
is produced from $G$ by a single elementary replacement,
which replaces $0 \rightarrow 1$ by $0 \rightarrow 2 \rightarrow 1$.
Assume this is the case.

Let the problem be decidable for every CFSM protocol with
the communication graph $G$, and let $\prot'$ be a protocol with
the communication graph $G'$.
Using the abstract flow control argument of sections 5 and 10
(with the highest priority at the node 2),
we can confine ourselves to the global states in which the channel
from 0 to 2 contains at most one symbol, and we do not lose any
reachable global states of the form $ ( S, C^0 )$.
Now we combine the state of the machine at 0,
the state of the machine at 2,
and the content of the channel $ 0 \rightarrow 2 $ into a single state;
this transforms $\prot'$ into a protocol with the communication graph $G$.
It follows that the problem is decidable for $G'$.

Conversely, assume that the problem is decidable for $G'$.
Every CFSM protocol with the communication graph $G$ can be transformed
into one with the communication graph $G'$ by including a repeater
(perfect transmission demon) at the node 2.
It follows that the problem is decidable for $G$.
\qed

\proof{ of \ref{res-7-2}}
Clearly it suffices to prove the undecidability for every graph $G$
for which \und\ is a circle.
When \und\ is a circle, there are two possibilities:
Either $G$ itself is a (directed) cycle or $G$ is acyclic as a directed graph.
Since every directed cycle is homeomorphic to the graph \twographi,
the case of $G$ being a cycle is taken care of by~\ref{res-7-1}(a)
(or~\ref{res-7-1}(c)) and~\ref{res-7-6}.

It remains to be proved that the stable composite state problem
is undecidable for every acyclic graph $G$ for which \und\ is a circle.
The proof is based on the undecidability of
{\em modified Post's correspondence problem\/} (MPCP).
Recall~\cite{bib:hop} that an {\em instance\/} of MPCP consists of two lists
$ x = ( x_0 , x_1 , \ldots , x_n ) $ and $ y = ( y_0 , y_1 , \ldots , y_n ) $
of strings over an alphabet $\Sigma$.
The instance {\em has a solution\/} if there is a sequence of integers
$ j_1 , j_2 , \ldots , j_k $ such that
\[
x_0 x_{j_1} \ldots x_{j_k} = y_0 y_{j_1} \ldots y_{j_k} \; ;
\]
The sequence $ j_1 , j_2 , \ldots , j_k $ is called a {\em solution\/}
for the instance of MPCP.
It is known that the problem ``Given an instance of MPCP, has it a solution?''
is undecidable (\cite{bib:hop},~8.5).

\begin{figure}[tbp]
\begin{center}
\begin{picture}(300,200)(-150,-70)
\thicklines

\put(0,120){\circle{40}}
\put(-10,110){\makebox(20,20)[c]{$0$}}

\put(120,0){\circle{40}}
\put(110,-10){\makebox(20,20)[c]{$2$}}

\put(-120,0){\circle{40}}
\put(-130,-10){\makebox(20,20)[c]{$2m$}}

\put(40,40){\circle{40}}
\put(30,30){\makebox(20,20)[c]{$1$}}

\put(-40,40){\circle{40}}
\put(-50,30){\makebox(20,20)[c]{$2m\!+\!1$}}

\put(-40,-40){\circle{40}}
\put(-50,-50){\makebox(20,20)[c]{$2m\!-\!1$}}

\put(40,-40){\circle{40}}
\put(30,-50){\makebox(20,20)[c]{$3$}}

\put(9,102){\vector(1,-2){22}}
\put(25,74){\makebox(20,20)[l]{$\alpha_0$}}

\put(-9,102){\vector(-1,-2){22}}
\put(-42,74){\makebox(20,20)[r]{$\alpha_{2m+1}$}}

\put(-102,9){\vector(2,1){44}}
\put(-95,20){\makebox(20,20)[r]{$\alpha_{2m}$}}

\put(-102,-9){\vector(2,-1){44}}
\put(-90,-40){\makebox(20,20)[r]{$\alpha_{2m-1}$}}

\put(102,9){\vector(-2,1){44}}
\put(82,15){\makebox(20,20)[l]{$\alpha_1$}}

\put(102,-9){\vector(-2,-1){44}}
\put(82,-35){\makebox(20,20)[l]{$\alpha_2$}}

\put(-15,-65){\circle*{3}}
\put(-5,-68){\circle*{3}}
\put(5,-68){\circle*{3}}
\put(15,-65){\circle*{3}}

\end{picture}

\addspu

{\bf Fig. 7.5.}
\end{center}
\end{figure}

Every acyclic graph $G$ for which \und\ is a circle is homeomorphic to
the graph in Fig.~7.5, for some $m \geq 0 $.
Hence the undecidability result follows from~\ref{res-7-6} and from this lemma:

\begin{lemma}
    \label{res-7-7}
\samepage
For the graph $G$ in Fig.~7.5 and for every instance of MPCP there exist
a CFSM protocol with the communication graph $G$ and a composite state~$S$
such that $S$ is stable if and only the instance of MPCP has a solution.
\end{lemma}

\proof{ of \ref{res-7-7}}
Let $\Sigma$ be the alphabet of the instance of MPCP.
For every edge $\xi$ in $G$, the channel alphabet $M_\xi$ is defined to be
$ \{ \; b_\xi \; | \; b \inn \Sigma \;\} $, where the symbols $ b_\xi $ are chosen so that
the sets $ M_\xi $ are pairwise disjoint.

All the finite state machines except the one at 0 are simple comparators:
Those at the even numbered nodes (except 0) send the same sequences of messages
to both channels, those at the odd numbered nodes receive the same sequences
from both channels.
For example, the machine at 1 has the initial state $h_1$ and a separate state
$p_b$ for each $ b \inn \Sigma $, with transitions \trans{h_1}{+b_{\alpha_0}}{p_b}
and \trans{p_b}{+b_{\alpha_1}}{h_1}.

\begin{figure}[tbp]
\begin{center}
\begin{picture}(405,215)(-285,-100)
\thicklines

\put(0,0){\circle{30}}
\put(-10,-10){\makebox(20,20)[c]{$q_0$}}

\put(-100,0){\circle{30}}

\put(-200,0){\circle{30}}
\put(-210,-10){\makebox(20,20)[c]{$h_0$}}

\put(0,100){\circle{30}}

\put(100,0){\circle{30}}

\put(-52,-82){\circle{30}}

\put(-250,0){\vector(1,0){34}}
\put(-285,-10){\makebox(30,20)[r]{start}}

\put(-184,0){\vector(1,0){68}}
\put(-170,0){\makebox(40,20){$-(x_0 )_{\alpha_{2m+1}}$}}

\put(-84,0){\vector(1,0){68}}
\put(-70,0){\makebox(40,20){$-(y_0 )_{\alpha_0}$}}

\put(-44,-68){\vector(1,2){30}}
\put(-66,-54){\makebox(30,20)[r]{$-(y_n )_{\alpha_0}$}}
\put(-5,-16){\vector(-1,-2){31}}
\put(-18,-60){\makebox(50,20)[l]{$-(x_n )_{\alpha_{2m+1}}$}}

\put(-5,15){\vector(0,1){70}}
\put(-46,50){\makebox(40,20)[r]{$-(x_0 )_{\alpha_{2m+1}}$}}
\put(5,85){\vector(0,-1){70}}
\put(8,50){\makebox(40,20)[l]{$-(y_0 )_{\alpha_0}$}}

\put(15,5){\vector(1,0){70}}
\put(30,5){\makebox(40,20)[c]{$-(x_1 )_{\alpha_{2m+1}}$}}
\put(85,-5){\vector(-1,0){70}}
\put(30,-25){\makebox(40,20)[c]{$-(y_1 )_{\alpha_0}$}}

\put(22,-78){\circle*{4}}
\put(42,-72){\circle*{4}}
\put(60,-62){\circle*{4}}

\end{picture}

\addspu

{\bf Fig. 7.6. A machine to simulate MPCP.}
\end{center}
\end{figure}
The machine at 0 is capable of sending, for every infinite sequence of indices
$ j_1 , j_2 , \ldots $, the sequence of messages
\[
( x_0 )_{\alpha_{2m+1}} ( x_{j_1} )_{\alpha_{2m+1}} \ldots
\]
on the channel $ \alpha_{2m+1} $, and the sequence
\[
( y_0 )_{\alpha_0} ( y_{j_1} )_{\alpha_0} \ldots
\]
on the channel $ \alpha_0 $.
A schematic transition diagram is in Fig.~7.6.
The composite state $ S = ( q_0 , h_1 , h_2 , \ldots , h_{2m+1} ) $
is stable if and only if there is solution of the MPCP.

This completes the proofs of~\ref{res-7-7} and~\ref{res-7-2}.
\qed


\section{Rational channels for cyclic protocols}
    \label{sec-8}

The results in the previous section show that general CFSM protocols can,
with the help of their infinite channels, simulate arbitrary
computation processes.
It is for this reason that the reachability problems are undecidable.
However, we are primarily interested in the protocols that use their channels
more simply.
Can we disqualify the CFSM protocols that, by using the channels as
an infinite memory, simulate general computations?
Can the ``simple channel property'' (or, more precisely, the property
of ``the channels being used in a simple manner'') be formalized?
One sufficient condition for this kind of channel simplicity is
the bounded channel property.
Two more general conditions are offered in this section.

The popular classification of verification techniques for communication protocols
distinguishes between reachability analysis and program proofs~\cite{bib:bo1}.
Traditionally, program proofs have been used to verify the protocol
properties that are not amenable to reachability analysis.
Our present aim is different:
The primitive assertion proving technique proposed below is more powerful
than the exhaustive reachability analysis, but it stays within the realm
of reachability properties.

Rather than treating the reachability analysis and program proofs as two opposites,
we shall regard the former as a simple special case of the latter.
(Bochmann alludes to this perspective in~\cite{bib:bo2}, p.~649.)
In this view, illustrated by the following example, the reachability analysis
of a bounded-channel CFSM protocol is a method for constructing and proving
a set of simple assertions attached to composite states.

\begin{example}
\rm
The purpose of the protocol is to limit the total number of messages
simultaneously in transit
(ie. the total number of buffers needed).
In the example, the limit is two.
(Any other limit can be used.
The larger the limit, the more states the finite state machines have.)
The protocol assumes error-free channels.
Data messages are transmitted in both directions.
There are three message types:
\begin{tabbing}
xxxxx\= \RELE xxxxx\= \kill
\>\DATA \>data message, \\
\>\ACK \>acknowledgement of \DATA, \\
\>\RELE \>releasing buffer.
\end{tabbing}
Initially, each channel is allocated one buffer.
Either transmitter can release a buffer, which is then used
for transmissions in the opposite direction.
The two finite state machines are identical.
Fig.~8.1 shows their transmission diagrams and the communication graph.

\begin{figure}[tbp]
\begin{center}
\includegraphics[height=5in]{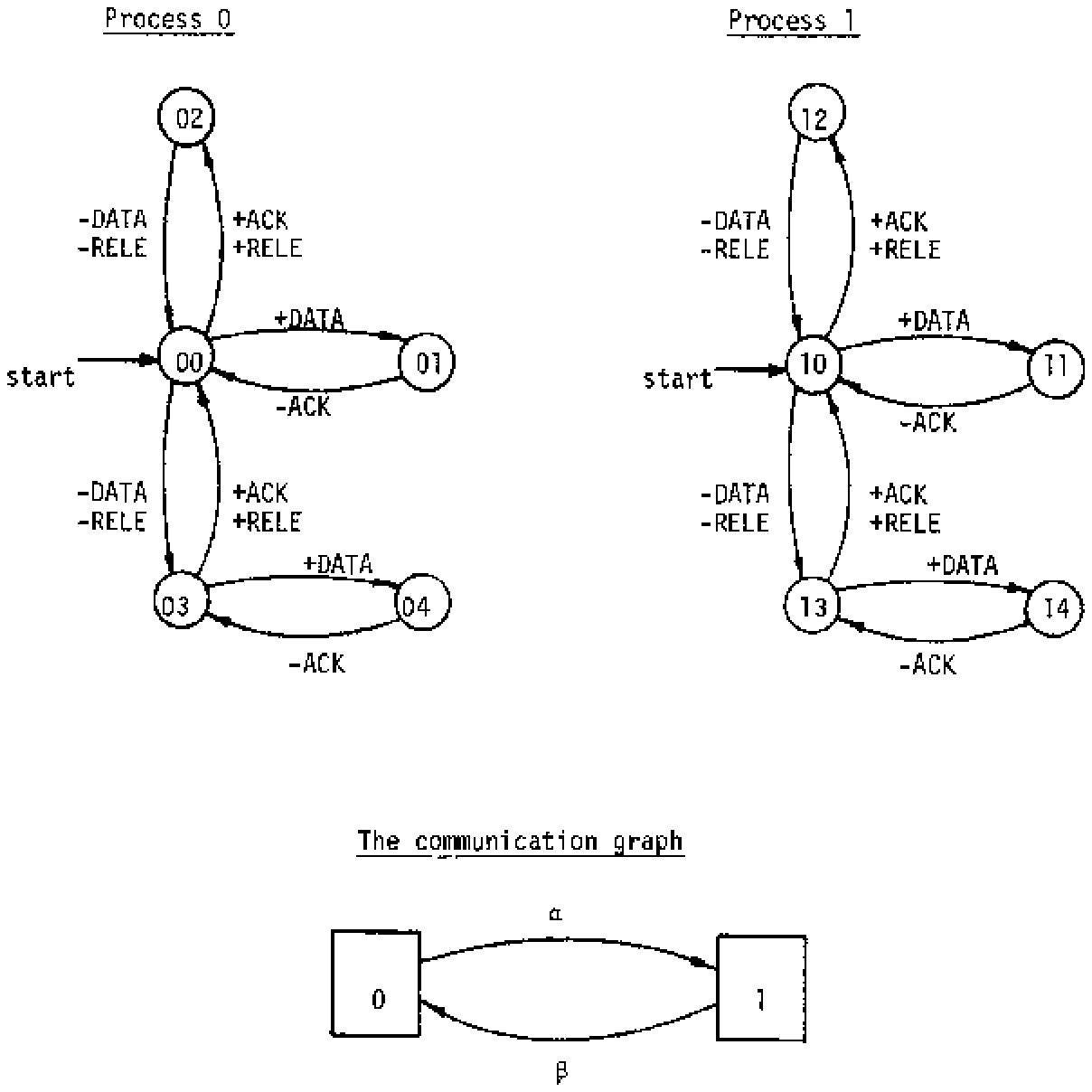}
\addspu\\[1cm]
{\bf Fig. 8.1. A simple flow control protocol.}
\end{center}
\end{figure}

Fig.~8.2 is the complete global state space of the protocol.
(We write D=\DATA, R=\RELE\ and A=\ACK.)
The global state space is finite;
from it one can read various reachability properties:
The total number of messages in transit is at most two,
the protocol is deadlock-free, etc.
Fig.~8.3 shows a different data structure, which contains the same
information as Fig.~8.2 (when Fig.~8.1 is known).
The table in Fig.~8.3 lists, for each composite state,
the set of all possible channel contents.
We can regard each entry in the table as an {\em assertion\/}.
For example, the entry
$ \{ (\DATA, \lambda), (\RELE, \lambda), (\ACK, \lambda),
(\lambda, \DATA), (\lambda, \RELE), (\lambda, \ACK) \} $
at $(03,10)$ asserts:
If the state of Process 0 is 03 and the state of Process 1 is 10,
then the channel content is
$ (\DATA, \lambda) $ or $ (\RELE, \lambda) $ or $ (\ACK, \lambda) $
or $ (\lambda, \DATA) $ or $ (\lambda, \RELE) $ or $ (\lambda, \ACK) $.

\begin{figure}[p]
\begin{center}
\include{fig82}
{\bf Fig. 8.2. The global state space.}
\end{center}
\end{figure}

\begin{figure}[t]
\begin{center}
\begin{tabular}{|c||l|l|l|l|l|} \hline
& 10 & 11 & 12 & 13 & 14 \\ \hline\hline

\addspu 00 &
(\la,\la) &
$ \emptyset $ &
$ \emptyset $ &
(\la,D), (\la,R), (\la,A), &
(\la,\la) \\
\addspd & & & & (D,\la), (R,\la), (A,\la) & \\ \hline

\addspu 01 &
$ \emptyset $ &
$ \emptyset $ &
$ \emptyset $ &
(\la,\la) &
$ \emptyset $ \\ \hline

\addspu 02 &
$ \emptyset $ &
$ \emptyset $ &
$ \emptyset $ &
(\la,\la) &
$ \emptyset $ \\ \hline

\addspu 03 &
(\la,D), &
(\la,\la) &
(\la,\la) &
(\la,DD), (\la,DR), (\la, DA), &
(\la,D), \\
& (\la,R), & & &
(\la,RD), (\la,RR), (\la,RA), &
(\la,R), \\
& (\la,A), & & &
(\la,AD), (\la,AR), (\la,AA), &
(\la,A), \\
& (D,\la), & & &
(D,D), (D,R), (D,A), &
(D,\la), \\
& (R,\la), & & &
(R,D), (R,R), (R,A), &
(R,\la), \\
& (A,\la) & & &
(A,D), (A,R), (A,A), &
(A,\la) \\
& & & &
(DD,\la), (DR,\la), (DA,\la), & \\
& & & &
(RD,\la), (RR,\la), (RA,\la), & \\
& & & &
(AD,\la), (AR,\la), (AA,\la) & \\ \hline

\addspu 04 &
(\la,\la) &
$ \emptyset $ &
$ \emptyset $ &
(\la,D), (\la,R), (\la,A), &
(\la,\la) \\
\addspd & & & & (D,\la), (R,\la), (A,\la) & \\ \hline
\end{tabular}

\addspu

{\bf Fig. 8.3. Another description of the global state space.}
\end{center}
\end{figure}

The assertions in Fig.~8.3 can be written more compactly.
E.g. the entry at $(03,13)$ is the relation
$ \{(x,y)\; | \; |x|+|y|=2 \} $,
the entry at $(00,13)$ is the relation $ \{(x,y)\; | \; |x|+|y|=1 \} $,
etc.
Quite simply, the protocol implements a distributed counter.
However,
an automatic assertion verifier would have to be considerably more
intelligent to understand such descriptions.

From the table in Fig.~8.3 we can read, for example,
that the composite state $(01,13)$ is stable, and that no message
can arrive at 02. (End of Example 8.1.)
\end{example}

In this view, the exhaustive reachability analysis is a method for
constructing and verifying the correctness of tables whose entries
are finite sets of channel contents.
One can argue that the table, or a portion of it, should be a part
of the protocol description, because it offers an additional insight
into the structure of the protocol.
This is especially true if the protocol has not the bounded channel property.
In that case the entries in the table are infinite sets,
and the complete table cannot be constructed by the exhaustive
reachability analysis.
If the table is supplied together with the CFSM description then
the analysis algorithm need not construct the table, it merely has to
verify its correctness (consistency).

The distinctive feature of the exhaustive reachability analysis is that
the domain of assertions (the language that they are formulated in)
is extremely simple, and therefore analysis can be efficiently automated.
On the other hand, the method has several limitations.
Here we address its inability to analyze protocols with unbounded channels.

Generally speaking, the way to overcome the limitations of any assertion
proving system is to extend the domain of assertions;
in doing so we trade simplicity for power.
A natural extension of the exhaustive reachability analysis is to use
more general relations, instead of finite ones, in the assertions.
Two important families of relations have been extensively studied
in the last ten years, the {\em recognizable\/} and the {\em rational\/}
relations; their basic properties can be found in~\cite{bib:ber} and~\cite{bib:eil}.
Every finite relation is recognizable and every recognizable relation
is rational.

We are going to extend the assertion domain by using recognizable and rational
relations in place of finite ones.
We gain power (ability to analyze protocols with unbounded channels), while
not losing all the simplicity:
The assertion verifier will have to be smarter but still fairly simple.

\begin{definition}
    \label{res-8-2}
\rm
\samepage
Let \prot\ be a CFSM protocol.
Say that \prot\ has the {\em rational channel property\/} if the relation
\[
\LL (S) = \{ \; C \; | \; \TTrans{\glo}{\gl} \;\} \;\subseteq \; \Timesa{\xi\in E} M^*_\xi
\]
is rational for each composite state $ S \inn \Timesa{j\in N} K_j $.
Say that \prot\ has the {\em recognizable channel property\/} if $ \LL (S) $ is
recognizable for each $S$.
\end{definition}

Thus the bounded channel property implies the recognizable channel property,
which in turn implies the rational channel property.

In this section we concentrate on cyclic protocols.
We return to general CFSM protocols in the next section.
As we have seen in Theorem~\ref{res-5-2}, cyclic protocols have the property that
unbounded channel growth can be confined to a single channel.

\begin{theorem}
    \label{res-8-3}
\samepage
For any cyclic CFSM protocol \prot\ the following four conditions are equivalent: \\
(a) \prot\ has the recognizable channel property; \\
(b) \prot\ has the rational channel property; \\
(c) for every $ \beta \inn E $ and for every composite state $S$, the set
\[
Q_\beta ( S ) \; = \; \{ \; x_\beta \inn M^*_\beta \; | \;
( x_\xi\! : \!\xi \inn E ) \inn \LL ( S ) \;\; \mbox{\rm and} \;\;
x_\xi = \lambda \;\; \mbox{\rm for} \;\; \xi \neq \beta \; \}
\]
is regular; \\
(d) there exists $ \;\beta \inn E $ such that the set $ Q_\beta ( S ) $ is regular for every $S$.
\end{theorem}

Thus the recognizable and the rational channel property coincide for cyclic protocols.
We shall see later that this is not the case in general.

The sets $ Q_\beta ( S ) $ of Theorem~\ref{res-8-3} are consistent, in this sense:
If
\TTrans{(S,(x_\xi\! : \!\xi\inn E))}{(S',(x'_\xi\! : \!\xi\inn E))},
$ x_\xi = x'_\xi = \lambda $ for $ \xi \neq \beta $, and $ x_\beta \inn Q_\beta (S) $
then $ x'_\beta \inn Q_\beta (S')$.
At the same time, there is an efficient algorithm to decide whether a given family
of {\em regular\/} sets $Q(S)$, indexed by $ S \inn \Timesa{j\in N} K_j $,
is consistent (with respect to $\beta$ and \prot).

Moreover, a consistent family $Q(S)$ such that $\lambda\inn Q(S^0 )$ and
$\lambda \not\in Q(S)$ constitutes a {\em proof\/} that $ ( S, C^0 ) $ is not reachable
from \glo (i.e. that $S$ is not a stable state).
Consequently, if a cyclic protocol has the rational channel property then for each
non-stable $S$ there is an {\em automatically verifiable\/} proof that $S$ is not stable.

The foregoing discussion is summed up in Definition~\ref{res-8-4}
and Theorems~\ref{res-8-5} and~\ref{res-8-6}.

\begin{definition}
    \label{res-8-4}
\rm
Let \prot\ be a CFSM protocol, $\beta\inn E$, and let $ Q ( S ) \subseteq M^*_\beta $
for every $ S \inn \Timesa{j\in N} K_j $.
Say that the sets $Q(S)$ are {\em consistent (with respect to \prot\ and~$\beta$)\/}
if
\[
\TTrans{(S,(x_\xi\! : \!\xi\inn E))}{(S',(x'_\xi\! : \!\xi\inn E))}, \;\;
x_\xi \! = \! x'_\xi \! = \! \lambda \;\;\mbox{\rm for }\; \xi \!\neq\! \beta ,
\;\;\mbox{\rm and }\; x_\beta \!\inn Q (S)
\]
imply $ x'_\beta \!\inn Q (S')$.
\end{definition}

\begin{theorem}
    \label{res-8-5}
There is an algorithm to decide whether any given family of regular sets \/ $Q(S)$
is consistent (with respect to a given cyclic \prot\ and a given $\beta$).
\end{theorem}

\begin{theorem}
    \label{res-8-6}
\samepage
Let \prot\ be a cyclic CFSM protocol with the rational channel property,
and let $ \beta \inn E $.
A composite state $S'$ is not stable if and only if there is a consistent
family of regular sets \/ $Q(S)$, $ S \inn \Timesa{j\in N} K_j $,
such that $ \lambda \inn Q ( S^0 ) $ and $ \lambda \not\in Q ( S') $.
\end{theorem}

The following corollary to~\ref{res-8-5} and~\ref{res-8-6} shows that
the rational channel property indeed prevents, in an essential way,
the cyclic protocol from using channels as a general infinite memory.

\begin{corollary}
    \label{res-8-7}
The deadlock problem is algorithmically decidable for cyclic CFSM protocols
with the rational channel property.
\end{corollary}

The algorithm in the proof of~\ref{res-8-7} (at the end of this section)
is awfully inefficient; it exhaustively searches for the proof of
deadlock-freedom.
However, once the proof is known, it can be efficiently verified.
Therefore it makes sense to require that the protocol designer supply the proof
(in the form of channel expressions) as a part of the protocol description.
The description of a protocol by means of CFSM augmented with channel expressions
will be exhibited in Example~\ref{res-8-9}.
The description is substantially abridged with the help of the simple result
in the forthcoming Theorem~\ref{res-8-8}.
It says that one need not supply the sets $Q(S)$ for all $S$;
it is sufficient to describe $Q(S)$ for sufficiently many $S$, and all the other
sets $Q(S)$ can be automatically computed.

\begin{theorem}
    \label{res-8-8}
Let \prot\ be a cyclic CFSM protocol and $\beta\inn E$.
For each $j\inn N$, let $ V_j \subseteq K_j $ be a set of states such that
$ h_j \inn V_j $ and if $ \trans{p_j}{-b}{q_j} $, $p_j , q_j \inn K_j $, then
$ q_j \inn V_j $.
Then there is an algorithm to decide whether any given family of regular sets
indexed by $ S \inn \Timesa{j\in N} V_j $ can be extended to a consistent family
of sets $ Q(S) $ indexed by $ S \inn \Timesa{j\in N} K_j $.
Moreover, if the family can be extended than the smallest such sets $Q(S)$ are regular
and can be automatically constructed.
\end{theorem}

The proofs of the results in this section come after the following example,
which illustrates the proposed proof method.

\begin{figure}[b]
\begin{center}
\fourgraph
\addspu\\[.5cm]
{\bf Fig. 8.4. The communication graph.}
\end{center}
\end{figure}

\begin{example}
    \label{res-8-9}
\rm
This is a variation of the alternating bit protocol described in section~\ref{sec-2}.
In the present version both stations take turns in transmitting and receiving data packets.
The communication graph is again as in Fig.~8.4.

Demon 2 and Demon 3 are identical.
Demon 2 is defined in Fig.~8.5;
Demon 3 differs only in state numbers (30, 31, \ldots\ instead of 20, 21, \ldots).
Processes 0 and 1 are defined in Fig.~8.6.
They differ only in the starting state.

\begin{figure}
\begin{center}
\includegraphics[height=3in]{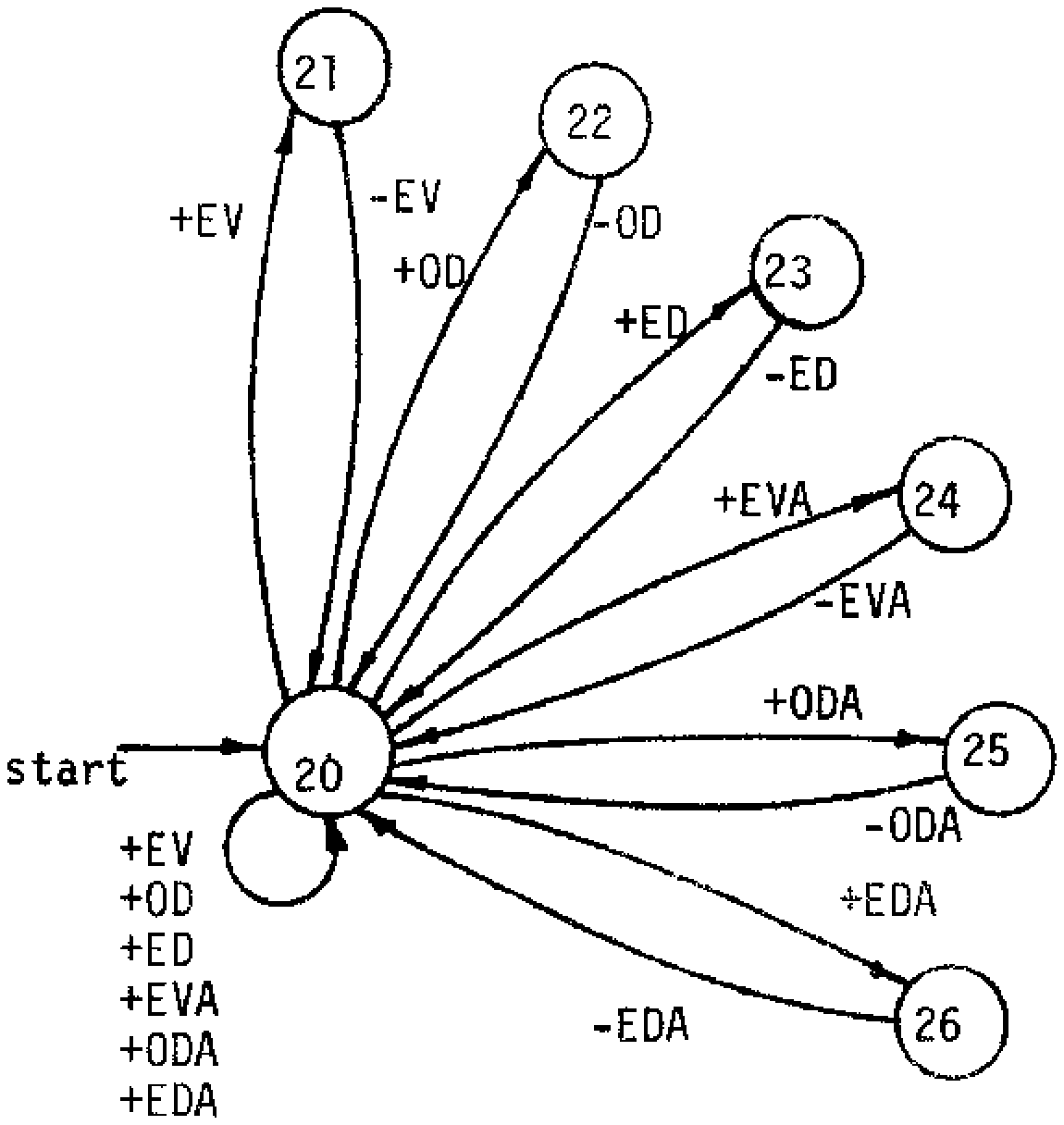}
\addspu\\[1cm]
{\bf Fig. 8.5. Demon 2.}
\end{center}
\end{figure}

Theorem~\ref{res-8-8} applies for these sets $V_j$: \\
$ V_0 = \{ 00, 01, 02, 04 \} $, \\
$ V_1 = \{ 10, 11, 12 , 14 \} $, \\
$ V_2 = \{ 20 \} $, \\
$ V_3 = \{ 30 \} $. \\
This reduces the number of the sets $Q(S)$ that have to be specified from
$6\times 6\times 7\times 7 = 1764$ to $4\times 4\times 1\times 1 = 16$.
The sets $ Q_\alpha (S)$ for $ S \inn \Timesb{3}{j=0} V_j $ are listed in Fig.~8.7.
(Recall that, in agreement with the notation in Theorem~\ref{res-8-3},
$ Q_\alpha (S) $ is the set of all possible contents of the channel from
Process 0 to Demon 2 when the other channels are empty.)
Each $ K_j$, $ j=0,1,2,3$, contains one receive state:
the protocol is deadlock-free if and only if the global state
$ ((04,14,20,30), C^0 ) $ is unreachable.
Since the $ (04,14,20,30)$ entry in Fig.~8.7 is the empty set,
the protocol is deadlock-free. \\
(End of Example~\ref{res-8-9}.)
\end{example}

\begin{figure}
\begin{center}
\includegraphics[width=5.44in]{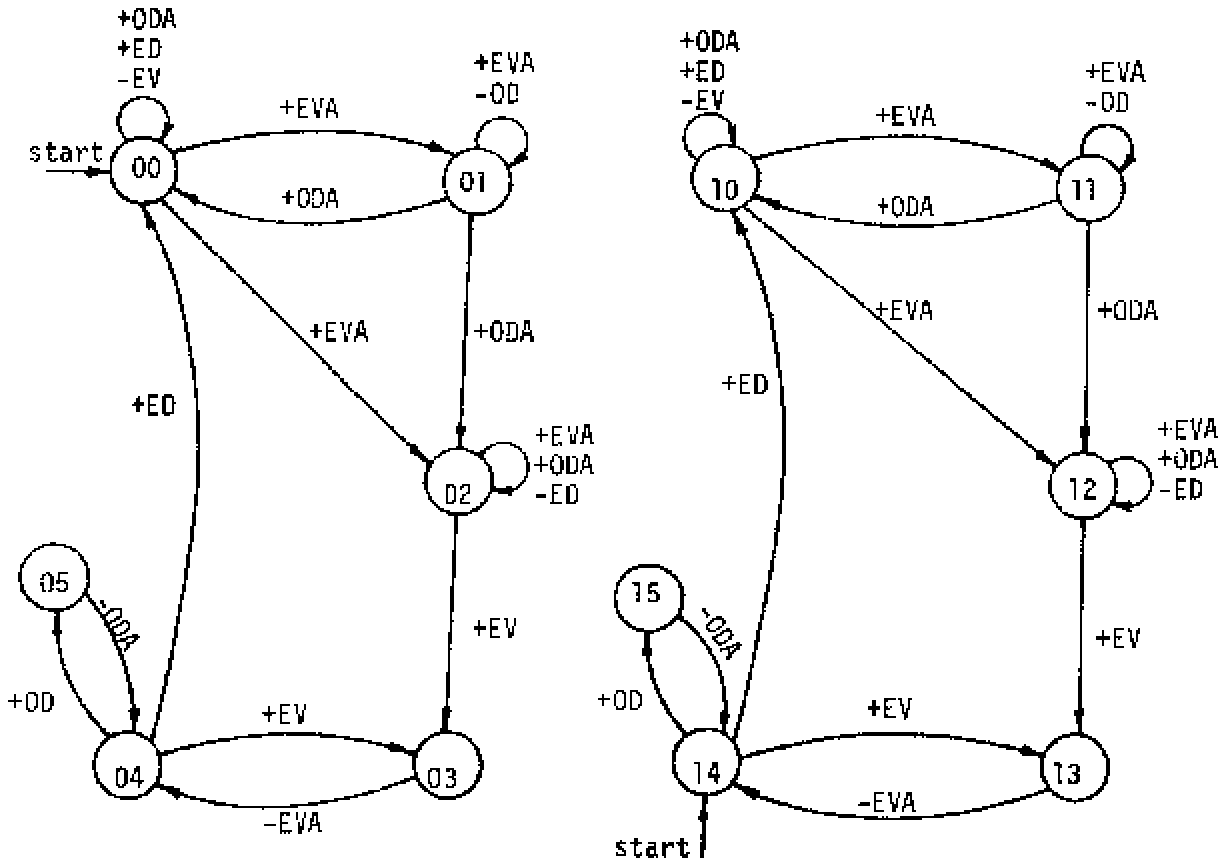}
\addspu\\[3mm]
{\bf Fig. 8.6. Another alternating bit protocol.}
\end{center}
\end{figure}

\begin{figure}
\begin{center}
\begin{tabular}{|c||c|c|c|c|} \hline
& \multicolumn{4}{c|}{$q$} \\ \cline{2-5}
\addspu $p$ & 10 & 11 & 12 & 14 \\ \cline{2-5}\hline

\addspu 00 & $\emptyset$ & $\emptyset$ &
$\EVA^* \EV^* \;\cup $ & $ \OD^* \EV^* $ \\
& & & $ \ODA^* \EV^* $ & \\ \hline

\addspu 01 & $\emptyset$ & $\emptyset$ &
$\emptyset$ & $ \EV^* \OD^* $ \\ \hline

\addspu 02 & $\ED^*$ & $\emptyset$ &
$\emptyset$ & $ \EV^* \ED^* \; \cup $ \\
& & & & $ \OD^* \ED^* $ \\ \hline

\addspu 04 & $ \ED^* \EVA^* \;\cup $ & $ \EVA^* \ODA^* $ &
$ \EVA^* \;\cup\; \ODA^* $ & $\emptyset$  \\
& $ \ODA^* \EVA^* $ & & & \\ \hline

\end{tabular}

\addspu

{\bf Fig. 8.7. $Q_\alpha ((p,q,20,30))$ for $ p \inn \{00,01,02,04\}$
    and $q\inn \{10,11,12,14\}$.}
\end{center}
\end{figure}

The proofs of the results in this section follow.
Several proofs use the ``priority argument'' informally;
it could be formalized as in the proof of~\ref{res-10-1}.

First we establish two lemmas that will be needed in the proof of~\ref{res-8-3}.

\begin{lemma}
    \label{res-8-10}
Let $M_1$ and $M_2$ be two alphabets.
If $ R \subseteq M^*_1 $ is a regular set and $ \LL \subseteq M^*_1 \times M^*_2 $
is a rational relation then the relation
\[
\LL \setminus R \; = \; \{ \; ( x, y ) \inn M^*_1 \times M^*_2 \;\; | \;\;
\exists z : zx \inn R \;\; \mbox{\rm\ and } \;\; (z,y) \inn \LL \; \}
\]
is recognizable.
\end{lemma}

\proof{}
Let $ F = ( K, M_1 , T, h , A ) $ be a deterministic finite automaton accepting $R$;
we use the notation of~\cite{bib:hop}.
For each $ p \inn K $, denote $ R_{hp} $ the language accepted by
$ ( K, M_1 , T, h , \{ p \} ) $, and $ R_{pA} $ the language accepted by
$ ( K, M_1 , T, p , A ) $.
Define
\[
\LL ( R_{hp} ) \; = \; \{ \; y \inn M^*_2 \; | \;
\exists x \inn R_{hp} : (x,y) \inn \LL \; \} \; .
\]
Now
\[
\LL \setminus R \; = \; \bigcup_{p \in K} R_{pA} \times \LL ( R_{hp} )
\]
and each $ \LL ( R_{hp} ) $ is regular.
It follows that $ \LL \setminus R $ is recognizable.
\qed

\begin{lemma}
    \label{res-8-11}
Let \prot\ be a cyclic CFSM protocol with the communication graph
$ G = (N,E) $ where $ E = \{ \alpha_0 , \alpha_1 , \ldots , \alpha_m \} $,
$ - \alpha_0 = + \alpha_m$,
$ - \alpha_1 = + \alpha_0$,
$\ldots$,
$ - \alpha_m = + \alpha_{m-1}$:
\begin{center}
\begin{picture}(90,80)
\thicklines
\put(24,72){\vector(1,0){24}}
\put(52,72){\vector(1,-1){20}}
\put(72,48){\vector(0,-1){24}}
\put(72,20){\vector(-1,-1){20}}
\put(0,24){\vector(0,1){24}}
\put(0,52){\vector(1,1){20}}
\put(12,12){\circle*{2}}
\put(20,6){\circle*{2}}
\put(28,0){\circle*{2}}
\put(36,79){\makebox(0,0){$\alpha_0$}}
\put(70,64){\makebox(0,0){$\alpha_1$}}
\put(82,36){\makebox(0,0){$\alpha_2$}}
\put(2,64){\makebox(0,0){$\alpha_m$}}
\end{picture}
\end{center}
If \gli\ is a reachable global state such that $ C' = \xchannel $,
$ x_{\alpha_i} = \lambda $ for $ k+1 < i \leq m $,
then there are a reachable global state \gl\ and a path $ \Gamma $
from \gl\ to \gli\ such that \\
(a) $ C = \ychannel $, $ y_{\alpha_i} = \lambda $ for $ k < i \leq m $; \\
(b) \Image{\Gamma} is a trivial path (of length 0) for each $ i \neq + \alpha_k $.
\end{lemma}

\proof{}
We use the same priority argument as in the proof of~\ref{res-5-2}.
There is a path from \glo\ to \gli;
rearrange it by giving the lowest priority to the node
$ + \alpha_k = - \alpha_{k+1} $.
Let $\Gamma$ be the longest suffix of the rearranged path for which (b) holds.
Let \gl\ be the starting global state of $ \Gamma $.
Then $ C = \ychannel $ must satisfy (a):
If $ y_{\alpha_{k+1}} \neq \lambda $ then $ \Gamma $ could be made one step longer;
if $ y_{\alpha_i} \neq \lambda $ for some $ i > k+1 $ then $ \Gamma $ could not
lead to \gli.
\qed

\proof{ of \ref{res-8-3}}
Clearly (a)$\Rightarrow$(b) and (c)$\Rightarrow$(d).
To prove the implication (b)$\Rightarrow$(c), observe that \\[-16pt]
\[
Q_\beta (S) \times
\Timesbi{\xi\in E}{\xi\neq\beta}
\{ \lambda \} \; = \; \LL ( S ) \cap
\{ \;\xchannel \; | \; x_\xi = \lambda \; \mbox{\rm\ for } \; \xi \neq \beta \; \} \; .
\]
The relation
$ \{ \xchannel \; | \; x_\xi = \lambda \; \mbox{\rm\ for } \; \xi \neq \beta \} $
is recognizable, hence the right hand side is rational (\cite{bib:ber}, p.~57).
Since $ Q_\beta (S) $ is a homomorphic image of the left hand side, it follows that
$ Q_\beta (S) $ is regular.

It remains to be shown that (d)$\Rightarrow$(a)
(this is the only part of the proof that uses the fact that \prot\ is cyclic).
Assume, without loss of generality, that
$ E = \{ \alpha_0 , \alpha_1 , \ldots , \alpha_m \} $,
$ - \alpha_0 = + \alpha_m$,
$ - \alpha_1 = + \alpha_0$,
$\ldots$,
$ - \alpha_m = + \alpha_{m-1}$, and
$ \beta = \alpha_0 $.
By induction on $k$ we show that the relation
\[
\LL_k (S) \; = \; \{ \; \xchannel \in \LL(S) \; | \;
x_{\alpha_i} = \lambda \; \mbox{\rm\ for } \; k+1 \leq i \leq m \; \}
\]
is recognizable for $ 0 \leq k \leq m $ and every $S$.
As $ \LL_m (S) = \LL(S) $, this proves (a).

Induction basis:
$ \LL_0 (S) = Q_\beta (S) \times
\Timesbi{\xi\in E}{\xi\neq\beta}
\{ \lambda \} $
and  $ Q_\beta (S) $ is regular, hence $ \LL_0 (S) $ is recognizable
(for every $S$).

Induction step:
Assume that $ 0 \leq k < m $ and $ \LL_k (S) $ is recognizable for every $S$.
Thus
\[
\LL_k (S) \; = \; \bigcup_{\nu=0}^{r(S)} \; \Timesb{m}{i=0} \; Q_{\nu i} (S) \; ,
\]
where every set $ Q_{\nu i} (S) $ is regular,
$ Q_{\nu i} (S) \subseteq M_{\alpha_i}^* $, and
$ Q_{\nu i} = \{ \lambda \} $ for $ k < i \leq m $.

For each $S'$, the relation $ \LL_{k+1} (S') $ can be expressed in terms of
the relations $ \LL_k (S) $,
$ S \inn \Timesa{j\in N} K_j $, and the finite state machine
$F_n = ( K_n , \Sigma_n , T_n , h_n ) $, where $ n = + \alpha_k = -\alpha_{k+1} $:
Write  $ S' = \pstate $ and for every $ q \inn K_n $ denote by
$ \RR (q) \subseteq M^*_{\alpha_k} \times M^*_{\alpha_{k+1}} $ the rational
relation defined by the transducer
\[
( K_n , M_{\alpha_k} , M_{\alpha_{k+1}} , T_n , q , \{ p_n \} ) \; .
\]
Let $ S' ( q ) = ( q_j : j \inn N ) $ where $ q_j = p_j $ for $ j \neq n $
and $ q_n = q $.
By Lemma~\ref{res-8-11},
\[
\LL_{k+1} ( S' ) \; = \;
\bigcup_{q \in K_n } \bigcup_{\nu=0}^{r(S'(q))}
\left[ \Timesai{m}{i=0}{i\neq k , k+1}
    \!\! Q_{\nu i } ( S' (q) ) \;\;\times\;\;
    ( \RR(q) \setminus Q_{\nu k} ( S' (q)))
\right]
\]
in the notation of Lemma~\ref{res-8-10}.
Hence $ \LL_{k+1} ( S' ) $ is recognizable by~\ref{res-8-10}.
\qed

The next two lemmas, \ref{res-8-12} and \ref{res-8-13},
are used in the proof of \ref{res-8-5}.

\begin{lemma}
    \label{res-8-12}
A family of sets\/ $Q(S)$ is consistent (with respect to a cyclic protocol \prot\
and an edge $\beta\inn E$) if and only if the following three conditions
are satisfied: \\
(a) If \Trans{(S,\xchannel)}{+b}{(S',(x'_\xi : \xi\inn E ))},
$ b \inn M_\beta $, $ x_\xi = x'_\xi = \lambda $ for $ \xi \neq \beta $,
and $ x_\beta \inn Q(S) $ then $ x'_\beta \inn Q(S') $. \\
(b) If \Trans{(S,\xchannel)}{-b}{(S',(x'_\xi : \xi\inn E ))},
$ b \inn M_\beta $, $ x_\xi = x'_\xi = \lambda $ for $ \xi \neq \beta $,
and $ x_\beta \inn Q(S) $ then $ x'_\beta \inn Q(S') $. \\
(c) If there is a path $\Gamma$ from $ ( S, C^0 ) $ to $ ( S' , C^0 ) $
whose no step is labelled $+b$ or $-b$, $ b \inn M_\beta $,
then $ Q(S) \subseteq Q'(S') $.
\end{lemma}

\proof{}
Observe that (c) is equivalent to the following, formally stronger, condition: \\
(d) If there is a path $\Gamma$ from $(S,\xchannel)$ to $(S',(x'_\xi : \xi\inn E ))$,
$ x_\xi = x'_\xi = \lambda $ for $ \xi \neq \beta $, $ x_\beta \inn Q(S)$ and no step
in $\Gamma$ is labelled $+b$ or $-b$, $ b \inn M_\beta $,
then $ x_\beta = x'_\beta \inn Q(S')$.

It is clear that (a), (b) and (d) each are necessary for the consistency of $Q(S)$.
To prove that the three conditions together are also sufficient, take any path
$\Gamma$ in the global state space,
say from $(S,\xchannel)$ to $(S',(x'_\xi : \xi\inn E ))$,
such that $ x_\xi = x'_\xi = \lambda $ for $ \xi \neq \beta $ and $ x_\beta \inn Q(S)$.
Using the priority argument again, rearrange $\Gamma$ so that
$ y_\xi = y'_\xi = \lambda $ for $ \xi \neq \beta $ whenever
\Trans{(S_1 , \ychannel)}{+b}{(S_2 ,(y'_\xi : \xi\inn E ))} or
\Trans{(S_1 , \ychannel)}{-b}{(S_2 ,(y'_\xi : \xi\inn E ))}
is a step in the rearranged path.
Thus the rearranged path is a concatenation of paths to each of which
either (a) or (b) or (d) applies.
It follows that $ x'_\beta \inn Q(S') $.
\qed

\begin{lemma}
    \label{res-8-13}
Let \prot\ be a cyclic CFSM protocol and $ \beta \inn E $.
Then there is an algorithm to find, for any composite state $S$,
every composite state $S'$ for which there is a path from
$ ( S, C^0 ) $ to $ ( S' , C^0 ) $ with no step labelled $+b$ or $-b$,
$ b \inn M_\beta $.
\end{lemma}

\proof{}
Construct the following directed graph $H$.
The nodes of $H$ are the composite states of \prot.
There is an edge in $H$ from $S_1$ to $S_2$ iff there exists
$\xi\inn E$, $\xi\neq\beta$, such that $ S_1 = \pstate $, $ S_2 = \qstate $,
$ p_j = q_j $ for $ j \neq +\xi, -\xi $, and
\trans{p_{-\xi}}{-b}{q_{-\xi}}, \trans{p_{+\xi}}{+b}{q_{+\xi}}
for some $ b \inn M_\xi$.
Now $S'$ can be reached from $S$ by a directed path in $H$ if and only if
there is a path $\Gamma$ from $ ( S , C^0 ) $ to $ ( S' , C^0 ) $
in the global state space such that no step of $\Gamma$ is labelled
$+b$ or $-b$, $ b \inn M_\beta $.
Hence the property can be decided by the standard reachability
(transitive closure) algorithm in the graph $H$.
\qed

\proof{ of \ref{res-8-5}}
To prove that there is an algorithm to decide the consistency of a family
of regular sets $Q(S)$, we construct algorithms to decide the properties
(a), (b) and (c) in Lemma~\ref{res-8-12}.

It is easy to check (a).
The condition says that if \trans{p}{+b}{q} in $ F_{+\beta} $,
$ S =\pstate $, $ S' = \qstate $, $ p_j = q_j $ for $ j \neq +\beta $,
$ p_{+\beta} = p $ and $ q_{+\beta} = q $, then
\[
\{ \; x \; | \; bx \inn Q(S) \; \} \;\subseteq\; Q(S') \; .
\]
The inclusion is algorithmically decidable for regular sets $Q(S)$ and $Q(S')$.

A similar algorithm decides (b).

The algorithm to decide (c) has two components.
The first, based on the algorithm in Lemma~\ref{res-8-13}, finds every pair of
composite states $S$ and $S'$ for which there is a path $\Gamma$ from
$ (S, C^0 ) $ to $ (S', C^0 ) $ whose no step is labelled $+b$ or $-b$,
$ b \inn M_\beta $.
The second component of the algorithm checks the inclusion  $ Q(S) \subseteq Q(S') $.
\qed

\proof{ of \ref{res-8-6}}
Let \prot\ be a cyclic CFSM protocol and $ \beta \inn E $.
If a composite state $S'$ is not stable then there is a consistent family
of regular sets, namely the sets $ Q_\beta ( S) $ of Theorem~\ref{res-8-3},
such that $ \lambda \inn Q_\beta ( S^0 ) $ and $ \lambda \not\in Q_\beta (S' ) $.

Conversely, if $S'$ is stable then \TTrans{\glo}{(S',C^0 )};
hence for any consistent family of sets $Q(S)$, regular or not,
such that $ \lambda \inn Q( S^0 ) $, we must have $ \lambda \inn Q (S' ) $.
\qed

\proof{ of \ref{res-8-7}}
An algorithm to decide the deadlock problem combines two semialgorithms,
one of which always terminates.

The first searches for a deadlock, using the exhaustive reachability analysis.
It terminates whenever the protocol allows a deadlock.

The second semialgorithm searches for a proof of deadlock-freedom in the form
of a consistent family of regular sets $Q(S)$ such that $ \lambda \inn Q(S^0 ) $
and $ \lambda \not\in Q(S) $ whenever $S$ consists solely of receive states.
It terminates if the protocol is deadlock-free.
\qed

\proof{ of \ref{res-8-8}}
Construct a finite state automaton $F$ with $\lambda$-transitions as follows:
The states of $F$ are the composite states of \prot.
There is a $\lambda$-transition from $S_1$ to $S_2$ in $F$ iff the graph $H$
in the proof of~\ref{res-8-13} has an edge from $S_1$ to $S_2$.
There is a transition from $ S_1 = \pstate $ to $ S_2 = \qstate $ labelled $b$,
$ b \inn M_\beta $, iff $ p_j = q_j $ for $ j \neq + \beta $ and
\trans{p_{+\beta}}{+b}{q_{+\beta}}.
Write \trans{S'}{w}{S}, $ w \inn M^*_\beta $, if the automaton $F$ can move from
$S'$ to $S$ by reading $w$.

For a given family of regular sets $Q(S)$, $ S \inn \Timesa{j \in N} V_j $, define
\[
Q(S) \; = \; \{ \; y \inn M^*_\beta \; | \;
\exists\, S' \inn \Timesa{j \in N} V_j \;\;\; \exists\, x \inn M^*_\beta \; : \;
\trans{S'}{x}{S} \;\; \mbox{\rm and } \; xy \inn Q(S') \; \}
\]
for every $ S \inn \Timesa{j \in N} K_j - \Timesa{j \in N} V_j $.

Both the given sets and the newly defined ones are regular.
In view of~\ref{res-8-5}, it is now sufficient to prove this lemma:

\begin{lemma}
    \label{res-8-14}
\samepage
If there is a consistent family $ Q'(S)$, $ S \inn \Timesa{j \in N} K_j $,
such that $ Q'(S) = Q(S) $ for every $ S \inn \Timesa{j \in N} V_j $, then \\
(a) $ Q(S) \subseteq Q'(S) $ for every $ S \in \Timesa{j \in N} K_j$ ; and \\
(b) the family $ Q(S) $, $ S \inn \Timesa{j \in N} K_j \,$, is consistent.
\end{lemma}

\proof{}
(a) Let $ x \inn Q(S) $, $ S \inn \Timesa{j \in N} K_j - \Timesa{j \in N} V_j $.
Write $ C = \xchannel $ where $ x_\beta = x $ and $ x_\xi = \lambda $ for $ \xi \neq \beta $.
From the definition of $Q(S)$ it follows that there are
$ S' \inn \Timesa{j \in N} V_j $ and $ C' = ( x'_\xi : \xi \inn E ) $
such that $ x'_\xi = \lambda $ for $ \xi \neq \beta $,
$ x'_\beta \inn Q(S') = Q'(S') $ and \TTrans{\gli}{\gl}.
Hence $ x \inn Q'(S) $ and, since $ x \inn Q(S) $ is arbitrary,
$ Q(S) \subseteq Q'(S) $.

(b) Let $ C = \xchannel $, $ C' = ( x'_\xi \! :\! \xi \inn E ) $,
$ x_\xi = x'_\xi = \lambda $ for $ \xi \neq \beta $, $ x'_\beta \inn Q(S') $
and \TTrans{\gli}{\gl}.
It is to be shown that $ x_\beta \inn Q(S) $.
We distinguish three cases:

\noindent
I. $ S \inn \Timesa{j \in N} V_j $; then the inclusion in (a) and the consistency
of $ Q'(S) $ imply that $ x_\beta \inn Q(S) $.

\noindent
II. $ S \inn \Timesa{j \in N} K_j - \Timesa{j \in N} V_j $
and $ S' \inn \Timesa{j \in N} V_j $.
Since \TTrans{\gli}{\gl}, there is a path $\Gamma$ from \gli\ to \gl\
in the global state space.
We assume, again, that $ E = \{ \alpha_0 , \alpha_1 , \ldots , \alpha_m \} $,
$ - \alpha_0 = + \alpha_m$,
$ - \alpha_1 = + \alpha_0$,
$\ldots$,
$ - \alpha_m = + \alpha_{m-1}$, and $ \beta = \alpha_0 $.
As before, we rearrange the path $\Gamma$ by using the highest priority at $+\alpha_m$,
the next at $+\alpha_{m-1}$, etc., with the lowest priority at $+\alpha_0$.
In the rearranged path, let $\Gamma_0$ be the longest prefix whose last step
is labelled $-b$, $b \inn M_\beta$, and let $ \Gamma_1 $ be the remaining suffix
of the path.
Thus $ \Gamma_1 $ is the longest suffix whose no step is labelled $-b$, $b \inn M_\beta$,
and the path $ \Gamma_0 \Gamma_1 $ is locally equal to $ \Gamma $.
The path $ \Gamma_0 $ leads from \gli\ to \glii, say, with
$ C'' = ( x''_\xi : \xi \inn E ) $.
From the choice of priorities it follows that $ S'' \inn \Timesa{j \in N} V_j $
and $ x''_\xi = \lambda $ for $ \xi \neq \beta $.
At the same time, $ \Gamma_1 $ defines a sequence of transitions from $S''$ to $S$
in the automaton $F$; let $ y \inn M^*_\beta $ be the corresponding input of $F$,
i.e. \trans{S''}{y}{S}.
Then $ y x_\beta = x''_\beta \inn Q(S'') $ and, therefore, $ x_\beta \inn Q(S) $.

\noindent
III. $ S, S' \inn \Timesa{j \in N} K_j - \Timesa{j \in N} V_j $.
By the definition of $ Q(S')$, there are $ S'' \inn \Timesa{j \in N} V_j $
and $ C'' = ( x''_\xi : \xi \inn E ) $ such that $ x''_\xi = \lambda $ for $ \xi \neq \beta $,
$ x''_\beta \inn Q ( C'' ) $ and \TTrans{\glii}{\gli}.
Hence \TTrans{\glii}{\gl} and the result follows from the already proved case II.

This completes the proofs of~\ref{res-8-14} and~\ref{res-8-8}.
\qed


\section{Recognizable channels for general protocols}
    \label{sec-9}

By Theorem~\ref{res-8-3}, the rational and the recognizable channel properties are
equivalent for cyclic protocols. We begin this section by showing that
the two properties differ in general.

\begin{example}
    \label{res-9-1}
\rm
The communication graph is \twographii; both $M_\alpha$ and $M_\beta$ contain
a single symbol:
$M_\alpha = \{d\}$, $M_\beta = \{b\}$.
The transition diagrams of the two finite state machines are in Fig.~9.1.
We have
\begin{eqnarray*}
\LL ( ( 00,10 )) & = & \{ ( d^n , b^n ) \; | \; n\geq 0 \} \\
\LL ( ( 00,11 )) & = & \{ ( d^n , b^{n+1} ) \; | \; n\geq 0 \} \\
\LL ( ( 01,10 )) & = & \{ ( d^{n+1} , b^n ) \; | \; n\geq 0 \} \\
\LL ( ( 01,11 )) & = & \{ ( d^n , b^n ) \; | \; n\geq 0 \}
\end{eqnarray*}
All these relations are rational, but none is recognizable. \\
(End of Example~\ref{res-9-1}.)
\end{example}

\begin{figure}[tbp]
\begin{center}
\begin{picture}(190,130)(-20,-15)
\thicklines

\put(0,0){\circle{30}}
\put(-10,-10){\makebox(20,20)[c]{$01$}}
\put(0,100){\circle{30}}
\put(-10,90){\makebox(20,20)[c]{$00$}}

\put(-8,86){\vector(0,-1){72}}
\put(-54,40){\makebox(40,20)[r]{$-d$}}
\put(8,14){\vector(0,1){72}}
\put(14,40){\makebox(40,20)[l]{$-b$}}

\put(170,0){\circle{30}}
\put(160,-10){\makebox(20,20)[c]{$11$}}
\put(170,100){\circle{30}}
\put(160,90){\makebox(20,20)[c]{$10$}}

\put(162,86){\vector(0,-1){72}}
\put(116,40){\makebox(40,20)[r]{$+d$}}
\put(178,14){\vector(0,1){72}}
\put(184,40){\makebox(40,20)[l]{$+b$}}
\end{picture}

\addspu

{\bf Fig. 9.1.}
\end{center}
\end{figure}

The results in section~\ref{sec-8} (particularly Corollary~\ref{res-8-7})
suggest the following problem.

\begin{problem}
    \label{res-9-2}
Is there an algorithm to decide whether an arbitrary CFSM protocol with
the rational channel property is deadlock-free?
\end{problem}

The present section gives a partial solution:
There is an algorithm to decide deadlock-freedom for the CFSM protocols
with the recognizable channel property.
(This also yields another proof of~\ref{res-8-7}.)
The key property of recognizable relations needed in this theory,
and not possessed by rational relations,
is the decidability of inclusion.

The following Definition~\ref{res-9-3} and Theorems~\ref{res-9-4}
through~\ref{res-9-7} are analogous to~\ref{res-8-4}, \ref{res-8-5},
\ref{res-8-6}, \ref{res-8-7} and~\ref{res-8-8}.
The results will be proved at the end of the section.

\begin{definition}
    \label{res-9-3}
\rm
Let \prot\ be a CFSM protocol, and let
$ \RR (S) \subseteq \Timesa{\xi\in E} M^*_\xi $
for $ S \inn \Timesa{j\in N} K_j $.
Say that the relations $ \RR (S)$ are {\em consistent\/}
(with respect to \prot) if
\TTrans{\gl}{\gli} and $ C \inn \RR(S)$ imply $ C' \inn \RR (S') $.
\end{definition}

\begin{theorem}
    \label{res-9-4}
There is an algorithm to decide whether any given family of recognizable
relations $\RR(S)$
is consistent (with respect to a given \prot).
\end{theorem}

\begin{theorem}
    \label{res-9-5}
\samepage
Let \prot\ be a CFSM protocol with the recognizable channel property.
A global state \gli\ is not reachable if and only if there is a consistent
family of recognizable relations $\RR(S)$, $ S \inn \Timesa{j\in N} K_j $,
such that $ C^0 \inn \RR ( S^0 ) $ and $ C' \not\in \RR ( S') $.
\end{theorem}

\begin{corollary}
    \label{res-9-6}
The simple reachability problems (such as the deadlock problem) are
algorithmically decidable for the CFSM protocols
with the recognizable channel property.
\end{corollary}

\begin{theorem}
    \label{res-9-7}
Let \prot\ be a CFSM protocol.
For each $j\inn N$ let $ V_j \subseteq K_j $ be a set of states such that
$ h_j \inn V_j $ and if $ \trans{p_j}{-b}{q_j} $, $p_j , q_j \inn K_j $, then
$ q_j \inn V_j $.
There is an algorithm to decide whether any given family of recognizable relations
indexed by $ S \inn \Timesa{j\in N} V_j $ can be extended to a consistent family
of relations $ \RR(S) $ indexed by $ S \inn \Timesa{j\in N} K_j $.
Moreover, if the family can be extended than the smallest such sets $\RR(S)$
are recognizable and can be automatically constructed.
\end{theorem}

Theorem~\ref{res-9-7} should be compared with the similar result in the next theorem,
which is analogous to placing intermediate assertions in program loops,
as in the Floyd-Hoare invariant assertion method~\cite{bib:man}.

Recall that a {\em feedback vertex set\/} in a directed graph is a set
of vertices that intersects every directed cycle in the graph.
Theorem~\ref{res-9-8} refers to feedback vertex sets in the {\em product graph\/}
(of the protocol \prot).
The nodes of the graph are the composite states of\/ \prot, and the edge
\trans{S}{}{S'} is in the graph iff there exists $i\inn N$ such that
$ S = \pstate $, $ S' = \qstate $, $ p_j = q_j $ for $ j \neq i $, and
the edge \trans{p_i}{}{q_i} is in the transition diagram of $F_i$.

\begin{theorem}
    \label{res-9-8}
\samepage
Let $V$ be a feedback vertex set in the product graph of a CFSM protocol \prot.
There is an algorithm to decide whether any given family of recognizable
sets\/ $\RR(S)$ indexed by $S\inn V$ can be extended to a consistent family
of sets\/ $\RR(S)$ indexed by $ S\inn\Timesa{j\in N} K_j $.
Moreover, if the given family can be extended then the smallest such sets $\RR(S)$,
$ S\inn\Timesa{j\in N} K_j - V $,
are recognizable and can be automatically constructed.
\end{theorem}

The results of this section are to be used to construct automatically verifiable
proofs of reachability properties for the CFSM protocols with the recognizable
channel property on general communication graphs,
in the same way as the results in section~\ref{sec-8} are used for cyclic protocols.
The proofs are again in the form of tables;
the entries are recognizable relations.
Theorems~\ref{res-9-7} and~\ref{res-9-8} help us limit the size of the tables.

The method in this section is in fact more general than the method of regular sets
in section~\ref{sec-8}.
Indeed, we can construct a proof that a general global state \gl\ is unreachable,
whereas previously we could only prove that $ (S,C^0 )$ is unreachable
(i.e. that $S$ is not stable).
We can even decide certain second-order reachability properties:

\begin{theorem}
    \label{res-9-9}
Let \prot\ be a CFSM protocol with the recognizable channel property.
Let $ b \inn M_\beta$, $ p_i \inn K_i $, $ i=+\beta$.
The message $b$ cannot arrive at $p_i$ if and only if there is
a consistent family of recognizable relations $ \RR(S)$,
$ S \inn \Timesa{j\in N} K_j $,
such that $ C^0 \inn \RR(S^0 ) $ and if
$ \xchannel \inn \RR(\pstate) $ then $ x_\beta $ does not begin with $b$.
\end{theorem}

\begin{corollary}
    \label{res-9-10}
The problem ``Can $b$ arrive at $p_i$?'' is algorithmically decidable
for the CFSM protocols with the recognizable channel property.
\end{corollary}

Now we prove \ref{res-9-4} through \ref{res-9-9}.

\proof{ of~\ref{res-9-4}}
Although the consistency of a family $\RR(S)$, $ S \inn \Timesa{j\in N} K_j $,
is defined in terms of the relation \TTrans{}{}, it can be equivalently
defined in terms of \Trans{}{}:
The relations $\RR(S)$ are consistent if and only if
\Trans{\gl}{}{\gli} and $ C \inn \RR(S) $ imply $ C' \inn \RR(S') $.
In other words, $ \RR(S)$ are consistent if and only if these two
conditions hold: \\
(a) If $ S = \pstate $, $ S' = \qstate $, $ i = + \beta $,
$ p_j = q_j $ for $ j \neq i $, and \trans{p_i}{+b}{q_i} in $F_i$, then
\[
\{ \; (x'_\xi \! :\! \xi \inn E ) \;\; | \;\; \exists \xchannel\inn\RR(S) :
x_\xi = x'_\xi \mbox{\rm\ for } \xi \neq \beta \mbox{\rm\ and }
x_\beta = bx'_\beta \; \} \;\subseteq\; \RR(S') \; .
\]
(b) If $ S = \pstate $, $ S' = \qstate $, $ i = - \beta $,
$ p_j = q_j $ for $ j \neq i $, and \trans{p_i}{-b}{q_i} in $F_i$, then
\[
\{ \; (x'_\xi \! :\! \xi \inn E ) \;\; | \;\; \exists \xchannel\inn\RR(S) :
x_\xi = x'_\xi \mbox{\rm\ for } \xi \neq \beta \mbox{\rm\ and }
x_\beta b = x'_\beta \; \} \;\subseteq\; \RR(S') \; .
\]
Since these inclusions are decidable for recognizable relations,
both (a) and (b) are decidable.
\qed

\proof{ of~\ref{res-9-5}}
The proof is similar to that of~\ref{res-8-6}.
If \gli\ is not reachable, then the relations $ \LL(S)$
of Definition~\ref{res-8-2} fulfill the condition.
Namely, $ \LL(S)$ are consistent, $ C^0 \inn \LL(S^0 ) $ and
$ C' \not\in \LL(S') $.

Conversely, if \gli\ is reachable then no consistent family of relations $\RR(S)$,
recognizable or not, satisfies $ C^0 \inn \RR(S^0 ) $ and
$ C' \not\in \RR(S') $.
\qed

\proof{ of~\ref{res-9-6}}
As in the proof of~\ref{res-8-7}, we combine two semialgorithms,
one of which always terminates.

Given a global state \gli, the first semialgorithm
searches for a path from \glo\ to \gli.
It terminates whenever \gli\ is reachable.

The second semialgorithm searches for a proof of non-reachability of \gli,
in the form of a consistent family of recognizable relations $ \RR(S)$
such that $ C^0 \inn \RR(S^0 ) $ and $ C' \not\in \RR(S') $.
Since the protocol has the recognizable channel property,
the semialgorithm terminates whenever \gli\ is not reachable.
\qed

\proof{ of~\ref{res-9-7}}
Define
\[
W^+ (q,p) \; = \; \{ \; b_0 b_1 \ldots b_n \; | \;
b_i \inn M_\xi \;\mbox{\rm\ for }\; 0 \leq i \leq n
\;\mbox{\rm\ and }\;
q \stackrel{+b_0 +b_1 \ldots +b_n}{\looongarrow} p
\; \}
\]
for $ q, p \inn K_{+\xi} $ , and
\[
\WW^+ (S',S) \; = \; \Timesa{j\in N} W^+ (q_j , p_j )
\]
for $ S' = \qstate $ and $ S = \pstate $.
For a given family of recognizable relations $ \RR (S) $,
$ S \inn \Timesa{j\in N} V_j $ , define
\[
\RR(S) \; = \; \bigcup_{S'\in\Timesa{j\in N} V_j}
\{ \; \ychannel \; | \; \exists \xchannel \inn \WW^+ (S', S)
\; : \; ( x_\xi y_\xi : \xi \inn E ) \inn \RR(S') \; \}
\]
for $ S \inn \Timesa{j\in N} K_j - \Timesa{j\in N} V_j $.
All the relations $ \RR(S) $, $ S \inn \Timesa{j\in N} K_j $,
are recognizable, and Theorem~\ref{res-9-7} follows from this lemma:

\begin{lemma}
    \label{res-9-11}
If there is a consistent family $ \RR'(S) $, $ S \inn \Timesa{j\in N} K_j\,$,
such that $ \RR'(S) = \RR(S) $ for every $ S \inn \Timesa{j\in N} V_j\,$,
then \\
(a) $ \RR(S) \subseteq \RR'(S) $ for every $ S \inn \Timesa{j\in N} K_j $; and \\
(b) the family $ \RR(S) $, $ S \inn \Timesa{j\in N} K_j\,$, is consistent.
\end{lemma}

\proof{ of~\ref{res-9-11}}
(a) Let $ C \inn \RR(S) $, $ S \inn \Timesa{j\in N} K_j - \Timesa{j\in N} V_j $.
By the definition of $ \RR(S)$, there is $ S' \inn \Timesa{j\in N} V_j $
such that \TTrans{\gli}{\gl} for some $ C' \inn \RR(S') $.
Since the relations $ \RR'(S) $ are consistent, it follows that
$ C \inn \RR'(S) $.
Hence $ \RR(S) \subseteq \RR'(S) $.

\noindent
(b) Let \TTrans{\gli}{\gl} and $ C' \inn \RR(S') $.
We want to prove that $ C \inn \RR(S) $.
We distinguish three cases:

\noindent
I. $ S \inn \Timesa{j \in N} V_j $.
Then the inclusion in (a) and the consistency
of $ \RR'(S) $ imply $ C \inn \RR(S) $.

\noindent
II. $ S \inn \Timesa{j \in N} K_j - \Timesa{j \in N} V_j $
and $ S' \inn \Timesa{j \in N} V_j $.
Since \TTrans{\gli}{\gl}, there is a path $\Gamma$ from \gli\ to \gl.
There are two paths $\Gamma'$ and $\Gamma''$ such that $\Gamma' \Gamma''$
is locally equal to $\Gamma$,
the end state \glii\ of $\Gamma'$ satisfies $ S'' \inn \Timesa{j\in N} V_j $,
and all the steps in $\Gamma''$ are receptions
(i.e. are labelled $+b$).
Since $\RR'(S)$ are consistent and $ \RR'(S') = \RR(S') $ and
$ \RR'(S'') = \RR(S'') $, it follows that $ C'' \inn \RR(S'') $.
The path $ \Gamma''$ defines a vector $ \xchannel \inn \WW^+ (S'' , S) $,
and with $ C = \ychannel $ we have
$ ( x_\xi y_\xi : \xi \inn E ) = C'' \inn \RR(S'') $.
By the definition of $ \RR(S) $ we get $ C \inn \RR(S) $.

\noindent
III. $ S, S' \inn \Timesa{j \in N} K_j - \Timesa{j \in N} V_j $.
By the definition of $ \RR(S')$, there is a global state \glii\ such that
$ S'' \inn \Timesa{j \in N} V_j\,$, \TTrans{\glii}{\gli} and
$ C'' \inn \RR(S'') $.
Hence \TTrans{\glii}{\gl} and we apply the already proved case II.
\qed

\proof{ of~\ref{res-9-8}}
We start with the given recognizable relations $\RR(S)$, $ S\inn V$,
and first define relations $ \RR(S)$, for $ S\not\in V $, as follows.
For $ S \inn \Timesa{j\in N} K_j - V $, let $ \RR(S) $ be the set of
all those $ C \inn \Timesa{\xi\in E} M^*_\xi $ for which there are
$ S' \inn V $, $ C' \inn \RR ( S' ) $, and a path from \gli\
to \gl\ such that no composite state $S''$ along the path (except $S'$)
belongs to $V$.
Since $V$ is a feedback vertex set, no such path can pass through the same
composite state twice.
Hence the length of all such paths is bounded, and therefore the sets $\RR(S)$,
$ S \inn \Timesa{j\in N} K_j - V $, are recognizable and automatically constructible.
The result now follows from this lemma:

\begin{lemma}
    \label{res-9-12}
If there is a consistent family $ \RR'(S)$, $ S \inn \Timesa{j\in N} K_j $,
such that $ \RR'(S) = \RR(S) $ for every $S\inn V$, then \\
(a) $ \RR(S) \subseteq \RR'(S) $ for every $S$; and \\
(b) the family $ \RR(S) $, $ S \inn \Timesa{j\in N} K_j\, $, is consistent.
\end{lemma}

\proofnodot{ of~\ref{res-9-12}} is similar to, but simpler than,
that of~\ref{res-9-11}.

\noindent
(a) Let $ C \inn \RR(S) $, $ S \inn \Timesa{j\in N} K_j - V $.
There are $ S' \inn V$ and $ C' \inn \RR(S') $ such that \TTrans{\gli}{\gl}.
Since the relations $ \RR'(S) $ are consistent, $ C \inn \RR'(S) $.
Hence $ \RR(S) \subseteq \RR'(S) $.

\noindent
(b) Let \TTrans{\gli}{\gl} and $ C' \inn \RR(S') $.
We want to prove that $ C \inn \RR(S) $.
We distinguish three cases:

\noindent
I. $ S \inn V$.
Then the inclusion in (a) and the consistency of $ \RR'(S) $ imply $ C \inn \RR(S) $.

\noindent
II. $ S \not\in V $ and $ S' \in V $.
There is a path $\Gamma$ from \gli\ to \gl.
Let $\Gamma''$, from \glii\ to \gl, be the shortest suffix of $\Gamma$ such that
$ S'' \inn V $.
Thus $ \Gamma = \Gamma' \Gamma'' $,
$ \Gamma' $ leads from \gli\ to \glii, and $ C'' \inn \RR(S'') $.
No composite state along $ \Gamma'' $ (except $S''$) belongs to $V$,
hence $ C \inn \RR(S) $ by the definition of $ \RR(S) $.

\noindent
III. $ S \not\in V $ and $ S' \not\in V $ .
By the definition of $ \RR(S') $, there is a global state \glii\ such that $ S'' \inn V $,
and a path from \glii\ to \gli.
Hence there is a path from \glii\ to \gl, and the result follows from case II.
\qed

\proof{ of~\ref{res-9-9}}
If $b$ cannot arrive at $p_i$ then the relations $ \LL(S) $ of Definition~\ref{res-8-2}
fulfill the condition.
Conversely, if there is a consistent family of recognizable relations
$ \RR(S) $, $ S \inn \Timesa{j\in N} K_j\,$, such that $ C^0 \inn \RR(S^0 ) $ and
$ x_\beta$ does not begin with $b$ whenever $ \xchannel \inn \RR(\pstate) $, then,
by Theorem~\ref{res-9-5}, no global state $ ( \pstate, \xchannel ) $ in which $ x_\beta $
begins with $b$ is reachable.
In other words, $b$ cannot arrive at $p_i$.
\qed

\proof{ of~\ref{res-9-10}}
Again it is sufficient to show that if $b$ cannot arrive at $p_i$ then there is
an algorithmically verifiable proof.
This follows from the previous results in this section and from the following:
There is an algorithm to decide, for every recognizable relation
$ \RR \subseteq \Timesa{\xi\in E} M^*_\xi \,$, every $ \beta \inn  E $ and every
$ b \inn M_\beta $, whether there is $ \xchannel \inn \RR $ such that $ x_\beta $
begins with $b$.
\qed


\section{Abstract flow control in general graphs}
    \label{sec-10}

We now return to the idea of abstract flow control, introduced in section~\ref{sec-5}
for cyclic graphs.
Recall that our first aim is to limit the number of locally equal paths
to be examined by the reachability algorithms.
This alone is easily achieved;
we can order all nodes of the communication graph by assigning them
distinct priorities, and thus select a {\em unique\/} path in every class
of locally equal paths.

However, not all such priority assignments are of equal value.
Our second aim is to choose locally equal paths that use a small number
of global states.
Two methods for making the choice, leading to two different priority schemes,
are described in this section.
Then the priority arguments are applied to give a partial solution of
the reachability problem for the rational channel CFSM protocols.

Let $\Gamma$ be a path in the global state space of a CFSM protocol.
Suppose that $ \Phi\gl $ is a proposition applicable to every global state \gl;
that is, $ \Phi\gl $ is a (true or false) statement for every \gl.
Say that $ \Phi\gl $ is true {\em frequently along \/} $\Gamma$
if $ \Phi\gl $ is true for at least one of every two consecutive global states
along $\Gamma$.
In particular, if $\beta$ is an edge in the communication graph and the statement
``if $ C = \xchannel $ then $ x_\beta = \lambda $'' is true frequently along $\Gamma$,
then the transmissions and receptions on the channel $\beta$ are tightly coupled
in the execution described by $\Gamma$;
in other words, every symbol sent on $\beta$ is received at once
(in the next step).

The first result uses collections of noncrossing boundaries in
the communication graph;
the concept is somewhat similar to the laminar collection of (or valuation on)
directed cuts in a directed graph, in the sense of Lucchesi and Younger~\cite{bib:luc}.
Let $ G = ( N, E ) $ be a directed graph.
For $ A \subseteq N $ denote
\begin{eqnarray*}
\partial^- (A) & = & \{ \; \xi\inn E \; | \; +\!\xi\inn A
    \;\mbox{\rm\ and }\; -\!\xi\!\not\in\! A \; \} \\
\partial^+ (A) & = & \{ \; \xi\inn E \; | \; -\!\xi\inn A
    \;\mbox{\rm\ and }\; +\!\xi\!\not\in\! A \; \}
\end{eqnarray*}
and call the sets $\partial^- (A)$ and $\partial^+ (A)$ the {\em negative\/} and
the {\em positive boundary\/} of $A$.

A set $\Psi$ of subsets of $N$ is {\em smooth\/} if for all $ A, B \inn \Psi$
we have (i) $ A \subseteq B $ or (ii) $ B \subseteq A $ or
(iii) $ A \cap B = \emptyset\; $ and
$ \;\partial^- (A\cup B) = \partial^- (A) \cup \partial^- (B) $.

\begin{theorem}
    \label{res-10-1}
Let $ G = ( N, E ) $ be the communication graph of a CFSM protocol and let
$\Psi$ be a smooth set of subsets of $N$.
For every path that ends in a global state with empty channels,
there exists a locally equal path along which the following is frequently true:
\[
\forall A\inn \Psi \;\; \exists \beta \inn \partial^- (A) \; : \;\;
\mbox{\rm if }\; C = \xchannel\;\; \mbox{\rm then }\; x_\beta = \lambda \; .
\]
\end{theorem}

\proof{ of~\ref{res-10-1}}
Order the sets in $\Psi$ in a sequence $ A_0 , A_1 , \ldots , A_n $ such that
if $ A_i \subseteq A_j $ then $ i \leq j $.
Set $ B_0 = A_0 $ and $\displaystyle B_k = A_k - \bigcup_{i=0}^{k-1} A_i$ for $ k > 0 $.
Let $\Gamma$ be a path ending in a global state with empty channels.
We rearrange $\Gamma$ by executing the processes in $B_0$ with the highest
priority, those in $B_1$ with the second highest, etc.

Formally, if $ \Gamma $ contains two adjacent steps
\begin{eqnarray*}
\Gamma_1 & : & \Trans{(S_1 , C_1 )}{e_1}{(S_2 , C_2 )} \\
\Gamma_2 & : & \Trans{(S_2 , C_2 )}{e_2}{(S_3 , C_3 )}
\end{eqnarray*}
such that \Imagei{{i_1}}{\Gamma_1} and \Imagei{{i_2}}{\Gamma_2} are nontrivial paths,
$ i_1 \inn B_{j_1} $, $ i_2 \inn B_{j_2} $, $ j_1 > j_2 $, and if it is {\em not\/}
the case that $ e_2 = +b $, $ b\inn M_\beta $, $ C_1 = \xchannel $ and $ x_\beta = \lambda $,
then we replace the subpath $ \Gamma_1 \Gamma_2 $ in $\Gamma$ by the path
\Trans{(S_1 , C_1 )}{e_2}{(S_4 , C_4 )} \Transi{e_1} $(S_3 , C_3 )$
for a suitable $ (S_4 , C_4 )$.
We repeat the same with the new path, etc., until no further transformation is possible.
Let $ \Gamma'$ be the path constructed by this process.
We wish to show that
\[
\forall A\inn \Psi \;\; \exists \beta \inn \partial^- (A) \; : \;\;
\mbox{\rm if }\; C = \xchannel\;\; \mbox{\rm then }\; x_\beta = \lambda
\]
frequently along $\Gamma'$.

If not then there are two consecutive global states \gl\ and \gli\ in $\Gamma'$ and
two sets $ A, A' \inn \Psi $ such that $ C = \xchannel $,
$ C' = ( x'_\xi : \xi\inn E ) $ and
\begin{eqnarray*}
\forall \xi \inn \partial^- (A) & : & x_\xi \neq \lambda \\
\forall \xi \inn \partial^- (A') & : & x'_\xi \neq \lambda
\end{eqnarray*}
First observe that we can assume, without loss of generality, that $ A = A'$.
Indeed, if the move from \gl\ to \gli\ is a reception on a channel
$ \beta \inn \partial^- (A) $ then
\[
\forall \xi \inn \partial^- (A') \; : \; x_\xi \neq \lambda ,
\]
and if the move from \gl\ to \gli\ is not a reception on a channel in $ \partial^- (A) $
then
\[
\forall \xi \inn \partial^- (A) \; : \; x'_\xi \neq \lambda .
\]

Now assume $A=A'$.
Since $\Gamma'$ ends in a global state with empty channels, there is a later step
\Transi{+b} in $\Gamma'$, for some $ b \inn M_\beta $,
$ \beta \inn \partial^- (A)$.
Taking the first such step, say \Trans{(S_1 , C_1)}{+b}{(S_2 , C_2 )}, we get
a contradiction with the construction of $\Gamma'$:
We have $ b \inn M_\beta $, $ \beta \inn \partial^- (A) $ and from the properties of
$\Psi$ it follows that $ +\beta\inn B_i$, $-\beta\inn B_j$, $i<j$.
Hence the step \Trans{(S_1 , C_1)}{+b}{(S_2 , C_2 )} could be exchanged with the
previous step in $\Gamma'$, contrary to the assumption that no further
transformation is applicable to $\Gamma'$.
\qed

Observe that Theorem~\ref{res-5-2} follows immediately from~\ref{res-10-1}.
Indeed, with the notation of~\ref{res-5-2}, there is a smooth set $\Psi$ of subsets
of $N$ such that $ \{\beta\} = \partial^+ (A)$ for every $ A\inn\Psi$ and
\[
\{ \; \partial^- (A) \; | \; A\inn\Psi \; \} \; = \;
\{ \; \{\xi\} \; | \; \xi \in E - \{\beta\} \; \} \; .
\]
Fig.~10.1 shows such a set $\Psi$ for a cyclic protocol whose graph has four nodes.
\begin{figure}[tbp]
\begin{center}
\includegraphics[height=3.3in]{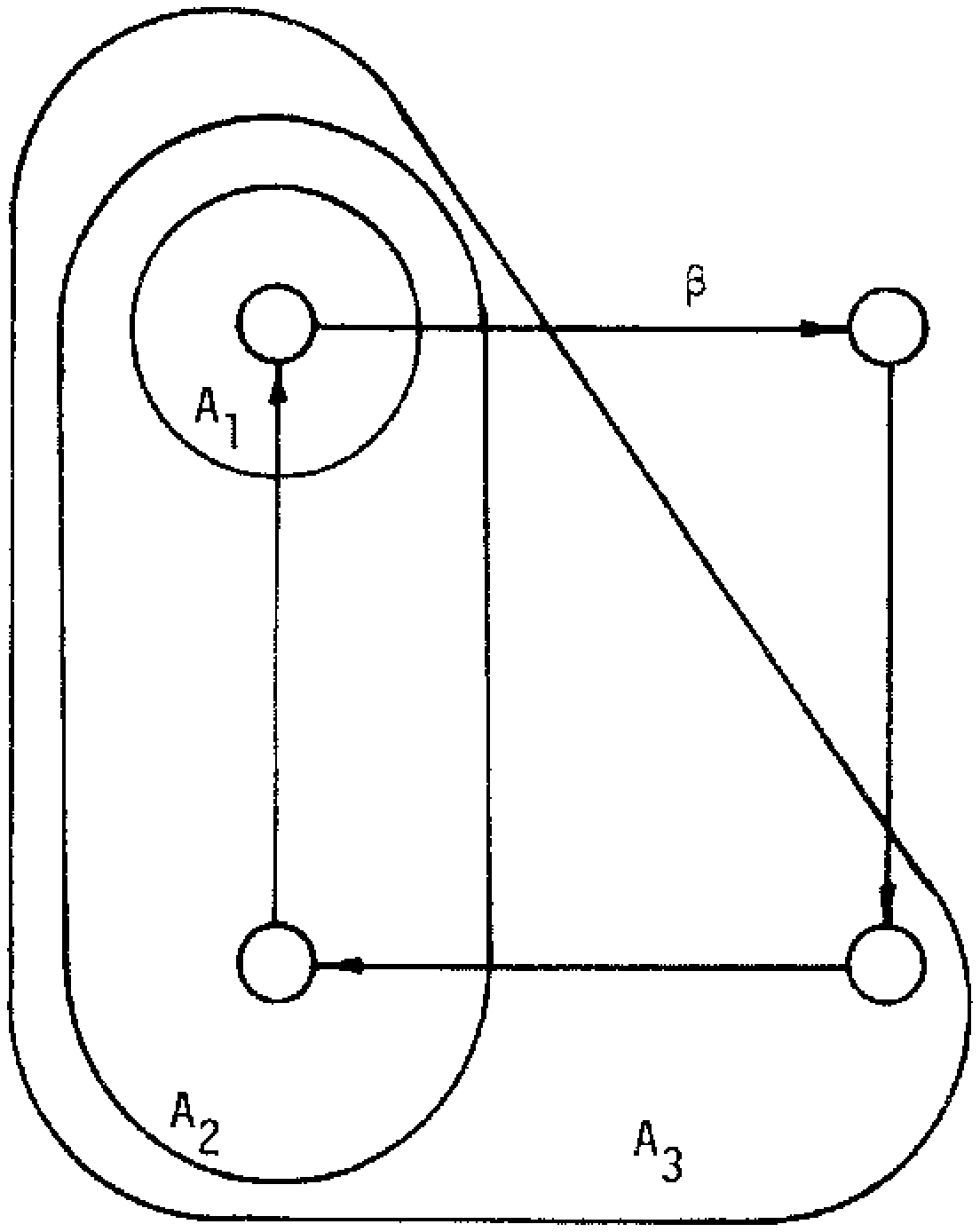}
\addspu\\
{\bf Fig. 10.1. The smooth set $ \{\; A_1 , A_2 , A_3 \;\}$.}
\end{center}
\end{figure}

The priorities in the proof of Theorem~\ref{res-10-1} are interpreted
in the ``standard'' way:
A node executes (i.e. its finite state machine makes a move) if and only if it is not blocked
(waiting for input) and all the nodes with higher priorities are blocked.

A different priority scheme arises when we apply Theorem~\ref{res-10-1} recursively,
in a divide-and-conquer manner.

\begin{example}
    \label{res-10-2}
\rm
Consider the following communication graph.
\begin{center}
\begin{picture}(200,100)(-100,0)
\thicklines

\put(0,10){\circle{10}}
\put(0,90){\circle{10}}
\put(-80,90){\circle{10}}
\put(80,90){\circle{10}}

\put(-70,90){\vector(1,0){60}}
\put(70,90){\vector(-1,0){60}}
\put(0,80){\vector(0,-1){60}}
\put(10,20){\vector(1,1){60}}
\put(-10,20){\vector(-1,1){60}}

\put(-50,92){\makebox(20,10)[c]{$\alpha$}}
\put(30,92){\makebox(20,10)[c]{$\beta$}}
\put(4,45){\makebox(20,10)[l]{$\gamma$}}

\end{picture}
\end{center}
Every execution that begins and ends with empty channels can be reordered
so that $\alpha$, $\beta$ and $\gamma$ are frequently empty.
However, such a reordering cannot be achieved by the standard priority scheme.

Instead, we first apply Theorem~\ref{res-10-1} to the set $\{\{+\gamma\}\}$,
to make $\gamma$ frequently empty.
Then we restrict all subsequent reorderings to the remaining nodes of the graph;
we next apply~\ref{res-10-1} to the set $ \Psi = \{\{-\beta , +\beta \}\} $;
this makes $\alpha$ frequently empty.
Then, in the graph with the two nodes $-\beta$ and $+\beta$, we apply~\ref{res-10-1}
to $ \Psi = \{\{+\beta\}\}$, to make $\beta$ frequently empty.

Another way of describing the new execution is to say that the nodes are ordered
$+\gamma$, $-\gamma$, $-\beta$, $-\alpha$, from the highest to the lowest priority.
However, the priorities now have a different meaning.
In the standard scheme, the unblocked process with the highest priority executes.
In the present scheme, that unblocked process executes on which the process with
the highest priority is (directly or indirectly) blocked.
In our example, the priorities are as follows:
\begin{center}
\begin{picture}(200,110)(-110,-10)
\thicklines

\put(0,10){\circle{10}}
\put(0,90){\circle{10}}
\put(-80,90){\circle{10}}
\put(80,90){\circle{10}}

\put(-70,90){\vector(1,0){60}}
\put(70,90){\vector(-1,0){60}}
\put(0,80){\vector(0,-1){60}}
\put(10,20){\vector(1,1){60}}
\put(-10,20){\vector(-1,1){60}}

\put(-10,-10){\makebox(20,10)[c]{$1$}}
\put(-10,100){\makebox(20,10)[c]{$2$}}
\put(-80,100){\makebox(20,10)[l]{$4$}}
\put(80,100){\makebox(20,10)[l]{$3$}}

\end{picture}
\end{center}
If 2 is blocked on 4 and both 4 and 3 are unblocked, then 4 (not 3)
executes; in the standard scheme, 3 would execute. \\
(End of Example~\ref{res-10-2}.)
\end{example}

Clearly the priority schemes, as well as any other abstract flow control methods,
improve the efficiency of the exhaustive reachability analysis by reducing
the number of global states that the analysis must enumerate.
It is difficult to make any quantitative claims about the efficiency gains
because, as Brand and Zafiropulo~\cite{bib:bra} note when they evaluate two analysis
methods,
``in both approaches a protocol can be analyzed successfully only if its behavior is
far from the worst case, as is true for protocols designed in practice.''
However, in the context of the theory developed in this paper we can prove
qualitative claims about the {\em existence\/} of algorithms (rather than their {\em cost\/}).

We have already seen (in section~\ref{sec-8}) how a priority scheme can be used to
construct an algorithm to solve the deadlock problem for the cyclic protocols with the
rational channel property.
In the remainder of this section we shall see, on two examples, that the same can be
done for some other communication graphs.

\begin{theorem}
    \label{res-10-3}
The problem ``Is a given composite state stable?'' is algorithmically decidable for
the CFSM protocols with the rational channel property and the communication graph
\twographii.
\end{theorem}

\begin{theorem}
    \label{res-10-4}
The problem ``Is a given composite state stable?'' is algorithmically decidable for
the CFSM protocols with the rational channel property and the communication graph
in Fig.~10.2(a).
\end{theorem}

The same result can be proved for the graphs in Fig.~10.2(b), (c), (d) and other
similar ones.
On the other hand, it is not known (to the author) whether the stable composite state problem
is decidable for the CFSM protocols with the rational channel property and
the communication graphs in Fig.~10.3(a) and (b).

\begin{figure}
\begin{center}
\includegraphics[width=5in]{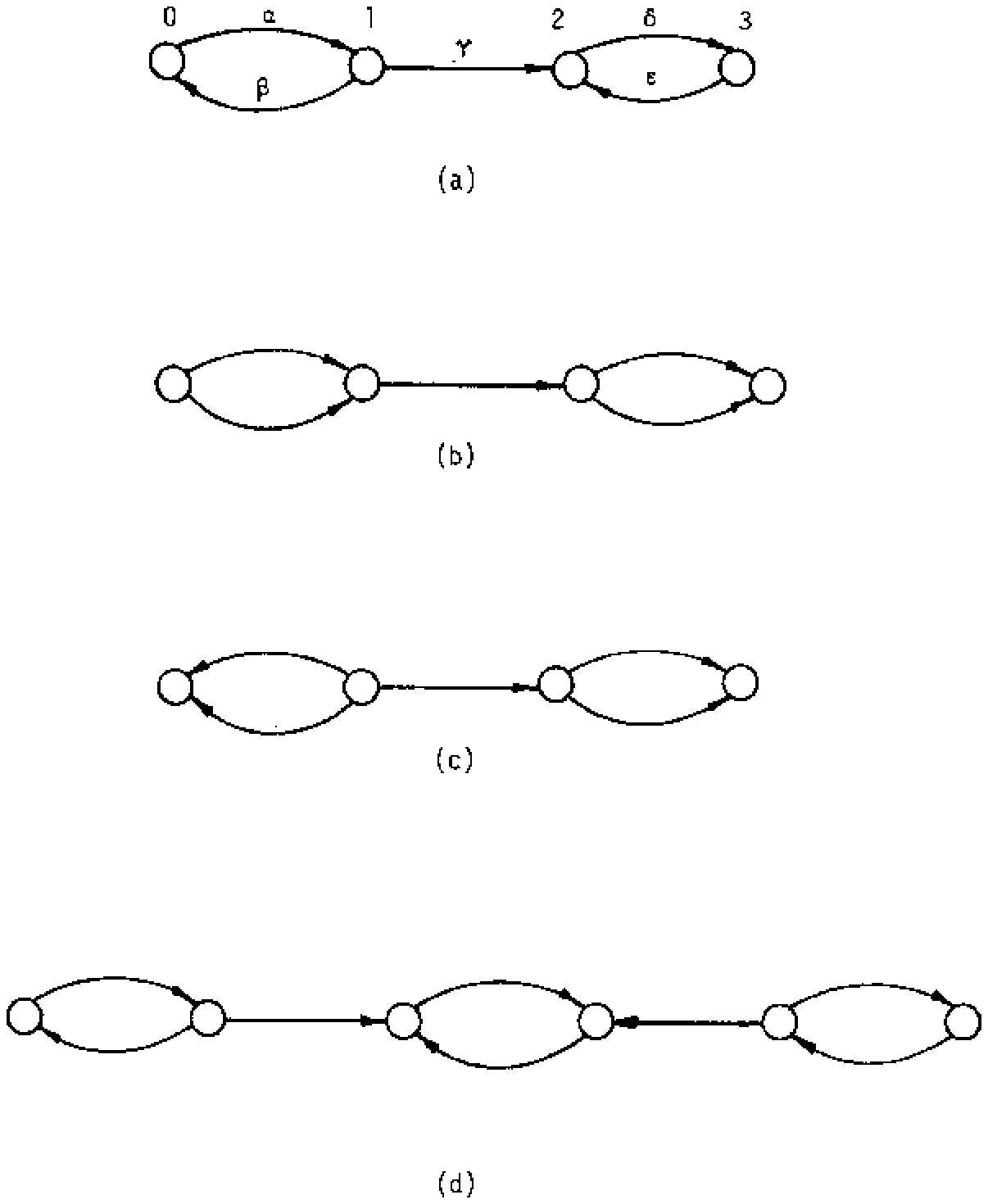}
\addspu\\[1cm]
{\bf Fig. 10.2.}
\end{center}
\end{figure}

\begin{figure}
\begin{center}
\includegraphics[width=3in]{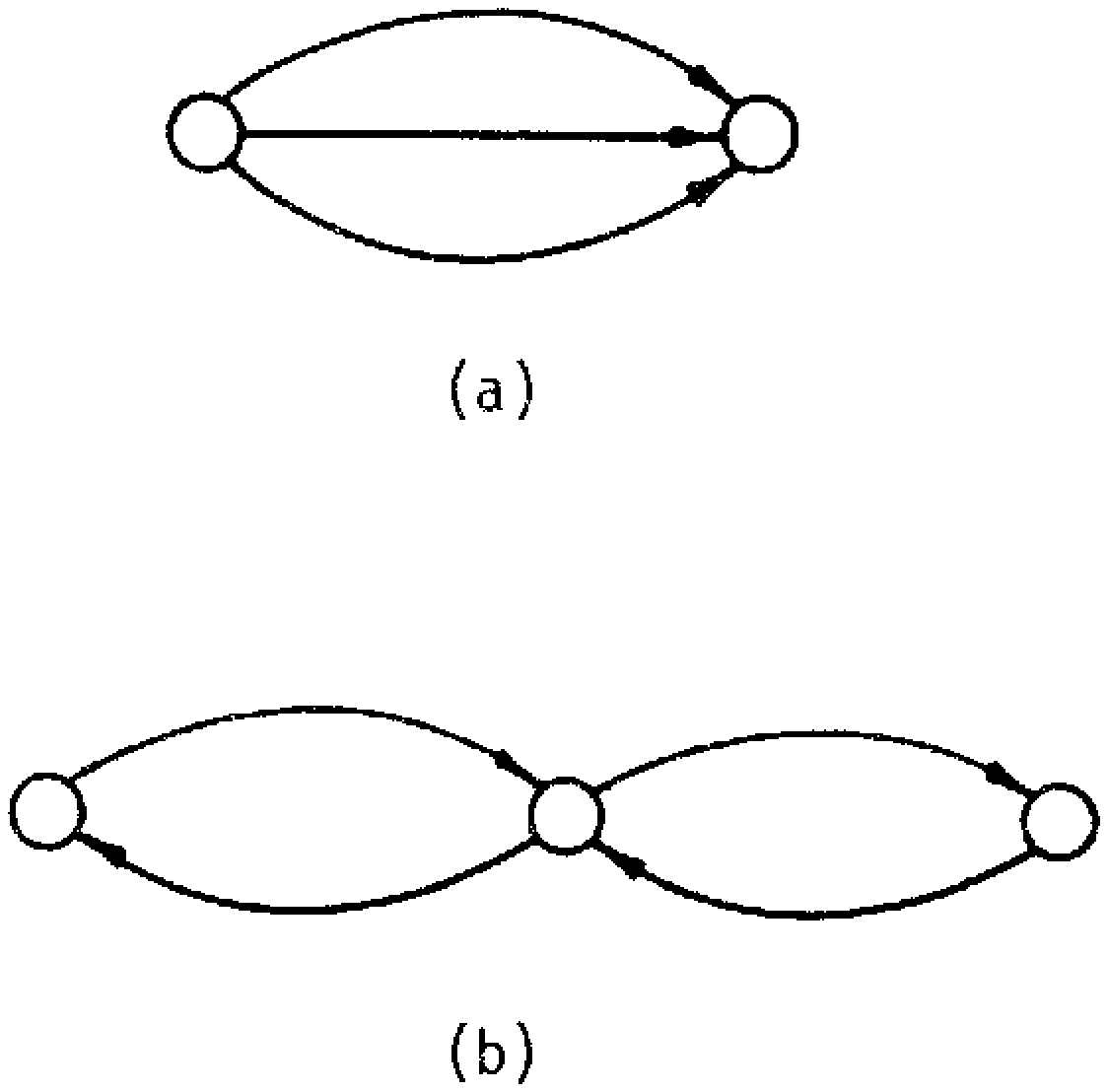}
\addspu\\
{\bf Fig. 10.3.}
\end{center}
\end{figure}

\begin{figure}
\begin{center}
\includegraphics[width=3in]{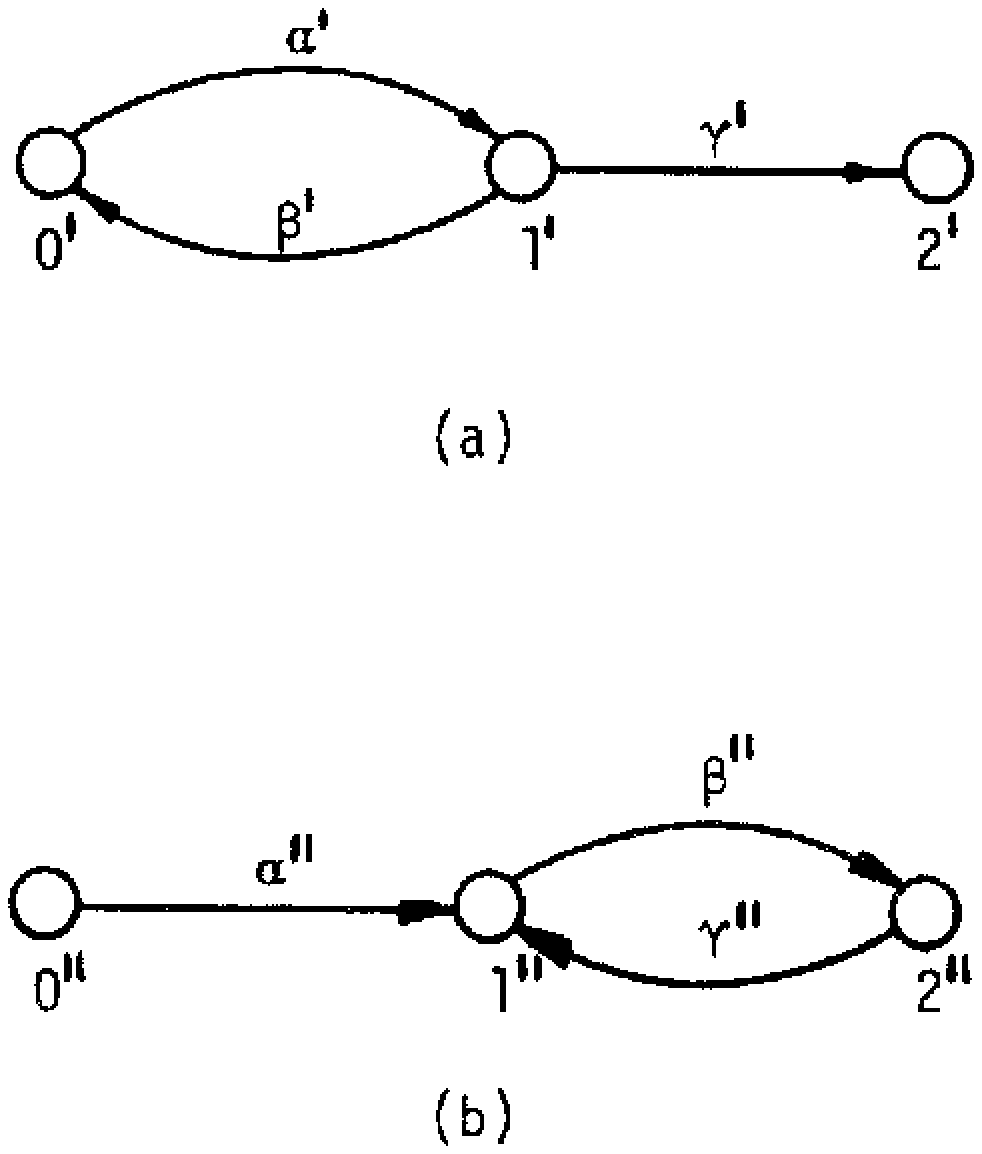}
\addspu\\
{\bf Fig. 10.4.}
\end{center}
\end{figure}

In the forthcoming proofs, we say that a family of relations $\RR(S)$,
$ S \inn \Timesa{j\in N} K_j $, is {\em consistent relative to a restriction\/}
if this condition holds:
if \Trans{\gl}{}{\gli}, $ C\inn\RR(S) $, and both \gl\ and \gli\ satisfy the restriction,
then $ C' \inn \RR(S') $.

\proof{ of~\ref{res-10-3}}
If a composite state is stable then its stability is verified by the exhaustive
reachability analysis.
Thus it suffices to construct a semialgorithm that verifies non-stability
and terminates whenever the composite state is not stable.
We show that there is an algorithmically verifiable proof of non-stability
for every non-stable composite state;
the semialgorithm then simply generates proof candidates until it finds
a correct one.

{\samepage

Choosing $ \Psi = \{\{ 1 \}\} $ in Theorem~\ref{res-10-1}, we  can restrict
our attention to the paths along which frequently $\alpha$ or $\beta$ is empty.
Thus for every non-stable composite state $S'$ there is a proof of non-stability
of $S'$, in the form of a family of relations $\RR(S)$, $ S\inn \Timesa{j\in N} K_j $,
that are consistent relative to the restriction
``$|x_\alpha | \leq 1$ or $|x_\beta | \leq 1$'' and such that
$ C^0 \inn \RR(S^0 ) $ and $ C^0 \not\in \RR(S') $.
The consistency is algorithmically verifiable when the relations are recognizable;
hence the result follows from this lemma:

}

\begin{lemma}
    \label{res-10-5}
If $\; \RR \subseteq M^*_\alpha \times M^*_\beta $ is a rational relation then
the relation
\[
\RR' = \{\; ( x_\alpha , x_\beta ) \inn \RR \; | \;
| x_\alpha | \leq 1 \;\mbox{\rm or }\; | x_\beta | \leq 1 \; \}
\]
is recognizable.
\end{lemma}

\proof{ of~\ref{res-10-5}}
For every $ x \inn M^*_\alpha $ the relation
\[
\RR^\beta (x) = \{\; (x_\alpha , x_\beta ) \inn \RR \; | \; x_\alpha = x \; \}
\]
is recognizable; similarly, for every $ y \inn M^*_\beta $ the relation
\[
\RR^\alpha (y) = \{\; (x_\alpha , x_\beta ) \inn \RR \; | \; x_\beta = y \; \}
\]
is recognizable.
Since the relation $\RR'$ is equal to
\[
\RR^\beta (\lambda) \;\cup\; \RR^\alpha (\lambda) \;\cup
\bigcup_{x_\alpha \in M_\alpha} \RR^\beta (x_\alpha ) \;\cup
\bigcup_{x_\beta \in M_\beta} \RR^\alpha (x_\beta ) \; ,
\]
it is recognizable.

This completes the proofs of~\ref{res-10-5} and~\ref{res-10-3}.
\qed

In the forthcoming proof of~\ref{res-10-4} we split the graph in Fig.~10.2(a)
into the two graphs in Fig.~10.4.
For the given CFSM protocol \prot\ (with the communication graph in Fig.~10.2(a))
and for an arbitrary deterministic (complete) finite automaton $F$ over the alphabet
$ M_\gamma\, $, we define two protocols $\prot' (F) $ and $\prot'' (F)$ as follows:
The protocol $ \prot'(F)$ has the communication graph of Fig.~10.4(a),
the finite state machines at the nodes $0'$ and $1'$ are the same as those
at 0 and 1 in \prot\ and the machine at $2'$ is $F$ (with every label in its
transition diagram prefixed by $+$).
The communication graph of $ \prot''(F) $ is as in Fig.~10.4(b), the finite
state machine at $0''$ is $F$ (with every label prefixed by $-$)
and the machines at $1''$ and $2''$ are the same as those at 2 and 3 in \prot.

{
\samepage

\begin{lemma}
    \label{res-10-6}
Let \prot\ be a CFSM protocol with the communication graph in Fig.~10.2(a),
and let $ ( p_0 , p_1 , p_2 , p_3 ) $ be a composite state of \prot.
Assume that there exist a deterministic finite automaton (over $M_\gamma$) and
a set $U$ of its states such that \\
(a) if $p$ is a state of $F$, $p \not\in U $, then $ ( p_0 , p_1 , p ) $ is not
stable for $ \prot'(F) $; and \\
(b) if $ p \inn U $ then $ ( p , p_2 , p_3 ) $ is not stable for $ \prot''(F)$. \\
Then $ ( p_0 , p_1 , p_2 , p_3 ) $ is not stable (for \prot).
\end{lemma}

}

\proof{ of~\ref{res-10-6}}
Suppose that $ S' = ( p_0 , p_1 , p_2 , p_3 ) $ is stable, i.e.
\TTrans{\glo}{(S' , C^0 )}.
We use higher priority for the nodes 0 and 1 to get two paths $\Gamma_0$ and $\Gamma_1$
and a channel content $ C' = \xchannel$ such that \\
(1) $ x_\xi = \lambda $ for $ \xi \neq \gamma $
(where $\gamma$ is the edge from 1 to 2 in Fig.~10.2(a)); \\
(2) $\Gamma_0$ leads from \glo\ to $ (( p_0 , p_1 , h_2 , h_3 ) , C' ) $; \\
(3) $\Gamma_1$ leads from $ (( p_0 , p_1 , h_2 , h_3 ) , C' ) $ to $(S' , C^0 )$; and \\
(4) the images \Imagei{2}{\Gamma_0}, \Imagei{3}{\Gamma_0}, \Imagei{0}{\Gamma_1},
\Imagei{1}{\Gamma_1} are all trivial paths.

Let $F$ be any deterministic finite automaton over $M_\gamma \,$,
and $U$ a set of its states.
Let $p$ be the state of $F$ to which $F$ moves from its initial state by reading $x_\gamma \,$.
The composite state $ ( p_0 , p_1 , p ) $ is stable for $ \prot'(F) $, and
$ ( p , p_2 , p_3 ) $ is stable for $ \prot''(F)$.
Thus (a) and (b) in~\ref{res-10-6} cannot be both true.
\qed

The crucial step in the proof of~\ref{res-10-4} is the following lemma,
which (together with~\ref{res-10-6}) shows that for every non-stable composite state
of \prot\ there is an algorithmically verifiable proof of its non-stability.

\begin{lemma}
    \label{res-10-7}
Let \prot\ be a CFSM protocol with the rational channel property and the communication
graph in Fig.~10.2(a).
If a composite state $ ( p_0 , p_1 , p_2 , p_3 ) $ of \prot\ is not stable then
there exist a deterministic finite automaton $F$ and a set $U$ of its states such that \\
(a) there is a family of recognizable relations $ \RR'(S') $ indexed by the composite
states $S'$ of the protocol $ \prot'(F) $, consistent relative to the restriction
``$\;| x_{\alpha'} | \leq 1$ and $| x_{\gamma'} | \leq 1$'', such that
$ C^0 \inn \RR'(S^0 ) $ and $ C^0 \not\in \RR' ( p_0 , p_1 , p )) $ for every state $p$
of $F$ not in $U$; \\
(b) there is a family of recognizable relations $ \RR''(S'') $ indexed by the composite
states $S''$ of the protocol $ \prot''(F) $, consistent relative to the restriction
``$\;| x_{\alpha''} | \leq 1$ and $| x_{\gamma''} | \leq 1$'', such that
$ C^0 \inn \RR''(S^0 ) $ and $ C^0 \not\in \RR'' ( p , p_2 , p_3 )) $ for every state $p\inn U$.
\end{lemma}

\proof{ of~\ref{res-10-7}}
Let $Q$ be the set $Q_\gamma (( p_0, p_1 , h_2 , h_3 )) $ of Theorem~\ref{res-8-3};
that is,
\[
Q \; = \; \{ \; x_\gamma \inn M^*_\gamma \; | \;
\TTrans{\glo}{(( p_0 , p_1 , h_2 , h_3 ) , \xchannel )}
\;\;\mbox{\rm and }\; x_\xi = \lambda \;\;\mbox{\rm for }\; \xi \neq \gamma \; \} .
\]
Since \prot\ has the rational channel property, $Q$ is regular.
There is a deterministic finite automaton $F$ to recognize $Q$;
let $U$ be the set of accepting states of $F$.
To define the relations $ \RR'((q_0 , q_1 , p))$ and $ \RR''((p, q_2 , q_3 )) $,
we use the relations $ \LL(S) $ of Definition~\ref{res-8-2}.
Denote $ h'_0 $ the initial state of $F$.

Define
\begin{eqnarray*}
\RR'(( q_0 , q_1 , p )) = & & \\
\{ \; ( x_{\alpha'} , x_{\beta'} , x_{\gamma'} ) & | &
\exists \, ( x_\alpha , x_\beta , x_\gamma , x_\delta , x_\varepsilon ) \inn
\LL (( q_0 , q_1 , h_2 , h_3 )) \;\;\exists \, y \inn M^*_\gamma \; : \\
& & x_{\alpha'} = x_\alpha \, , x_{\beta'} = x_\beta \, ,
x_\delta = x_\varepsilon = \lambda \, , | x_{\alpha'} | \leq 1 \, , | x_{\gamma'} | \leq 1 \, , \\
& &  \trans{h'_{0}}{y}{p} \;\;\;\mbox{\rm in }\; F
\;\mbox{\rm and }\; y x_{\gamma'} = x_\gamma \; \}
\end{eqnarray*}
for every composite state $ ( q_0 , q_1 , p ) $ of  $ \prot'(F) $.
If $ ( p , q_2 , q_3 ) $ is a composite state of $ \prot''(F) $ such that a state in $U$
can be reached from $p$ in $F$ then define
\begin{eqnarray*}
\RR''(( p , q_2 , q_3 )) = & & \\
\{ \; ( x_{\alpha''} , x_{\beta''} , x_{\gamma''} ) & | &
\forall \, y\inn M^*_\gamma \;\;\forall \, p' \inn U \; : \; \trans{p}{y}{p'} \; \Longrightarrow \\
& & \exists \, ( x_\alpha , x_\beta , x_\gamma , x_\delta , x_\varepsilon ) \inn
\LL (( p_0 , p_1 , q_2 , q_3 )) \; :  x_\alpha = x_\beta = \lambda \, , \\
& & x_{\beta''} = x_\delta \, , x_{\gamma''} = x_\varepsilon \, ,
| x_{\gamma''} | \leq 1 \, , | x_{\alpha''} | \leq 1 \, ,
\;\mbox{\rm and }\; x_{\alpha''} y = x_\gamma \; \} \, ,
\end{eqnarray*}
and if no state in $U$ can be reached from $p$, define
\[
\RR''(( p , q_2 , q_3 )) =
\{ \; ( x_{\alpha''} , x_{\beta''} , x_{\gamma''} ) \;\;\; | \;\;\;
 | x_{\alpha''} | \leq 1 \;\mbox{\rm and }\; | x_{\gamma''} | \leq 1 \; \} \; .
\]
Since \prot\ has the rational channel property, $ \LL(S) $ are rational,
and therefore $ \RR'(S') $ and $ \RR''(S'') $ are recognizable.

The consistency of $ \RR'(S') $ and $ \RR''(S'') $ follows from the consistency of $ \LL(S) $
and from the definition of $F$ and $U$.
It also follows from the definition of $F$ and $U$ that if
$ C^0 \inn \RR' ((p_0 , p_1 , p)) $ then  $ p \inn U $,
and that if $ C^0 \inn \RR'' (( p , p_ 2 , p_3 )) $ then $ p \not\in U $.
This completes the proof of~\ref{res-10-7}.
\qed

\proof{ of~\ref{res-10-4}}
As in the proof of~\ref{res-10-3}, it suffices to show that for every non-stable composite
state there is an algorithmically verifiable proof of its non-stability.
By~\ref{res-10-7} and~\ref{res-10-6} there is such a proof, consisting of $F$, $U$,
the family $\RR'(S') $ and the family $\RR''(S'')$.

Indeed, if $p$ is a state of $F$ not in $U$ then the family $ \RR'(S') $ is a proof
that $ ( p_0 , p_1 , p ) $ is not stable for $ \prot' (F) $
(by the priority argument applied to the graph in Fig.~10.4(a), every stable
composite state is reachable by a path along which frequently $\alpha'$ and $\gamma'$
are empty).
Similarly, the priority argument applied to the graph in Fig.~10.4(b) shows that
the family $ \RR''(S'') $ is a proof that $ ( p, p_2 , p_3 ) $ is not stable for
$ \prot''(F) $ whenever $ p \inn U $.
\qed


\section{Recapitulation and conclusions}
    \label{sec-11}

{\samepage

The theory of communicating finite state machines, or, more precisely,
of finite state machines connected by unbounded queues, is emerging as a valuable tool
for the specification and correctness analysis of communication protocols operating
over channels with indefinite delays.
Although the CFSM model is very simple, it is rich enough to encompass certain
basic protocol properties, which are expressed as reachability properties
in the global state space.

}

The reachability properties cannot be automatically verified in the class of all
CFSM protocols; in other words, the reachability problems are (algorithmically)
undecidable.
However, since the usefulness of the model is greatly enhanced by its amenability to
automated analysis, it is well worthwhile to look for classes of CFSM protocols
in which the problems are decidable.
Traditionally, the emphasis has been on the class of the protocols with the bounded
channel property.

The present paper advances our understanding of the question
``What makes the reachability problems in the CFSM theory undecidable?''
The paper contributes three new concepts to the theory:
Affinity of SR-machines, simple-channel properties, and abstract flow control.

The results about affine SR-machines point out close ties between the traditional automata
theory and the theory of CFSM protocols.
It is also shown (in section~\ref{sec-6}) that, although many
interesting properties of communicating SR-machines are undecidable,
some become decidable under additional restrictions
(affinity in this case).

Similarly, the results about simple-channel (recognizable channel and rational channel)
properties demonstrate how {\em some\/} protocols with unbounded channels can be
automatically analyzed, although the problems are undecidable for {\em general\/} protocols.
The simple-channel restrictions formally express the observation that common protocols
do not make use of the full generality of the CFSM model.
``Protocols with unbounded channels usually use them in a simple manner,
which makes them worth considering'' (\cite{bib:bra},~p.~10).
The results in this paper suggest a new formalism for protocol description
(CFSM augmented with channel expressions) together with algorithms for automated
analysis of the protocols so described.

It should be pointed out that a proof of, say, deadlock-freedom in the form of a table
of recognizable relations can be potentially advantageous even for a protocol with the
bounded channel property.
Indeed, it can happen that the reachable global states are separated from the deadlocked ones
by a consistent family of recognizable relations that are described by short expressions,
while at the same time the complete list of all reachable global states is very large.

The theory of ``recognizable proofs'' (i.e. proofs based on recognizable relations)
is all ready for use; the theory of ``rational proofs'', on the other hand,
is not well understood.
The key open question is whether reachability problems are algorithmically decidable
for protocols with the rational channel property.
The problem is answered in the affirmative for cyclic protocols in section~\ref{sec-8},
and for several other simple communication graphs in section~\ref{sec-10}.

The aim of the abstract flow control, as defined and studied in this paper,
is to limit the redundancy in the global state space, thereby improving the efficiency
of the algorithms that decide the reachability properties.
Abstract flow control methods should exploit the topology of the communication graph,
as do the two priority schemes proposed in section~\ref{sec-10}.

In section~\ref{sec-10} it is shown how the priority schemes lead to qualitative gains:
They allow us to construct algorithms for solving reachability problems for the rational-channel
CFSM protocols with some communication graphs.
Abstract flow control methods yield quantitative gains as well, but these are difficult
to estimate in any meaningful way for general protocols.
Perhaps a fruitful approach would be to study {\em algorithms\/} for finding optimal abstract
flow control methods, or, for the sake of concreteness, optimal priority assignments.
For example, one can formulate the optimization problem of finding
(for an arbitrary communication graph)
the priority assignment that minimizes a cost function, which measures the number of
``needlessly reachable'' global states.
But that, as Kipling says, is another story.

{\bf Acknowledgement.}
The work reported here is a part of my Ph.D. thesis supervised by K. Culik II.
I wish to thank him for encouragement and many fruitful discussions.

\newpage
\appendix

\noindent
\section{Appendix: Post's tag systems}

\vspace{.5cm}
The tag systems are first mentioned by Post~\cite{bib:pos} as a source of possibly
undecidable problems.
The undecidability is actually proved by Minsky (\cite{bib:mi1}, \cite{bib:mi2}).

A {\em tag system\/} is a 3-tuple $\TT = \tags$ where $\Sigma$ is a finite alphabet,
$g$ is a function from $\Sigma$ to $\Sigma^*$ and $ w_0 \inn \Sigma^*$.
Define
\begin{eqnarray*}
| g |^- & = & \min \;\{ \; |g(b)| \;\; | \;\; b \inn \Sigma \; \} \; ,\\
| g |^+ & = & \max \;\{ \; |g(b)| \;\; | \;\; b \inn \Sigma \; \} \; .
\end{eqnarray*}
For every positive integer ({\em deletion number\/}), the tag system defines
a function from $\Sigma^*$ to $\Sigma^*$;
in what follows we only consider the function corresponding to the deletion number 2.
The function, denoted $f_\TT$, is defined by

(a) if $ |w| \leq 1 $ then $ f_\TT (w) = \lambda $; and

(b) if $ w = b_0 b_1 \ldots b_n $, $ n\geq 1 $, then
$ f_\TT (w) = b_2 \ldots b_n g(b_0 ) $.

\noindent
The {\em sequence of\/} $\TT$, denoted $ \{ s_n (\TT )\}_{n=0}^\infty $, is defined by
$ s_0 (\TT ) = w_0 $, and $ s_{n+1} ( \TT ) = f_\TT ( s_n (\TT )) $, $ n\geq 0 $.

\begin{theorem}
    \label{res-a-1}
There is no algorithm to decide, for every tag system $ \TT = \tags $
with $ |g|^- = 1 $ and $ |g|^+ = 3 $,
whether $ s_n (\TT) = \lambda $ for some $n$.
\end{theorem}

\proofnodot:
See Theorem~5 in~\cite{bib:wan}.

\begin{theorem}
    \label{res-a-2}
There is no algorithm to decide, for every tag system $ \TT = \tags $
with $ |g|^- = 1 $ and $ |g|^+ = 3 $,
whether $ | s_n ( \TT ) | \leq c $ for some constant $c$ and every $n$.
\end{theorem}

\proof{}
If there were such an algorithm, we could construct an algorithm to decide the problem
$ s_n ( \TT ) = \lambda $ of Theorem~\ref{res-a-1} as follows:
For a given $\TT$, first decide whether $ s_n ( \TT ) \leq c $ for some $c$ and all $n$.
If this is not the case then $ s_n ( \TT ) \neq \lambda $ for all $n$.
If, on the other hand, the sequence of $\TT$ is bounded then generate the successive
strings $ s_n ( \TT ) $ until $ s_{m_0} ( \TT ) = s_{m_1} ( \TT ) $
for some $m_0$ and $m_1$, $ m_0 \neq m_1 $;
now if $ s_{m_0} ( \TT ) = \lambda $ then the problem is decided, and
if $ s_{m_0} ( \TT ) \neq \lambda $ then $ s_n ( \TT ) \neq \lambda $ for all $n$.
\qed

\begin{theorem}
There is no algorithm to decide, for every tag system $ \TT = \tags $ such that
$ |g|^- = 1 $, $ |g|^+ = 3 $ and $ s_n ( \TT ) \neq \lambda $ for all $n$,
whether $ | s_n ( \TT ) | \leq c $ for some constant $c$ and every $n$.
\end{theorem}

{\samepage

\proof{}
For every tag system $ \TT = \tags $ choose a symbol $ \# \not\in \Sigma $ and define
\begin{eqnarray*}
\Sigma' & = & \Sigma \cup \{ \# \} \, , \\
w'_0 & = & w_0 \# \# \, , \\
g' (b) & = & g(b) \;\;\;\mbox{\rm for }\;\; b \inn \Sigma \, , \\
g'( \# ) & = & \# \# \, .
\end{eqnarray*}
}
$\!\!$Then the tag system $\TT' = ( \Sigma' , g' , w'_0 ) $ is bounded
(i.e. $ | s_n ( \TT' ) | \leq c $ for some $c$ and all~$n$)
if and only if $\,\TT$ is.
Moreover, $ s_n ( \TT' ) \neq \lambda $ for all $n$, because
every $ s_n ( \TT' ) $ contains the subsequence $ \# \# $.

Thus if we had an algorithm to decide the boundedness for every
$ \TT' = ( \Sigma' , g' , w'_0 ) $ such that $ |g|^- = 1 $, $ |g|^+ = 3 $
and $ s_n ( \TT' ) \neq \lambda $ for all $n$,
then we would also have an algorithm to decide boundedness for every
$ \TT = \tags $ such that $ |g|^- = 1 $ and $ |g|^+ = 3 $,
in contradiction to~\ref{res-a-2}.
\qed




\newpage

\begin{thebibliography}{a}
\bibitem[Ber]{bib:ber}
    J. Berstel:
    Transductions and context-free languages,
    B. G. Teubner Stuttgart (1979).
\bibitem[Bir]{bib:bir}
    M. Bird:
    The equivalence problem for deterministic two-tape automata,
    J. Comput. System Sci. 7 (1973) 218-236.
\bibitem[Bo1]{bib:bo1}
    G. V. Bochmann and C. Sunshine:
    Formal methods in communication protocol design,
    I.E.E.E. Trans. Comm. COM-28 (1980), 624-631.
\bibitem[Bo2]{bib:bo2}
    G. V. Bochmann:
    A general transition model for protocols and communication services,
    I.E.E.E. Trans. Comm. COM-28 (1980), 643-650.
\bibitem[Bra]{bib:bra}
    D. Brand and P. Zafiropulo:
    On communicating finite state machines,
    IBM RZ 1053 (1981).
\bibitem[Eil]{bib:eil}
    S. Eilenberg:
    Automata, languages and machines,
    Vol. A, Academic Press (1974).
\bibitem[Gou]{bib:gou}
    M. G. Gouda:
    Protocol machines -- towards a logical theory of communication protocols,
    Univ. of Waterloo Ph. D. thesis (1977).
\bibitem[Hop]{bib:hop}
    J. E. Hopcroft and J. D. Ullman:
    Introduction to automata theory, languages and computation,
    Addison-Wesley (1979).
\bibitem[Kan]{bib:kan}
    R. Kannan and R. J. Lipton:
    The orbit problem is decidable,
    Proc. 12th Annual ACM Symp. on Theory of Computing (1980), 252-261.
\bibitem[Luc]{bib:luc}
    C. L. Lucchesi and D. H. Younger:
    A minimax theorem for directed graphs,
    J. London Math. Soc. (2) 17 (1978), 369-374.
\bibitem[Man]{bib:man}
    Z. Manna and R. Waldinger:
    The logic of computer programming,
    IEEE Trans. on Software Engineering, SE-4 (1978), 199-229.
\bibitem[May]{bib:may}
    E. Mayr:
    An algorithm for the general Petri net reachability problem,
    Proc. 13th Annual ACM Symp. on Theory of Computing (1981), 238-246.
\bibitem[Mi1]{bib:mi1}
    M. Minsky:
    Recursive unsolvability of Post's problem of ``tag'' and other
    topics in theory of Turing machines,
    Ann. Math. 74 (1961), 437-455.
\bibitem[Mi2]{bib:mi2}
    M. Minsky:
    Computation -- finite and infinite machines,
    Prentice-Hall (1967).
\bibitem[Pos]{bib:pos}
    E. Post:
    Formal reduction of the combinatorial decision problems,
    Amer. J. Math. 65 (1943), 196-215.
\bibitem[Rub]{bib:rub}
    J. Rubin and C. H. West:
    An improved protocol validation technique,
    IBM RZ 1024 (1980).
\bibitem[Wan]{bib:wan}
    H. Wang:
    Tag systems and lag systems,
    Math. Annalen 152 (1963), 65-74.
\bibitem[Zaf]{bib:zaf}
    P. Zafiropulo, C. H. West, H. Rudin, D. D. Cowan and D. Brand:
    Towards analyzing and synthesizing protocols,
    I.E.E.E. Trans. Comm. COM-28 (1980), 651-661.
\end{thebibliography}
\end{document}